\documentclass[preprint,floats,eqsecnum,aps,tightenlines,showpacs]{revtex4}
\usepackage{epsfig}
 
\begin{document}

\preprint{
 \parbox{1.5in}{\leftline{JLAB-THY-02-14}
                 \leftline{WM-02-104}
                 %\leftline{nucl-th/02????} 
}  }  
\title{\bf \ \\ Covariant description of inelastic electron--deuteron
scattering:\\ predictions of the relativistic impulse
approximation}
\author{
 J.~Adam, Jr.$^{1,2}$, Franz~Gross$^{1,3}$,
Sabine~Jeschonnek$^{1,4}$,
  Paul Ulmer$^{5}$ and  J.~W.~Van Orden$^{1,5}$ }
\address{
$^1$Jefferson Lab,
12000 Jefferson Avenue, Newport News, VA 23606\\
$^2$ Nuclear Physics Institute,  {\v R}e{\v z} near Prague,
CZ-25068,  Czech Republic \\
$^3$ Department of Physics,
College of William and Mary, Williamsburg, VA 23185 \\
$^4$The Ohio
State University, Physics Department, Lima, OH 45804\\
$^5$Department of Physics, Old Dominion University, Norfolk,
VA 23529}

\date{\today}

\begin{abstract}
Using the covariant spectator theory and the transversity formalism,
the unpolarized, coincidence cross section for deuteron
electrodisintegration, $d(e,e'p)n$, is studied.  The relativistic
kinematics are reviewed, and simple theoretical formulae
for  the relativistic impulse approximation (RIA) are derived and
discussed. Numerical predictions for the scattering in the high
$Q^2$ region obtained from the RIA and five other  approximations 
are presented and compared.  We conclude that  measurements of the
unpolarized coincidence cross section and the asymmetry $A_\phi$, 
to an accuracy that will distinguish between different theoretical
models, is feasible over most of the wide kinematic range accessible
at Jefferson Lab.    

\end{abstract}

\pacs{21.45.+v,25.10.+s,25.30.Fj}

%\keywords{some keywords}

%\maketitle 

%%%%%%%%%%%%%%%%%%%%%%%%%%%%%%%%%%%%%%%

\phantom{0}
\vspace{5.6in}
\section{Introduction}

Inelastic scattering of electrons from the deuteron is an 
important  source of information about the  nuclear current,
deuteron structure, and the $NN$ force. The exclusive
scattering cross section, $d(e,e'p)n$, was first measured  
almost forty years ago \cite{first}, and since then it has been
measured  under a wide variety of kinematic conditions
\cite{coincidence}.  There is a  substantial body of data for 
this  reaction, including  cross section measurements
\cite{bernheim,turck,breuker,blomqvist} as well as separations  of
various response functions
\cite{vdschaarlt,tamae,vdschaarlandt,frommberger,ducret,bulten,
jordan,pellegrino,kasdorp,oops} which differentiate between absorption
of longitudinal and transverse photons.

In this paper we survey results that might be expected from a 
new generation of $d(e,e'p)n$ coincidence measurements proposed
for Jefferson Laboratory (JLab).   At JLab it is possible 
to carry out a comprehensive program of measurements at both
high $Q^2$ and large $W$ (where $W$ is the invariant mass of
the final $np$ state).  A broad program of such measurements
offers the best hope of independently determining effects of
final state interactions and  the nuclear current, permitting
the extraction of important new information about the short
range $NN$ interaction.

\vspace{-9in}
\maketitle 

Electrodisintegration of the deuteron has been studied theoretically
by many groups.   Recently, Arenh\"ovel, Beck, and Wilbois \cite{Beck}
have emphasized that the relativistic effects in inelastic
scattering can be very large, even at modest momenta, and it is
therefore particularly important to have a fully relativistic theory
available for the analysis of the higher momenta data that will be
measured at JLab.   Relativistic calculations of this reaction
date back to the early work of Durand \cite{durand} and McGee
\cite{mcgee} and lead up to more recent work by Tjon \cite{T92}.  One
of the goals of this paper is to present a fully modern, covariant
treatment of this process suitable for the analysis of JLab data.

This paper imbeds the dynamical calculation in the general formalism
developed in Ref.~\cite{dg89}, where a covariant, systematic treatment
of most of the polarization observables that can be measured in the
$d(e,e'p)n$ reaction were classified and defined.  There it was found
that the use of transversity amplitudes (closely related to helicity
amplitudes) gave a very efficient description of the reaction. 
Transversity amplitudes were discussed by Moravcsik \cite{morav}, who
found that they build in the constraints imposed by parity and
rotational invariance in the most efficient way.  In a transversity
basis, the constraints imposed by these symmetries insure that half
of the possible amplitudes vanish identically, so that the
cumbersome linear relations needed in other formalisms
\cite{ALT} are unnecessary.  This economy will be
essential some time in the future when large data sets exist, and it
may be important to know whether or not a proposed new measurement
will really be independent of amplitudes already measured.         
 
The details of the calculation are carried out using the covariant
spectator theory, which has been successfully applied to the
description of $NN$ scattering \cite{GVOH92} and the electromagnetic
form factors of the deuteron \cite{dff}.  One feature of this theory is
that the deuteron bound state is described by the covariant $dnp$
vertex with one nucleon on mass-shell, and this is precisely the
amplitude that is needed for the relativistic impulse approximation
(RIA), making the theory well suited to the analysis of the $d(e,e'p)n$
coincidence reaction
\cite{conference}.  In this first application of the covariant
spectator theory using the transversity basis, we present the RIA
calculation only.  This  provides the opportunity to work out several
new technical details for the simplest case, and to compare to other
approximations.  The inclusion of final state interactions and
interaction currents will be the subject of future work.

A second purpose of this paper is to estimate the size of the
unpolarized $d(e,e'p)n$ cross sections expected over the broad range
of $Q^2$ and $W$ accessible to JLab.  In preparing this survey we
found that relativistic and nonrelativistic predictions for
$d(e,e'p)n$ at high $Q^2$, where the cross section is most sensitive to
the theory, often differ by as much as an order of magnitude,
extending the observations of Arenh\"ovel, Beck, and
Wilbois \cite{Beck}.  Since nonrelativistic calculations cannot
be taken seriously at such high energies, we report our
results for a variety of relativistic or semi-relativistic
models only.  In this first exploratory study the
goal is to provide only a rough survey of the landscape.  The  
simplicity of the RIA allows a uniform treatment over the
entire kinematic range, but is, of course, very incomplete.  It is our
intention to follow up this study with complete calculations for
cases where the theoretical effects look especially interesting. 

Our notation for the cross section and the spectator theory for
the RIA are reviewed briefly in  Sec.~II, numerical results
presented in Sec.~III, and conclusions given in Sec.~IV.  Many
theoretical details are given in the several Appendices, which
are an important part of this paper.

\section{Theory}

In this section we define the coincidence cross section and 
the RIA matrix element.  All other theoretical
details can be found in the Appendices.  

\subsection{The Cross Section}    

%%%%%%%%%%%%%%%%%%%%%%%%%%
% scattering diagram  
\begin{figure} 
\centerline{\epsfysize=3.0in \epsfbox{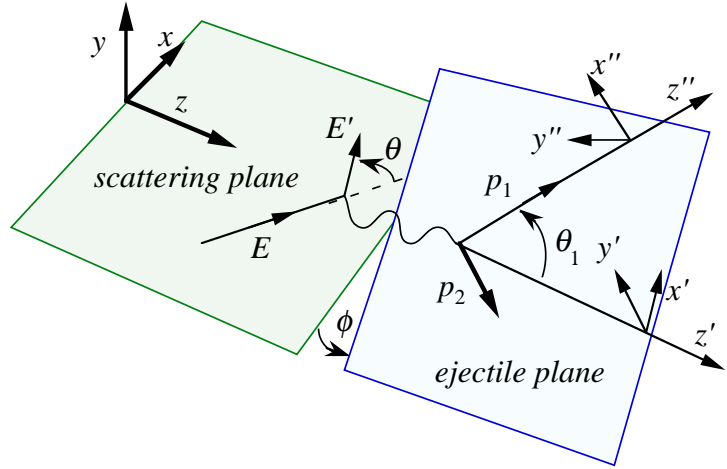}}
%\vspace*{0.2in}
\caption{\footnotesize\baselineskip=12pt The kinematics of electron
scattering when the final hadronic state  is broken into two fragments
with momenta $p_1$ and $p_2$.}
\label{fig_cross}
\end{figure}
%%%%%%%%%%%%%%%%%%%%%%%%%%%

Figure~\ref{fig_cross} shows the kinematics for the process  
$e+d\to e'+p+n$  (using the notation of Ref.~\cite{dg89}). 
The incident and scattered electron momenta form a  plane
called the ``scattering plane'' while the momenta of the
proton  and neutron  in the final state form a second plane
called the ``ejectile plane.''  The virtual photon momentum is
common to the two planes and is chosen as the direction of the
$z$-axis. The two planes, which are represented by the
$(x,y,z)$ and $(x',y',z')$  coordinate systems, are oriented at a
relative azimuthal angle of $\phi$. A  rotation of
the response tensor (defined below) from the unprimed to the
primed frame can be  used to extract all of the $\phi$
dependence from the tensor. Using this along with the explicit
form of the electron tensor, the cross section can be shown to
be of the form (c.f.~Eq.~(95) of Ref.~\cite{dg89}) 
\begin{eqnarray}
{d^5\sigma\over d\Omega^{\prime}dE^{\prime}d\Sigma} &=& 
{\sigma_M\over 4\pi  M_d}\frac{Q^2}{{q}_{_L}^2}
\biggl\{\tilde{R}^{({\rm I})}_L +   s_T\tilde{R}^{({\rm I})}_T 
-\frac{1}{2}\left[\cos 2\phi \tilde{R}^{({\rm I})}_{TT}
+ \sin 2\phi 
\tilde{R}^{({\rm II})}_{TT}\right] \nonumber\\
&& + s_{LT}\left[\cos\phi \tilde{R}^{({\rm I})}_{LT} + 
\sin\phi\tilde{R}^{({\rm II})}_{LT}\right] + 2 h \;  
s_{T'}\tilde{R}^{({\rm II})}_{T^{\prime}} \nonumber\\
&& + 2h\;s_{LT'}\left[\sin\phi\tilde{R}^{({\rm I})}_{LT^{\prime}}
+\cos\phi\tilde{R}^{({\rm II})}_{LT^{\prime}}\right]\biggl\}
\label{crosssec}
\end{eqnarray}
where
\begin{equation}
\sigma_M = \left[
\frac{\alpha\cos\frac{1}{2}\theta}{2E\sin^2\frac{1}{2}\theta} 
\right]^2
\end{equation}
is the Mott cross section and $d\Sigma$ is defined below.  The
quantities
$\theta$,
$E$,
$E'$, and 
$\Omega'$ are the electron scattering angle, the energies of the
initial and final electron, and the solid angle of the
final electron, all in the lab frame. The deuteron mass
is $M_d$ and $h=\pm1/2$ is the helicity of the incident  electron. The
electron kinematical factors are 
\begin{eqnarray}
s_T&=\frac{1}{2}+\xi^2 \qquad\qquad s_{LT}=&-\frac{1}{\sqrt{2}}\left(
1+\xi^2\right)^\frac{1}{2}
\nonumber\\ s_{T'}&=\xi\left( 1+\xi^2\right)^\frac{1}{2}\qquad 
s_{LT'}=&-\frac{1}{\sqrt{2}}\xi
\end{eqnarray}
where 
\begin{equation}
\xi=\frac{{q}_{_L}}{Q}\tan\frac{\theta}{2}
\end{equation}
with 
\begin{equation}
q^2=-Q^2= \nu^2-{q}_{_L}^2= \nu_0^2-{q}_0^2 \label{4q}
\end{equation}
the square of the virtual photon four-momentum, with 
$\{\nu,{\bf q}_L\}$ and $\{\nu_0,{\bf q}_0\}$ the energy and
three-momentum of the virtual photon in the lab and c.m. systems,
respectively, and $q_0=|{\bf q}_0|$, etc. 

There are two inertial reference frames that are of interest in the 
calculation of deuteron electrodisintegration: the laboratory 
frame which coincides with  the rest frame of the target
deuteron, and the  center of momentum  (c.m.) frame in which
the total three-momentum of the final state proton-neutron pair
(or of the initial virtual photon and the target deuteron) is
zero.   One of the virtues of Eq.~(\ref{crosssec}) is that the
response  functions
$\tilde{R}$ are {\it covariant\/}, and hence (\ref{crosssec}) 
can be used to describe the cross section in either the
c.m.~of the outgoing $np$ pair or the laboratory frame by the
replacement of
$d\Sigma$ by
\begin{equation}
d\Sigma \vert_{cm}  = {\rm p}~d\Omega^*
\end{equation}
in the c.m. frame or 
\begin{equation}
d\Sigma \vert_{lab}  = {\rm p}_{1}~d\Omega_{1}~{\cal R}
\end{equation}
in the laboratory frame.  [Except for special notation used in
Eq.~(\ref{4q}), we use a roman character for the magnitude of a
three-momentum, so that p$_1$=$|{\bf p}_1|$, to distinguish it from
the corresponding four-momentum, $p_1$.]  The factor
\begin{equation}
{\cal R} = {W\over  M_d}\quad \frac{1}{\left( 1 + 
\frac{\textstyle\nu {\rm p}_1 - E_1 q \cos\theta_1}{\textstyle M_d
{\rm p}_1} 
\right)_L}
\label{defrecoil}
\end{equation}
is the recoil factor, where $W$ is the invariant mass of the 
outgoing pair and the subscript $L$ means that each variable 
in the parentheses is to be replaced by its value in the lab
frame (for  example, $q\to q_{_L}$ and ${\rm p}_1$ is a function of
$\theta_1$, the angle between the outgoing proton and
the $\hat z$ axis in the lab system).  We will sometimes use
an asterisk ($^*$) to denote a variable in the c.m.~system.  Notation
for some of the most important variables is summarized in
Table~\ref{tab:variables}.

%%%%%%%%%%%%%%%%%%%%%%%%%%%%
\begin{table}
\begin{minipage}{4in}
\caption{\label{tab:variables}Notation for frequently used
variables. }
\begin{ruledtabular}
\begin{tabular}{lll}
variable & Lab & c.m.\\
\tableline
photon energy & $\nu$ & $\nu_0$ \\
magnitude of photon 3-momentum & ${ q}_{_L}$ & ${ q}_0$ \\
deuteron 4-momentum & $P$ & $P^*$\\
deuteron energy & $M_d$ & $D_0$ \\
proton 4-momentum & $p_1$ & $p^*_1$\\
neutron 4-momentum & $p_2$ & $p^*_2$\\
proton angle & $\theta_1$ & ${\theta}^*$\\
neutron angle & $\theta_2$ & ${\theta}^*+\pi$\\
magnitude of proton 3-momentum & p$_{1}$ & p \\ 
magnitude of neutron 3-momentum & p$_2$ & p \\
\end{tabular}
\end{ruledtabular}
\end{minipage}
\end{table}
%%%%%%%%%%%%%%%%%%%%%%%%%%%% 

The c.m. frame (referred to as the ``antilab'' frame in
Ref.~\cite{Beck}) is of interest for  theoretical reasons because
integrating over the final state kinematical variables is
particularly convenient in this frame and the partial wave expansion
of the final state is normally carried out in this frame.  While this
partial wave expansion is particularly convenient at low and medium
energies of a few hundred MeV, we would like to point out that the 
partial wave approach becomes extremely tedious and/or impractical at
GeV energies.  At such energies Glauber theory \cite{Glauber}, or the
new so-called ``three-dimensional'' methods of calculating the
$NN$ amplitude directly without partial wave expansions
\cite{nopartial}, are better. In any case, since the final scattering
state is by far the most complicated ingredient in the calculation of
the transition matrix elements, it is important to be able to carry
out calculations in this frame and to translate them to the lab
frame. The necessity of boosting the calculation from the c.m.~frame
to the lab frame requires that the  Lorentz properties of the matrix
elements be understood. This goal is conveniently accomplished by
using the Jacob and Wick helicity formalism \cite{jw} provided that
it can be shown that the various ingredients in the calculation of
the matrix elements, such as the wave functions, are covariant. We
will  assume for the moment that this is the case and will
demonstrate later that it is true for the particular calculations
which are described in this paper.

The nine response functions of (\ref{crosssec}) are related to 
sums over the squares of matrix elements of the deuteron
current.  In the helicity basis, with $\lambda_\gamma$ the
helicity of the virtual photon, $\lambda_1$ and $\lambda_2$
the helicities of particles 1 and 2 in the final state, and
$\lambda_d$ the helicity of the initial deuteron, the current
operator is written
$\left<\lambda_1\lambda_2\left|J_{\lambda_\gamma}(q)
\right|\lambda_d\right>$. Following the conventions of Jacob 
and Wick
\cite{jw} we choose  particle 1 in the final state to be the proton
and particle 2 to be the neutron.  The current operator conserves
parity, which means that the matrix elements satisfy the condition
\begin{equation}
\left<\lambda_1\lambda_2\left|J_{\lambda_\gamma}(q)
\right|\lambda_d\right> = \pm
\left<-\!\lambda_1,-\!\lambda_2\left|J_{-\lambda_\gamma}(q)
\right|-\!\lambda_d\right>\, ,
\end{equation}
where the phase depends on the helicities (see Ref~\cite{dg89}).  For
this reason it is convenient to introduce symmetric and
antisymmetric combinations of the $|\lambda_\gamma|=1$ amplitudes
\begin{eqnarray}
J^{\lambda_d}_{s\,\lambda_1\,\lambda_2}(p_1,p_2,q)\equiv
\left<\lambda_1\lambda_2\left|J_s(q)
\right|\lambda_d\right> &=& {1\over2}\Bigl\{
\left<\lambda_1\lambda_2\left|J_{1}(q)\right|\lambda_d\right>-
\left<\lambda_1\lambda_2\left|J_{-1}(q)\right|\lambda_d\right>\Big\}
\nonumber\\
J^{\lambda_d}_{a\,\lambda_1\,\lambda_2}(p_1,p_2,q)\equiv
\left<\lambda_1\lambda_2\left|J_a(q)
\right|\lambda_d\right> &=& {1\over2}\Bigl\{
\left<\lambda_1\lambda_2\left|J_{1}(q)\right|\lambda_d\right>+
\left<\lambda_1\lambda_2\left|J_{-1}(q)\right|\lambda_d\right>\Big\}
\, ,\qquad \label{gamps}
\end{eqnarray}
where, because of the phases, $J_s$ is {\it symmetric\/} under 
the $Y$ parity transformation (parity followed by rotation by
$\pi$ about the $y$ axis) and $J_a$ is {\it antisymmetric\/}
(and we note for future reference that
$J^{\lambda_d}_{0\,\lambda_1\,\lambda_2}\equiv
\left<\lambda_1\lambda_2\left|J_0(q)
\right|\lambda_d\right>$ is also {\it symmetric\/}). 
We then define the deuteron response tensors
$R^{({\rm I})}_{gg'}$ and $R^{({\rm II})}_{gg'}$
\begin{eqnarray}
R^{({\rm I})}_{gg'}&=&\frac{m^2}{2\pi^2W}\sum_{{\tiny\begin{array}{c}
\lambda'_1\lambda_1\lambda_2\cr \lambda'_d
\lambda_d\end{array}}}\sum_{\rho=\pm}
\Bigg\{(\rho_N^\rho)_{\lambda'_1\lambda_1}\,
J^{\lambda_d}_{g\,\lambda_1\lambda_2}(p_1,p_2,q)\,
(\rho^\rho_D)_{\lambda_d\lambda'_d}
\,J^{\dagger\;\lambda'_d}_{g'\,\lambda'_1\lambda_2}(p_1,p_2,q)
\Bigg\}
\nonumber\\
R^{({\rm II})}_{gg'}&=&\frac{m^2}{2\pi^2W}\sum_{{\tiny\begin{array}{c}
\lambda'_1\lambda_1\lambda_2\cr \lambda'_d
\lambda_d\rho\end{array}}} \sum_{\rho=\pm}
\Bigg\{(\rho_N^\rho)_{\lambda'_1\lambda_1}
\,J^{\lambda_d}_{g\,\lambda_1\lambda_2}(p_1,p_2,q)\,
(\rho^{(-\rho)}_D)_{\lambda_d\lambda'_d}
\,J^{\dagger\;\lambda'_d}_{g'\,\lambda'_1\lambda_2}(p_1,p_2,q)
\Bigg\}\, ,\qquad
\label{resptens}
\end{eqnarray}
where $g$ and $g'=\{0,s,a\}$, and $\rho_N^\rho$ and $\rho_D^\rho$ are 
the spin density matrices for one nucleon in the final state
or the deuteron target, with $\rho^+$ being the part of the
density matrix {\it symmetric \/} under $Y$ parity and
$\rho^-$ the part {\it antisymmetric\/} under $Y$ parity.  Symmetry
under the $Y$ parity operation then insures that  those observables of
type (II) must include one, and only one factor of the
antisymmetric current $J_a$ (further details can be found in
Ref.~\cite{dg89}).  The relation between the nine response
functions that appear in Eq.~(\ref{crosssec}) and the tensors
defined in (\ref{resptens}) are given in Table~\ref{tab:tens}.  The
normalization of Eq.~(\ref{resptens}) and the density matrices is
consistent, for unpolarized reactions, to summing over final
state spins and averaging over the initial deuteron spin.  

%%%%%%%%%%%%%%%%%%%%%%%%%%%%
\begin{table}
\begin{minipage}{3in}
\caption{\label{tab:tens}Response functions. }
\begin{ruledtabular}
\begin{tabular}{ll}
%\tableline
$\tilde{R}^{({\rm I})}_L=R_{00}$ & \\
$ \tilde{R}^{({\rm I})}_T=2(R_{aa}+R_{ss}) $  & 
$\tilde{R}^{({\rm II})}_{T'}=4{\rm Re}\;R_{sa}$   \\
$\tilde{R}^{({\rm I})}_{TT}=2(R_{aa}-R_{ss})$ & 
$\tilde{R}^{({\rm II})}_{TT}=-4{\rm Im}\;R_{sa}$  \\
$\tilde{R}^{({\rm I})}_{LT}=4 {\rm Re}\; R_{0s}$  & 
$\tilde{R}^{({\rm II})}_{LT}=4{\rm Im}\; R_{0a}$   \\ 
$ \tilde{R}^{({\rm I})}_{LT'}=4 {\rm Im}\; R_{0s}$  & 
$\tilde{R}^{({\rm II})}_{LT'}=4{\rm Re}\;R_{0a}$ \\
\end{tabular}
\end{ruledtabular}
\end{minipage}
\end{table}
%%%%%%%%%%%%%%%%%%%%%%%%%%%% 

For unpolarized particles,  
\begin{eqnarray}
(\rho_N^+)_{\lambda'_1\lambda_1}&=&
{1\over2}\delta_{\lambda'_1\lambda_1}
\qquad (\rho_N^-)_{\lambda'_1\lambda_1}=0 \nonumber\\
(\rho_D^+)_{\lambda_d\lambda'_d}&=&
{1\over3}\delta_{\lambda_d\lambda'_d}
\qquad (\rho_D^-)_{\lambda_d\lambda'_d}=0\, ,
\end{eqnarray}
so the observables of type (II) are zero.  If we also limit discussion
to unpolarized electrons, the terms proportional to the electron
helicity $h$ average to zero, and the cross section depends on only 4
response functions:  
\begin{eqnarray}
{d^5\sigma\over d\Omega^{\prime}dE^{\prime}d\Sigma} = 
{\sigma_M\over 4\pi  M_d}\frac{Q^2}{{q}_{_L}^2}\biggl\{
\tilde{R}^{({\rm I})}_L +  
s_T\tilde{R}^{({\rm I})}_T  -\frac{1}{2}\cos 2\phi \;\tilde{R}^{({\rm
I})}_{TT} + s_{LT} \cos\phi\; \tilde{R}^{({\rm I})}_{LT} \biggl\}
\label{crosssec2}
\end{eqnarray}
A second, independent combination of the same four response functions 
gives the asymmetry
\begin{eqnarray}
A_\phi=\frac{{\displaystyle {d^5\sigma \over
d\Omega^{\prime}dE^{\prime}d\Sigma}(\phi=0) - {d^5\sigma
\over d\Omega^{\prime}dE^{\prime}d\Sigma}(\phi=\pi)}}{
{\displaystyle 
{d^5\sigma \over d\Omega^{\prime}dE^{\prime}d\Sigma}(\phi=0) +
{d^5\sigma  \over
d\Omega^{\prime}dE^{\prime}d\Sigma}(\phi=\pi)}}
=\frac{s_{LT} \tilde{R}^{({\rm I})}_{LT}}{\tilde{R}^{({\rm
I})}_L +   s_T\tilde{R}^{({\rm I})}_T 
-\frac{1}{2}\tilde{R}^{({\rm I})}_{TT}} \label{Aphi}
\end{eqnarray} 
where the electron kinematics is held fixed and the outgoing 
proton is measured forward to the direction of the virtual
photon momentum
${\bf q}$ (at $\phi=0$) and backward (at $\phi=\pi$).  The 
longitudinal contributions $\tilde{R}^{({\rm I})}_L -
\frac{1}{2}\tilde{R}^{({\rm I})}_{TT}$ can be separated from
the transverse response 
$\tilde{R}^{({\rm I})}_T$ by measuring the cross section for 
the same kinematics at forward and backward electron
scattering angles, but the transverse interference term 
$\tilde{R}^{({\rm I})}_{TT}$ can be separated from 
$\tilde{R}^{({\rm I})}_L$ only by an out-of plane measurement
(for example, $\phi=\pi/2$).

These four unpolarized structure functions are only a small  
fraction of  the structure functions which can be measured.  
With {\it polarized\/}  electrons, targets, and recoiling
nucleons many more can be studied \cite{dg89,ALT}, but these
observables tend to be very sensitive to final state
interactions and interaction currents.  In this first paper we
have omitted final state interactions and interaction currents,
and hence also limit the discussion to unpolarized
observables.   

%%%%%%%%%%%%%%%%%%%%%%%%%%
% Fig. 2
%
\begin{figure}[t]
\begin{center}
\mbox{
%\vspace*{-6in}
%   \epsfxsize=6.0in
   \epsfysize=3.3in
\epsfbox{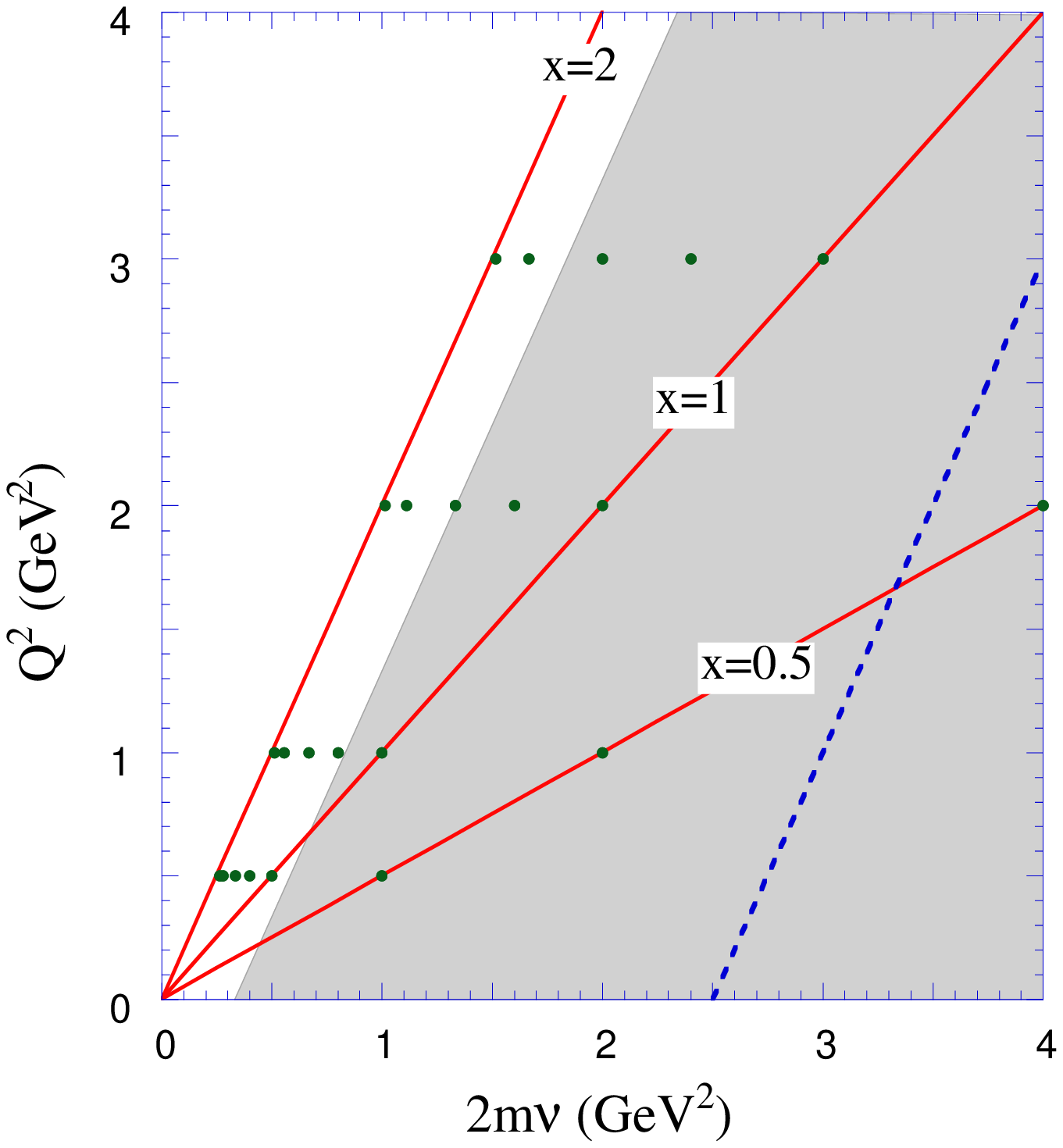}
   \epsfysize=3.3in
\epsfbox{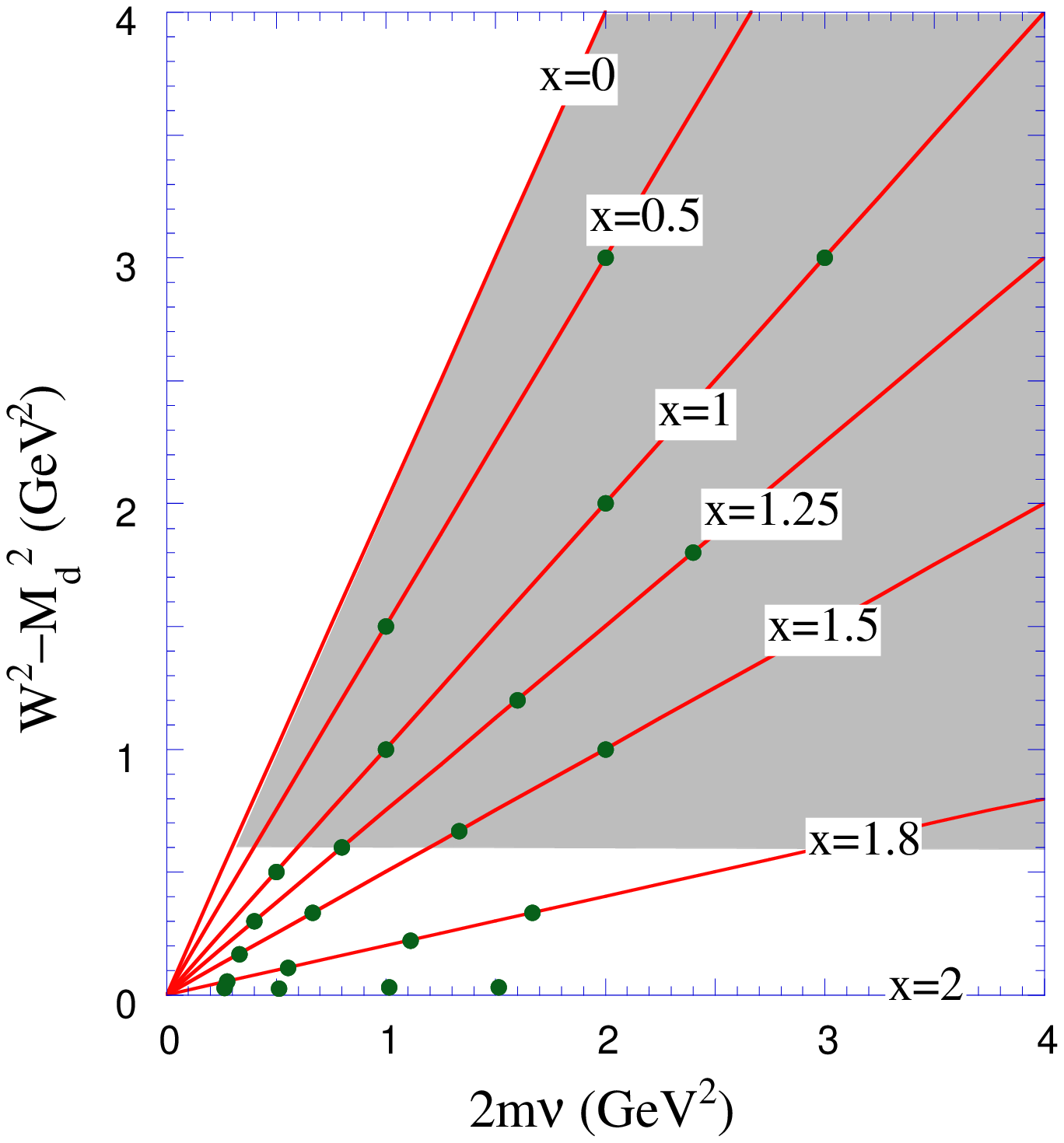}
%\vspace*{-6in}
}
\end{center}
\caption{\footnotesize\baselineskip=12pt Left panel is the $Q^2\nu$
plane, and the right panel  the $W^2\nu$ plane.  In each panel lines
with constant $x$ are shown, and the cases analyzed in the following
section are shown as dots.  The shaded area in each panel is the
region where pion production is kinematically possible.  Note that
pions cannot be produced near the line $x=2$, but that
inelasticity sets in even at small $Q^2$ along the
quasielastic ridge ($x=1$) and at smaller values of $x$.  The
dashed line in the left panel corresponds to $W^2$ = 9$m^2$,
just above the region shown in the right panel.}
\label{x4}
\end{figure}
%%%%%%%%%%%%%%%%%%%%%%%%%%

%%%%%%%%%%%%%%%%%%%%%%%%%%%%%%%%%%%%%%%%%%%%%%%%%%
\subsection{Kinematics} 
\label{Sec:kin}
 
The response functions $\tilde R$ depend on three
variables: $Q^2$, $\nu$, and the angle $\theta_1$ between 
${\bf p}_1$  and ${\bf q}$, where ${\bf p}_1$ is the
three-momentum of the particle detected in coincidence with
the final electron (assumed here to be the proton).  The
variables $Q^2$  and $\nu$ are fixed by the virtual photon, and we
choose $\theta^*$ (the lab value of $\theta_1$) rather than
$\theta_1$ because it is independent of $Q^2$ and
$\nu$ and always varies between 0 and $\pi$.  In place of $\nu$, it
is often convenient to use $W^2$ or $x=Q^2/2m\nu$, the Bjorken
scaling variable.  The mass of the final state, $W$, is related to
$\nu$ (or $x$) by
\begin{eqnarray}
W^2=&& M_d^2 +2 M_d \nu -Q^2\nonumber\\
=&& M_d^2 +2M_d\nu\left(1-\frac{mx}{M_d} \right)
\end{eqnarray}
The region of allowed values of $Q^2$ and $\nu$ is shown in
Fig.~\ref{x4}.  If the scattering is  elastic, so that the
deuteron remains bound after the scattering,
$x\simeq2$ , and this defines one boundary of the allowed  
scattering region.  It is sometimes assumed that pions must
necessarily be produced as $Q^2$ increases, but as long as $x$
remains close to 2, the final state remains below the pion production
threshold up to very large values of $Q^2$, and one may try to
explain the large $Q^2$ behavior of  these inelastic processes using a
theory with no pion rescattering in the final state.  The line $x=1$
is the quasielastic peak; when $Q^2$ is large the region between $x=1$
and $x=0$ is the region where $y$ (or $x$) scaling is observed.  
If $x$ is small, pions will be produced more and more easily as $Q^2$
increases (penetrating further and further into the shaded region in
Fig.~\ref{x4}), and explicit treatment of the pion degrees of
freedom will be necessary.

%%%%%%%%%%%%%%%%%%%%%%%%%%
% PWBA diagrams 
%
\begin{figure}
\vspace*{-0.4in}
\centerline{\epsfysize=3.9in \epsfbox{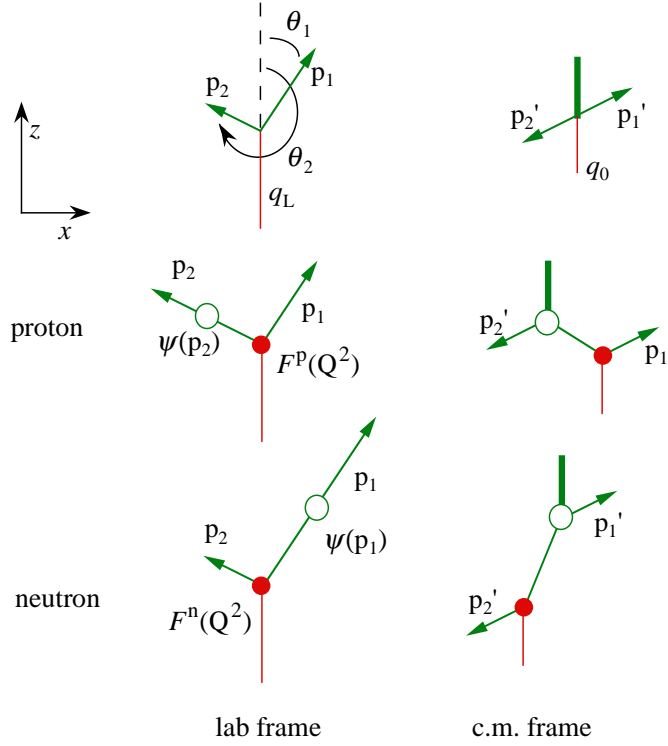}} 
\vspace*{0.4in}
\caption{\footnotesize\baselineskip=12pt The relativistic impulse
approximation (RIA) to  deuteron electrodisintegration in the lab
frame and the c.m. frame. The open circle denotes the deuteron wave
function, the filled circle the nucleon form factor, and the nucleon
propagating between the two is off-shell.  Note that the wave
functions $\psi$ always have one particle off-shell. }
\label{fig_diag}
\end{figure}
%%%%%%%%%%%%%%%%%%%%%%%%%%% 

%%%%%%%%%%%%%%%%%%%%%%%%%%
% momentum circles diagrams
%
\begin{figure}
\vspace*{-0.4in}
\centerline{\epsfysize=7.5in \epsfbox{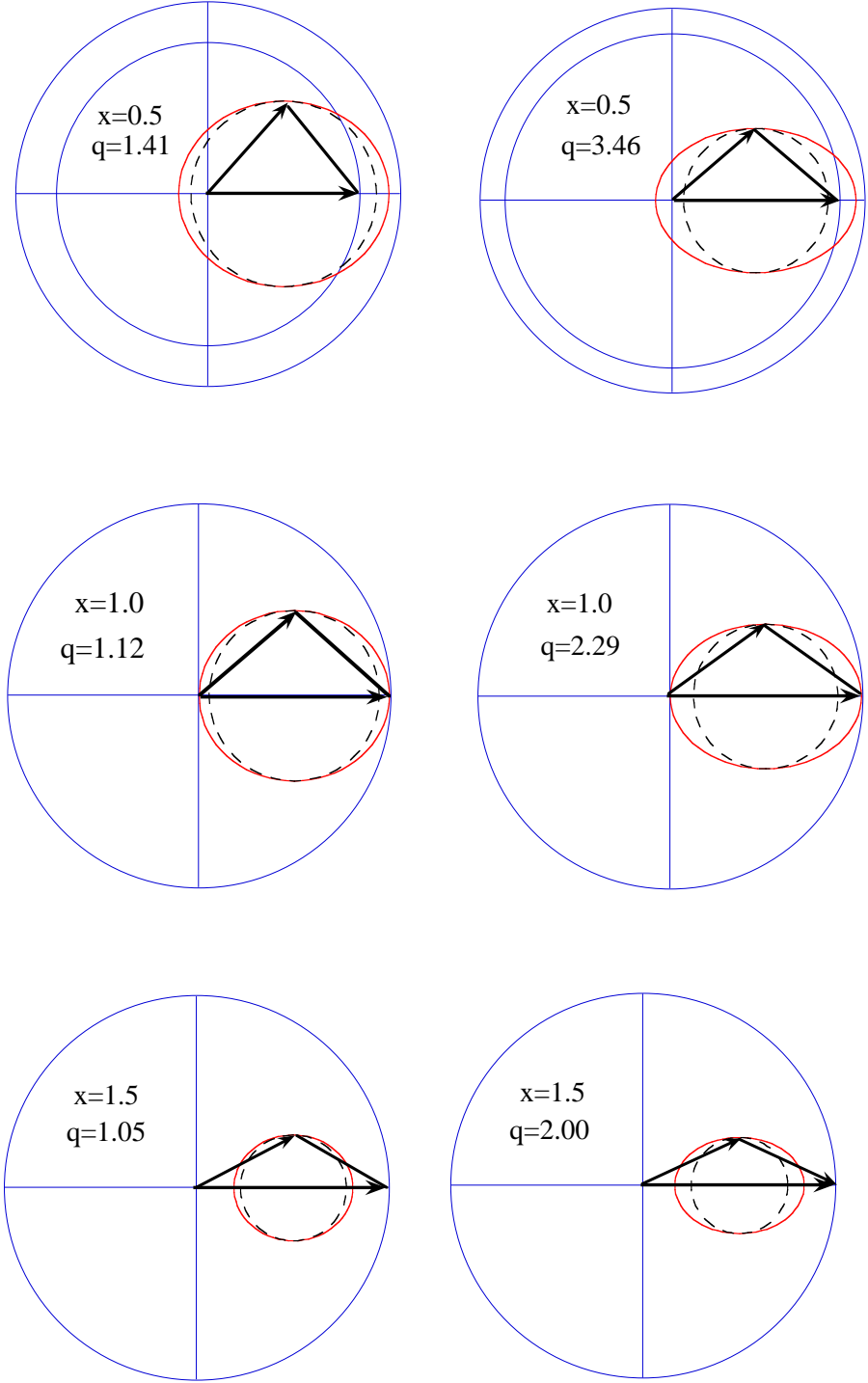}}
\vspace*{0.4in}
\caption{\footnotesize\baselineskip=12pt Polar plots showing the locus
of the momentum vector ${\bf p}_1$ in the lab system (solid lines are
relativistic; dashed lines nonrelativistic).  The horizontal axis in
each  panel is $\hat z$; the vertical is $\hat x$.  The three left
hand panels have $Q^2=1$ GeV$^2$ and the various values of $x$
and $q$ shown on each panel; the right hand panels are for
$Q^2=3$ GeV$^2$.  In all panels the ${\bf q}$ vector points to
the right along the $\hat z$ axis and sets the scale. The
other two vectors are ${\bf p}_1$ and
${\bf p}_2$ for the symmetry case discussed in the text.} 
\label{circles}
\end{figure}
%%%%%%%%%%%%%%%%%%%%%%%%%%% 

The variables $Q^2$, $\nu$ (or $x$ or $W^2$), and
$\theta^*$  are convenient for thinking about final
state interactions.  However the RIA depends primarily on only
two variables, $Q^2$ and ${\rm p}_{\rm miss}$, where ${\rm p}_{\rm
miss}$ is the value of the {\it spectator\/} momentum in the lab
system.  The spectator momentum ${\rm p}_{\rm miss}$ may be either
${\rm p}_1$ or ${\rm p}_2$ depending on which of the two nucleons was
struck by the virtual photon (in the absence of final state
interactions or interaction currents, this is all that can
happen), and the cross section is therefore the coherent sum
of two terms.  Symbolically, the RIA current is 
\begin{equation}
J_{\rm RIA}(p_1,p_2,q)=\left|\,F^p(Q^2)\psi({\rm p}_2) \pm
F^n(Q^2)\psi({\rm p}_1) \,
\right|^2 \, , \label{PWBA0}
\end{equation}
where explicit formulae for the magnitudes of the rest frame three
momenta, ${\rm p}_1$ and ${\rm p}_2$, are given in 
Eq.~(\ref{p1and2mags}) below.  The two terms contributing to this sum
are illustrated in Fig.~\ref{fig_diag}.  Since $\psi({\rm p})$ is
normally a rapidly decreasing function of ${\rm p}$, these two terms
are normally dominated by the one with the smallest ${\rm p}_{\rm
miss}$.  
 
The momenta ${\bf p}_{1}$ and ${\bf p}_{2}$ are most easily
obtained by boosting from the c.m. frame. Their 
$x$ and $z$ components are
\begin{eqnarray}
{\bf p}_{i}^x =&& \pm {\rm p}\,\sin\theta^*\nonumber\\
{\bf p}_{i}^z =&&\frac{q_{_L}}{2} \pm {\rm p} \frac{E_W}{W}
\cos\theta^* 
\label{p1and2}
\end{eqnarray}
where the upper (lower) sign is for $i=1$ ($i=2$), ${\rm p}$ is
the magnitude of the nucleon momenta in the c.m.~frame, $E_W$
is the energy of the outgoing pair in the lab frame, with  
\begin{eqnarray}
{\rm p}=&& \sqrt{\frac{[W^2-(m_1+m_2)^2][W^2-(m_1-m_2)^2]}{4W^2}}
\simeq \frac{1}{2}\sqrt{W^2-4m^2}\nonumber\\
E_W =&&\sqrt{W^2+q_{_L}^2}=M_d+\nu\, , \label{cmmom}
\end{eqnarray}
and the other variables were previously defined (recall
Table~\ref{tab:variables}).  Hence
\begin{eqnarray}
{\rm p}^2_{i}=\left(\frac{q_{_L}}{2} \pm {\rm p} \frac{E_W}{W}
\cos\theta^*\right)^2 + {\rm p}^2\sin^2\theta^* 
\label{p1and2mags}
\end{eqnarray}

The behavior of the magnitudes of ${\rm p}_{1}$ and ${\rm p}_{2}$,
and  the angles $\theta_{1}$ and $\theta_{2}$, for six choices
of $Q^2$ and $x$, can be inferred from  Fig.~\ref{circles}. 
The solid lines in each panel are the locus of points swept
out by Eq.~(\ref{p1and2}), and the dashed lines by
(\ref{p1and2}) with $E_W=W$ (for a Galilean boost).  For $x\ge
1$ the two vectors ${\bf p}_1$ and ${\bf p}_2$ always lie in the
first or third quadrant, but for $x<1$ the vectors may lie in any
quadrant.

The restriction of both momenta ${\bf p}_1$ and ${\bf p}_2$ to the
first and third quadrant, which happens for $x>1$, produces a
curious singularity in the lab cross section.  Under these
conditions, the lab angle $\theta_1$ will reach a maximum value
less than 90$^\circ$ at a c.m.~angle $\theta^*=\theta^*_{\rm crit}$. 
At this point 
\begin{eqnarray}
\frac{d\theta_1}{d\theta^*}\bigg|_{\theta^*_{\rm crit}} =0\, .
\label{crit1}
\end{eqnarray}
The differential cross section in the c.m.~is always finite, but the
lab cross section, defined by the transformation
\begin{eqnarray}
d\sigma=d\theta^*\,|{ M}_{\rm c.m.}(\theta^*_{\rm crit})|^2=
d\theta_1\,|{ M}_{\rm lab}(\theta^*_{\rm crit})|^2 \equiv
d\theta_1\,\frac{|{ M}_{\rm
c.m.}(\theta^*_{\rm crit})|^2}{d\theta_1/d\theta^*}\, ,
\label{crit2}
\end{eqnarray}
has a singularity at the critical point because of the vanishing of the
Jacobian (\ref{crit1}).  This singularity is analyzed in detail in
Appendix \ref{Appen:C}.  In this paper we present c.m.~cross sections
only, so we do not encounter this singularity.  

\begin{figure}
\vspace*{-0.4in}
\centerline{\epsfysize=3in \epsfbox{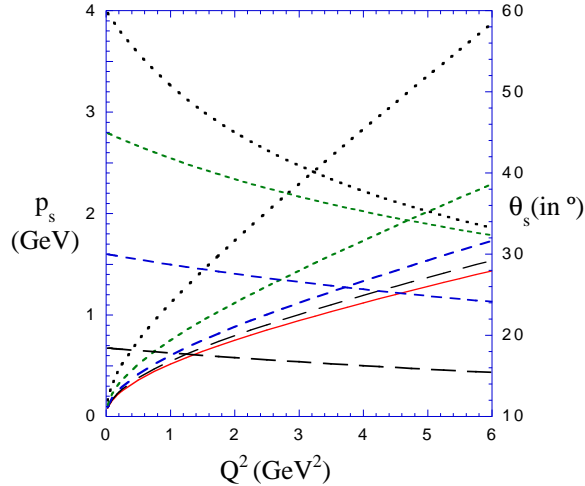}}
\vspace*{0.4in}
\caption{\footnotesize\baselineskip=12pt The symmetry momenta ${\rm
p}_s$ and the symmetry angles $\theta_s$ as functions of $Q^2$ for five
values of $x$: $x=2$ (solid line), 1.8 (long dashes), 1.5 (medium
dashes), 1.0 (dashes) and 0.5 (dots).  The momenta rise with $Q^2$
and the angles fall with $Q^2$.  For $x=2$ the angle $\theta_s=0$ and
is not shown.   }
\label{psym} 
\end{figure}

For the special case when $\theta^* =\pi/2$ (where
the  relativistic ellipse touches the nonrelativistic circle) 
the magnitudes of ${\bf p}_1$ and ${\bf p}_2$ are equal, and
the RIA depends uniquely on the wave function at only one
momentum point.  We will refer to this as the {\it symmetry
point\/}. Since the angle between  ${\bf p}_1$ and ${\bf p}_2$ is
90$^\circ$ in the nonrelativistic limit, this is referred to as
perpendicular kinematics.  Were there no final state interactions or
interaction currents, the symmetry point would be an optimal place to
measure the wave function.  The symmetry point momenta ${\rm p}_s=|{\bf
p}_{1}|=|{\bf p}_{2}|$ and angles
$\theta_s=\theta_{1}=-\theta_{2}$ are
\begin{eqnarray}
{\rm p}_s=&& \sqrt{{\rm p}^2+q^2}\simeq
\frac{Q}{4mx}\sqrt{8m^2x+Q^2}\nonumber\\ 
\theta_s
=&&\tan^{-1}\left[\frac{2{\rm p}}{q} \right] \simeq \tan^{-1}
\left[\sqrt{\frac{4m^2x(2-x)}{4m^2x^2+Q^2}}\right]\, .
\end{eqnarray}
These are shown in Fig.~\ref{psym} as a function of $Q^2$
for several fixed values of $x$.  The figure shows that if we wish
to measure the wave function at large ${\rm p}_s$ (near one GeV) and at
large $x$ where pion production is not large, we must go to large
$Q^2$ (about 2 to 3 GeV$^2$).

%%%%%%%%%%%%%%%%%%%%%%%%%%
% symmetry momentum and angles
%
%%%%%%%%%%%%%%%%%%%%%%%%%%% 

We now turn to a discussion of the RIA.

%%%%%%%%%%%%%%%%%%%%%%%%%%%%%%%%%%%%%%%%%%%%%%%%%%
\subsection{Matrix Element for the Relativistic Impulse
Approximation}

The RIA approximation used in this paper is 
based on the simple pole diagrams shown in
Fig.~\ref{fig_diag}.  We use a (standard but unfamiliar)
notation in which matrix elements of an operator between {\it
two outgoing\/} Dirac particles are written in the form 
\begin{equation}
\left<{\cal O}\right>= \bar u({\bf p}_1,\lambda_1)\,{\cal
O}\,    {\cal C} \,\bar u^T({\bf p}_2,\lambda_2) \, ,
\label{matrix} 
\end{equation}
where ${\cal C}=-i\gamma^0\gamma^2$ is the Dirac charge 
conjugation matrix.  This notation is very
convenient because ${\cal C}\,\bar u^T$ transforms like an
incoming $v$ spinor (but is {\it not\/} to be interpreted as
an antiparticle in this application), and therefore the most
general operator
${\cal O}$ can be constructed from the standard 16 
independent Dirac bilinear covariant operators.  The
matrix representation (\ref{matrix})  is equivalent to a
direct product representation 
\begin{equation} 
\bar u_\alpha({\bf p}_1,\lambda_1) \,\left[{\cal O} \,{\cal
C}\right]_{\alpha\beta}\, \bar u^T_\beta({\bf p}_2,\lambda_2)
\leftrightarrow
\bar u_\alpha({\bf p}_1,\lambda_1)\bar u_\beta({\bf
p}_2,\lambda_2)
\left[{\cal O} \,{\cal C}\right]_{\alpha\beta}  \, ,
\label{matrix2} 
\end{equation}  
but is more convenient for relativistic calculations.  [Note that
the RHS and LHS of this equation are {\it identical\/} as long as the
Dirac indices are shown explicitly, but only the LHS can be turned
into Eq.~(\ref{matrix}) by dropping explicit reference to the
indices.  Beware that the order of the momenta in (\ref{matrix})
and (\ref{matrix2}) is opposite from that used in a previous
reference \cite{bg} where $\bar u({\bf p}_2,\lambda_2)$ was 
multiplied from the left and $u({\bf p}_1,\lambda_1)$ from the right;
see  Appendix~\ref{Appen:A}.]     

Including the isospin factor 
$\left< \frac{1}{2}\frac{1}{2},\frac{1}{2}-\frac{1}{2} 
\left|\right. 0 0\right>= \frac{1}{\sqrt{2}}$ 
the Feynman amplitudes for the RIA in  {\it any\/} frame can
be written  
\begin{eqnarray}
\left< \lambda_1\lambda_2\left|
J_{g}(q)\right|\lambda_d\right>
&=&\frac{1}{\sqrt{2}\,N_d}\left[\bar u_1({\bf p}_1,\lambda_1)
j_{g}^{(1)}(p_1,p_1-q)
\psi^{(2)}_{\lambda_2,\lambda_d}(p_2,P)\right.\nonumber\\ 
&&\quad\left. - \;\bar u_2({\bf p}_2,\lambda_2)
j_{g}^{(2)}(p_2,p_2-q)
\psi^{(1)}_{\lambda_1,\lambda_d}(p_1,P)  \right]\, ,\label{feynman}
\end{eqnarray}
where $\lambda_i$ are the nucleon helicities and $\lambda_d$
the helicity of the deuteron, and $N_d$ is a normalization constant
defined and discussed below.  The subscript on the nucleon helicity
spinor, $\bar u_i$ [suppressed in Eq.~(\ref{matrix})], refers to
whether it is particle 1 or particle 2, in the sense of Jacob and Wick
\cite{jw} (see the discussion in Appendix~\ref{Appen:A}).  The
nucleon current is      
\begin{eqnarray}
j^{(i)}_{g}(p,p-q)=\varepsilon_{g}^\mu 
j^{(i)}_\mu (p,p-q) \, ,
\label{current}
\end{eqnarray}
where $p$ and $p-q$ are nucleon four-momenta with
$p$ on-shell ($p^2=m^2$) and $p-q$ off-shell, the superscript $i$ = 1
(proton) or 2 (neutron), and the virtual  photon has polarization
vector $\varepsilon_{g}$, where $g=\{0,s,a\}$ with $s$ and $a$ the
linear combination of photon helicities introduced in
Eq.~(\ref{gamps}) [for more details, see Eq.~(\ref{B14})].  The
relativistic deuteron wave  function \cite{bg} for a nucleon with
momentum $P-p$ off-shell and a nucleon with momentum $p$ and helicity
$\lambda$ on-shell (so that $p^2=m^2$) is {\it defined\/} to be
$\psi_{\lambda,\lambda_d}(p,P)$, and is related to the normalized
$dnp$ vertex function $\Gamma$ by   
\begin{eqnarray}
\psi^{(i)}_{\lambda,\lambda_d}(p,P)\equiv
\frac{m+\not{\!P}-\not{\!p}}{m^2-(P-p)^2}\;N_d\,
\Gamma_\mu(p,P)\, {\cal C}\,\bar{u}_i^T({\bf
p},\lambda)\, \xi^\mu_{\lambda_d}(P)
\, . \label{wf} 
\end{eqnarray}
where the normalization constant
\begin{eqnarray}
N_d=\left[2M_d\,(2\pi)^3\right]^{-1/2}
\end{eqnarray}
is chosen to give the defined wave function $\psi$ a convenient
normalization [see Eq.~(\ref{norm})].  The superscript
$(i)$ labels the choice of helicity convention (particle 1 or 2) for
the on-shell particle,
$\Gamma_\mu(p,P)$ is the normalized $dnp$ deuteron vertex (with the
off-shell particle on the left) first defined by Blankenbecler and
Cook \cite{BbC}, and $\xi^\mu_{\lambda_d}(P)$ is the deuteron
polarization vector for a state with helicity $\lambda_d$ and
four-momentum $P$.  We use the notation of Ref \cite{bg} for
$\Gamma$.   Note that the normalization constant in (\ref{feynman})
and (\ref{wf}) cancel; the  Feynman amplitude depends only on the
normalization of $\Gamma$ and not on the convention used to normalize
$\psi$.  For further details, see Appendix \ref{Appen:A}. 

\subsection{The issue of gauge invariance}

The RIA is not gauge invariant by itself.   This issue must be
dealt with before we can proceed with the calculation.  Here we
discuss how this is done. 

Using the method of
Ref.~\cite{RG}, the RIA, together with final state interactions (FSI)
and interaction currents (IntC), is part of a gauge invariant
calculation.  Once all of these pieces have been calculated and
assembled, the result will be gauge invariant.   Here we describe a
convenient prescription that is (i) covariant, (ii) renders each of
the individual contributions (RIA, FSI, and IntC) {\it separately\/}
gauge invariant without altering their sum, and (iii) modifies each of
the individual contributions as little as possible.  The method was
introduced in Ref.~\cite{bg98}, where it was also shown that the
prescription guarantees that the RIA also gives the correct asymptotic
result for deep inelastic scattering.  

If the individual contributions
to the total current are denoted $J_{\rm RIA}$, $J_{\rm FSI}$, and
$J_{\rm IntC}$, then the prescription calls for each to be modified
by the replacement 
\begin{equation}
\tilde J^\mu_{\rm X} = J^\mu_{\rm X} - \frac{q^\mu}{q^2}\;q\cdot
J_{\rm X}
\end{equation}
where X is any of the RIA, FSI, or IntC terms.  Since the total
current is gauge invariant, $q\cdot J_{\rm total}=0$, and
\begin{equation}
\tilde J^\mu_{\rm total} = J^\mu_{\rm total} 
\end{equation}
so the prescription does not modify the total current.  Furthermore,
since the photon helicity vectors are all orthogonal to $q$,
$\varepsilon_{_\lambda}\cdot q=0$, 
\begin{equation}
\varepsilon_{_\lambda}\cdot
\tilde J_{\rm x}=\varepsilon_{_\lambda}\cdot J_{\rm x}
\end{equation}
and the prescription has {\it no effect on the contribution of each of
the terms in the current\/}.  This prescription meets all three of the
conditions listed above.  

Unfortunately, there is no uniquely correct way to modify the RIA so
that it is gauge invariant.  The choice proposed here is only one
of many possibilities.  

\vspace{0.5in}   

\subsection{Calculation of the structure functions}  

The structure functions are obtained by squaring the matrix element
(\ref{feynman}) and summing over spins.  There will be three terms:
the two ``diagonal'' terms coming from the square of the proton term
and the square of the neutron term, and the interference term.  The
diagonal terms can be calculated by expanding the density matrices
\begin{eqnarray}
{\cal N}^{(i)}(p)=\sum_{\lambda_d\lambda}
\psi^{(i)}_{\lambda\lambda_d}(p,P)\otimes
\psi^{(i)\,\dagger}_{\lambda\lambda_d}(p,P)
\end{eqnarray}
in terms of independent Dirac spin invariants, and then performing the
sum over the off-shell particle degrees of freedom using Feynman trace
techniques.  The final result, given in Ref.~\cite{jvo}, is a sum of
squares of invariant functions and scalar products of four vectors,
and is manifestly covariant.   

In this paper we present an alternative method in which the
structure functions are calculated by first expanding the
off-shell nucleon in terms of on-shell nucleon degrees of
freedom, and then computing the squares of the matrix elements. 
It is possible that this method will simplify the calculation of
polarization observables planned for future work.  Unfortunately, the
results obtained using this method are not manifestly covariant (but
they are, nevertheless, covariant), and, unless one is extremely
careful, it is easy to make sign mistakes by dropping one of the many
phases that arise when transforming helicity amplitudes.  As a check
of the results presented here, we have shown explicitly that our
final analytical result for the diagonal term is {\it identical\/} to
the result obtained in Ref.~\cite{jvo}.

The discussion of this method begins by noting that the physical
content of the matrix element  (\ref{feynman}) can be displayed by
decomposing  the off-shell nucleon into positive energy ($u$ spinor)
and negative energy ($v$ spinor) states.  For example, if we
choose a four-momentum
$k=\{E_k, {\bf k}\}$, then the states
$u({\bf k}, \lambda)$ and $\gamma^5u({\bf k}, \lambda)$ (which
we use in place of the $v$ spinors), with helicity
$\lambda=\pm{1\over2}$, are complete, and
\begin{eqnarray}
\openone = \sum_\lambda\left\{u({\bf k}, \lambda) \,\bar 
u({\bf k},\lambda) + \gamma^5\, u({\bf k}, \lambda)\, \bar
u({\bf k}, \lambda) \gamma^5 \right\} \, .
\label{unitdecompose}
\end{eqnarray}
(The definitions and normalization of the helicity states are 
discussed in Appendix \ref{Appen:A}.)  It is important to
realize that while this decomposition can be  carried out {\it
in any frame\/} using nucleon states with {\it any on-shell
four-momentum\/}, the result may {\it appear\/} very
different depending on the frame and the spinor states used to
do the decomposition (even though the final numerical result
will always be independent of these choices).       

In this subsection we record the results for the current
Eq.~(\ref{feynman}) if the decomposition is  made in terms of the
states of the {\it spectator\/} nucleon (with four-momentum
$p_2$ for the proton term and  four-momentum
$p_1$ for the neutron term).  The final result in the c.m.~frame
[see Eq.~(\ref{curr3})] is 
\begin{eqnarray}
&&\left< \lambda_1\lambda_2\left|
J_{g}(q^*)\right|\lambda_d\right>=
\sqrt{3\over16\pi}\,{1\over N_d}\sum_{\lambda\,\rho}
\sum_{\lambda_1'\lambda_2'}
\nonumber\\ &&\qquad\quad\;\;
\Biggl\{\eta_\rho(2\lambda'_1)\,
j^{(1)\,\rho}_{\lambda_1\,\lambda,\, g}({\rm
p},\theta^*,q_0)\,
\phi^\rho_{|\Lambda|}({\rm p}_2)\,
d^{(1)}_{\Lambda,\,\lambda_d}(\theta_2-\pi)
\;d^{(1/2)}_{\lambda'_2\,\lambda_2}(\omega_2)
\;d^{(1/2)}_{\lambda'_1\,\lambda}(\omega_2)
\nonumber\\
&&\qquad\qquad-\eta_\rho(2\lambda_2')\,
j^{(2)\,\rho}_{\lambda_2\,\lambda,\, g}({\rm p},\theta^*,q_0)\,
\phi^\rho_{|\Lambda|}({\rm p}_1)\,
d^{(1)}_{-\Lambda,\,\lambda_d}(\theta_1)
\;d^{(1/2)}_{\lambda'_1\,\lambda_1}(\omega_1)
\;d^{(1/2)}_{\lambda'_2\,\lambda}(\omega_1) \Biggr\} \, ,
\qquad
\label{feynman2}
\end{eqnarray}
where the $d$'s are the rotation matrices,
$\Lambda=\lambda_1'+\lambda'_2$, $\omega_j$ are the Wigner rotation
angles resulting from the boost of the spectator nucleons with
four-momentum $p_j$ from the lab to the c.m.~frame, and ${\rm p}_j$
and $\theta_j$ are the magnitudes of the on-shell spectator
three-momenta and polar angles in the rest frame of the deuteron
[i.e. the lab frame; recall Table~\ref{tab:variables}].  The phase
$\eta_\rho(x)$ is
\begin{eqnarray}
\eta_\rho(x)=\cases{1 &if $\rho=+$ \cr
-x &if $\rho=-$} \label{phase1}
\end{eqnarray}
and the matrix elements of the single nucleon current, in the
c.m.~system, are 
\begin{eqnarray}
j^{(i)\,\rho}_{\lambda_i\lambda,\,g}({\rm p},\theta^*,q_0)
=\cases{\bar u_i({\bf p}^*_i,\lambda_i)\,
j_{g}^{(i)} (p^*_i,p^*_i-q^*)\, u_j({\bf p}^*_j,\lambda)
&if $\rho=+$ \cr
\bar u_i({\bf p}^*_i,\lambda_i)\,
j_{g}^{(i)} (p^*_i,p^*_i-q^*)\,\gamma^5\, u_j({\bf p}^*_j,\lambda) 
&if $\rho=-$\, ,}
\end{eqnarray}
with $j=1$ or 2, but $j\ne i$.
The deuteron matrix elements are defined in the deuteron rest
frame using the expansion (\ref{unitdecompose}), and are written
\begin{eqnarray}
\psi^{(i)}_{\lambda'_i,\lambda_d}({p}_i,{P}) &=&
%\frac{\sqrt{3}}{\sqrt{8\pi}}
\sqrt{\frac{3}{8\pi}}
 \sum_{\lambda'_j} \Biggl\{ u_i({\bf {p}}_i,\lambda'_j)
\;\phi^+_{|\Lambda|}({\rm p}_i) \nonumber\\ &&\qquad -
2\lambda'_j\,\gamma^5 u_i({\bf {p}}_i,\lambda'_j) \; 
\phi^-_{|\Lambda|}({{\rm p}}_i)
\Biggr\}\,\times\cases {d^{(1)}_{-\Lambda\,\lambda_d}({\theta}_1) &
if $i=1$ \cr
d^{(1)}_{\Lambda\,\lambda_d}({\theta}_2-\pi) & if $i=2$\, . \cr}
\end{eqnarray}
The $\phi$'s are combinations of the four scalar deuteron wave
functions defined by Eq.~(\ref{wf}).  These are the
S-state $u$, the D-state $w$, and the two
P-state wave functions $v_t$ and $v_s$, and expressions for 
the $\phi$'s are given in Table~\ref{tab:wave}.  The final result
(\ref{feynman2}) was obtained by boosting this result to the c.m.\
frame, as shown in Appendix \ref{Appen:A}.

%%%%%%%%%%%%%%%%%%%%%%%%%%%%
\begin{table}
\begin{minipage}{6in}
\caption{\label{tab:wave}Wave function combinations that 
enter the current. }
\begin{ruledtabular}
\begin{tabular}{ll}
%\tableline
$\phi_0^+({\rm p})$   =  ${\displaystyle{1\over\sqrt{3}}{E\over
m}}\left(u({\rm p})+\sqrt{2}\,w({\rm p})\right)$ & $\phi_0^-({\rm
p})$   =  ${\displaystyle{1\over\sqrt{3}}
\left\{{{\rm
p}\over m}\left(u({\rm
p})+\sqrt{2}w({\rm p})\right)-\sqrt{3}\,v_s({\rm p})\right\}}$
\\ & \\
$\phi_1^+({\rm p})$   =  ${\displaystyle{1\over\sqrt{3}}{E\over
m}}\left(\sqrt{2}\,u({\rm p})-w({\rm p})\right)$ & 
$\phi_1^-({\rm p})$  =  
${\displaystyle{1\over\sqrt{3}}
\left\{{{\rm p}\over m}
\left(\sqrt{2}\,u({\rm p})-w({\rm p})\right) +\sqrt{3}\,v_t({\rm
p})\right\}}$\\ 
\end{tabular}
\end{ruledtabular}
\end{minipage}
\end{table}
%%%%%%%%%%%%%%%%%%%%%%%%%%%%

This form of the Born term makes it easy to examine polarization
observables, and gives a simple form for the unpolarized response
tensors (\ref{resptens}).  Squaring the proton term [with $(i)=(1)$]
and summing over spins (averaging over the initial deuteron
polarization) gives 
\begin{eqnarray}
R_{gg'}&=&{m^2\over12\pi^2W}\left<J_g\,J_{g'}^\dagger\right>=
{m^2\over12\pi^2W}\sum_{\lambda_1\lambda_2\lambda_d} 
J^{(1)\,\lambda_d}_{g\,\lambda_1\,\lambda_2}(p^*_1,p^*_2; P^*) 
\,J^{(1)\lambda_d\,\dagger}_{g'\,\lambda_1\,\lambda_2}
(p^*_1,p^*_2; P^*)
\nonumber\\
&=& {m^2M_d\over W} \Biggl\{
J^+_{gg'}
\left[\left\{\phi_0^+({\rm p}_2)\right\}^2
+\left\{\phi_1^+({\rm p}_2)\right\}^2\right]  +J^-_{gg'}
\left[\left\{\phi_0^-({\rm p}_2)\right\}^2+\left\{\phi_1^-
({\rm p}_2)\right\}^2\right] 
\nonumber\\
&&\qquad\quad+\Bigl(J^c_{gg'}\cos\omega_2+J^s_{gg'}\sin\omega_2
\Big) \left[\phi_0^+({\rm p}_2)\phi_0^-({\rm p}_2)
+\phi_1^+({\rm p}_2)\phi_1^-({\rm p}_2)\right]
\Bigg\} \label{bornanalytic}
\end{eqnarray}
where the currents are
\begin{eqnarray}
J^+_{gg'}&=&j_{1g}^{(1)+}\,j_{1g'}^{(1)+} +
j_{2g}^{(1)+}j_{2g'}^{(1)+}\nonumber\\
J^-_{gg'}&=&j_{1g}^{(1)-}\,j_{1g'}^{(1)-} +
j_{2g}^{(1)-}\,j_{2g'}^{(1)-}\nonumber\\
J^c_{gg'}&=&j_{1g}^{(1)+}\,j_{1g'}^{(1)-} + 
j_{1g'}^{(1)+}\,j_{1g}^{(1)-} -j_{2g}^{(1)+}\,j_{2g'}^{(1)-} -
j_{2g'}^{(1)+}\,j_{2g}^{(1)-}
\nonumber\\ 
J^s_{gg'}&=&j_{1g}^{(1)+}\,j_{2g'}^{(1)-} + 
j_{2g'}^{(1)+}\,j_{1g}^{(1)-} +j_{2g}^{(1)+}\,j_{1g'}^{(1)-} +
j_{1g'}^{(1)+}\,j_{2g}^{(1)-}\, . \label{Bigjs}
\end{eqnarray}
Recall that $g$ and $g'$ can be either $0,s$, or $a$.  The 
12 individual current matrix elements are given in
Table~\ref{tab:current}. These exact expressions are easily 
evaluated, and the response functions determined from
Table~\ref{tab:tens}.  The square of the neutron term [with $(i)=(2)$]
is obtained by replacing $1\leftrightarrow2$, and the result for
the interference term is given in Appendix \ref{Appen:B}. 

%%%%%%%%%%%%%%%%%%%%%%%%%%%%
\begin{table}
\begin{minipage}{5.0in}
\caption{\label{tab:current}\footnotesize\baselineskip=12pt Matrix
elements of the current. All variables are in the c.m.~frame,
and $E=\sqrt{m^2+{\rm p}^2}$, ${\rm p}_\perp={\rm p}\sin\theta^*$, and
${\rm p}_z= {\rm p}\cos\theta^*$.  For the proton current, $j^{(1)}$,
substitute proton form factors for $F_1$ and $F_2$ and set the phase
$\delta$ = +.  For the neutron current,
$j^{(2)}$, substitute neutron form factors and set $\delta = -$.}
\begin{ruledtabular}
\begin{tabular}{ll}
$j_1^{+0}$  =  ${\displaystyle {1\over Q}\left(
F_{1}{q_0}-\delta\,2\tau F_2{{\rm p}_z}\right)}$
 & $j_2^{+0}$  =  ${\displaystyle F_1{\nu_0 \,{\rm p}_\perp\over
mQ}}$\\ & \\
$j_1^{+s}$  =  ${\displaystyle
-\delta\,{1\over\sqrt{2}}\,F_2{\nu_0\, {\rm p}_\perp\over2m^2}}$
 &  $j_2^{+s}$ =  ${\displaystyle{1\over\sqrt{2}\,m}}\,
\left( F_1 {\rm p}_z+ \delta\,{1\over2} F_2 \,{q_0}\right)$ \\ & \\
$j_1^{+a}$  = 
$\delta\,{\displaystyle{1\over\sqrt{2}}\,F_2\sin\theta^*\,
{q_0E\over2m^2}}$
 &  $j_2^{+a}$ =  ${\displaystyle{1\over\sqrt{2}\,m}}\,
\left( \delta\,F_1 {\rm p}+ {1\over2} F_2 \,{q_0}\cos\theta^*\right)$ 
\\ &
\\ 
\hline & \\
$j_1^{-0}$  =  $-\delta\,{\displaystyle{\cos\theta^*\over Q}\,
\left( F_1\nu_0-2\tau F_2 E\right)}$ & 
$j_2^{-0}$  = 
${\displaystyle{\sin\theta^*\over mQ}\,
\left( F_1\nu_0E-2\tau F_2 m^2\right)}$ 
\\ & \\
$j_1^{-s}$ =  $\delta\,{\displaystyle{\sin\theta^*\over \sqrt{2}}\,
\left( F_1+ F_2 {\nu_0 E\over2m^2}\right)}$ & 
$j_2^{-s}$ =  ${\displaystyle{\cos\theta^*\over \sqrt{2}\,m}}\,
\left( F_1E+ {1\over2}F_2 {\nu_0}\right)$ \\ & \\
$j_1^{-a}$  =  ${\displaystyle
-\delta\, {1\over\sqrt{2}}\,F_2{q_0\,{\rm p}_\perp\over2m^2}}$  & 
$j_2^{-a}$ =  ${\delta\, \displaystyle{1\over \sqrt{2}\,m}}\,
\left( F_1E+ {1\over2}F_2 {\nu_0}\right)$ \\
\end{tabular}
\end{ruledtabular}
\end{minipage}
\end{table}
%%%%%%%%%%%%%%%%%%%%%%%%%%%% 

\subsection{The cross section in the quasielastic limit}  

We may use expression (\ref{bornanalytic}) to look at the
cross section at the quasielastic peak, where ${\rm p}_2=0$ and
$x\simeq1$ (we assume here that $M_d=2m$).  Near ${\rm p}_2=0$ 
the minus components of the wave functions are both
suppressed, and the leading contribution to the cross
section comes only from the term proportional to
\begin{equation} 
\left(\phi_0^+\right)^2+\left(\phi_1^+\right)^2 =
{E^2\over m^2} \left[u^2+w^2\right]\simeq
\left[u^2+w^2\right] \equiv
4\pi\,n({\rm p}_2) \, ,
\end{equation}
where the momentum density is approximately normalized to
\begin{equation}
\int n({\rm p})\; d^3p ={1\over4\pi} \int d^3p \left[u^2({\rm
p}) +w^2({\rm p})\right] \approx 1
\end{equation}
[the exact relativistic normalization is given in Eq.~(\ref{norm})].
At the quasielastic peak $\theta^*=0$ and the
c.m.~momentum
${\rm p}$ that enters the current matrix elements given in Table
\ref{tab:current} is fixed.  From Eq.~(\ref{p1and2}) and the
condition $x=1$ we obtain
\begin{equation}
{\rm p}=\frac{W q_{_L}}{2E_W} = \frac{W^2 q_0}{2E_W M_d} \simeq
{q_0\over2} \, .
\end{equation}
This gives $R_{TT}=0$ and $R_{LT}=0$, and the following
simple formula for the coincidence cross section
\begin{eqnarray}
{d^5\sigma\over d\Omega^{\prime}dE^{\prime}d\Sigma} =
\sigma_M\,\frac{m^2}{W} \;n(0)
\Biggl\{ \frac{G_E^2(Q^2) + \tau G_M^2(Q^2)}{1+\tau}  +
2\tau G_M^2(Q^2) \tan^2\theta/2
\Biggl\}
\label{crosssec3}
\end{eqnarray}
This can be compared to the cross section for scattering from
a free proton, which is
\begin{eqnarray}
{d^2\sigma\over d\Omega^{\prime}} =
\sigma_M\frac{E'}{E} 
\Biggl\{ \frac{G_E^2(Q^2) + \tau G_M^2(Q^2)}{1+\tau}  +
2\tau G_M^2(Q^2) \tan^2\theta/2
\Biggl\}
\label{crosssec4}
\end{eqnarray}
In both of these formulae, $\tau=Q^2/(4m^2)$, and $G_E$ and
$G_M$ are the familiar electric and magnetic form factors,
related to $F_1$ and $F_2$:
\begin{eqnarray}
G_E&=& F_1-\tau F_2\nonumber\\
G_M&=& F_1+F_2 \, .
\end{eqnarray}

%%%%%%%%%%%%%%%%%%%%%%%%%%%%%%%%%%%%%%%%%%%%%%%%%%%%%%%%%%%%%%%%%%%%%%%%%%

\section{predictions of the relativistic impulse
approximation}

\subsection{How good is the RIA?}

%-------------------------------------------------------------
% Figure 4   Bernheim data
%
\begin{figure}
\vspace*{0.8in}
\centerline{\epsfysize=5.0in\epsffile{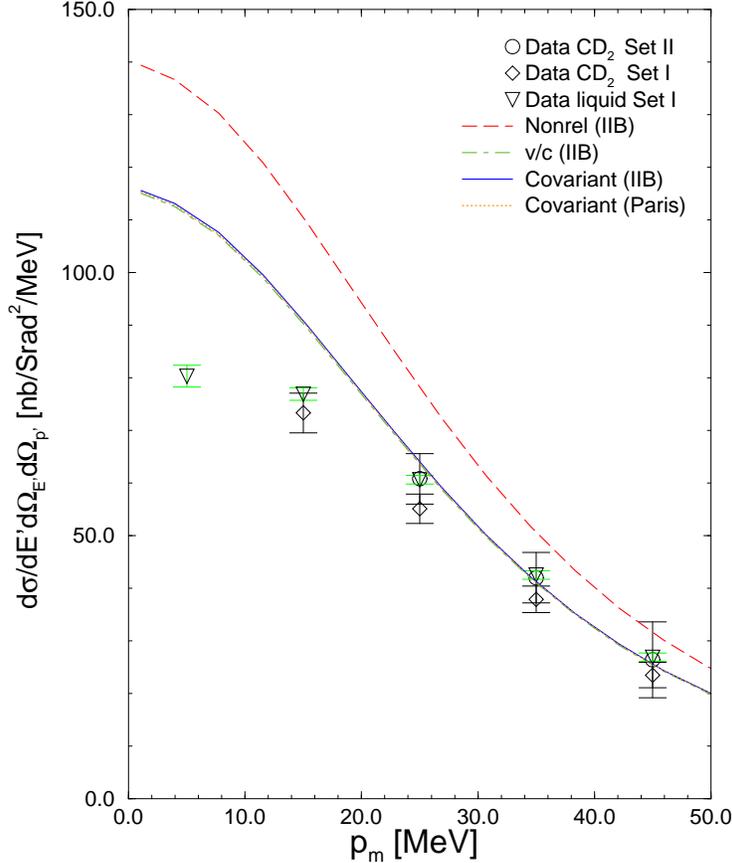}}
%\vspace{0.5in}
\caption{\footnotesize\baselineskip=12pt The Bernheim data at low
missing momentum.  Note that the relativistic effects (mostly from the
current operator) are  significant.  [Three of the lowest $p_m$ data
points, possibly contaminated by bremstrahluung, have been omited from
the figure.] }
\label{figB1}  
\end{figure}
%-----------------------------------------------------

%-------------------------------------------------------------
% Figure 5   Bernheim data over full range
%
\begin{figure}
% \vspace*{0.8in}
%\epsfysize=3.5in 
\centerline{\epsfysize=3.5in\epsffile{%/u/home/jadam/draft_qed/
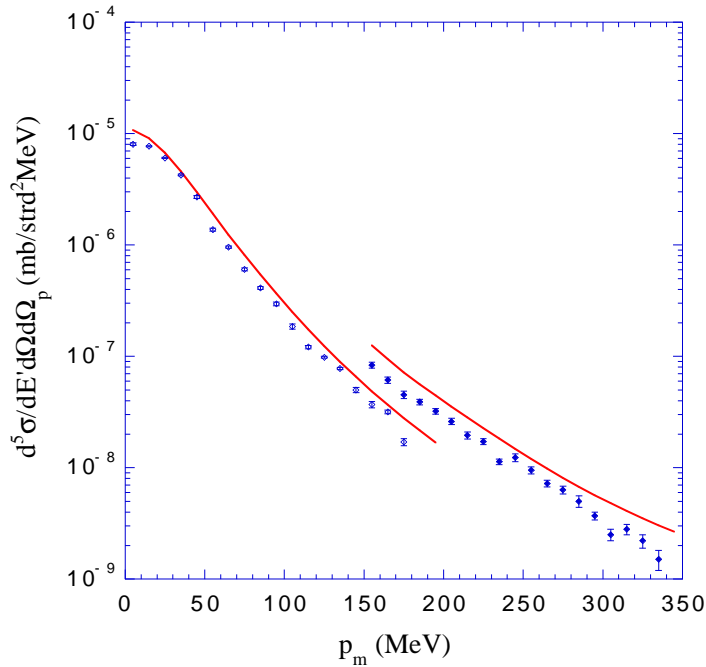}}
%\vspace{0.5in}
\caption{\footnotesize\baselineskip=12pt The acceptance averaged
relativistic RIA calculation  compared to the Bernheim data.}
\label{figB2}  
\end{figure}
%-----------------------------------------------------

%-------------------------------------------------------------
% Figure 8   Bernheim ratio full range
%
\begin{figure}
%\vspace*{0.8in}
%\epsfysize=3.5in 
\centerline{\epsfysize=3.5in\epsffile{%/u/home/jadam/draft_qed/
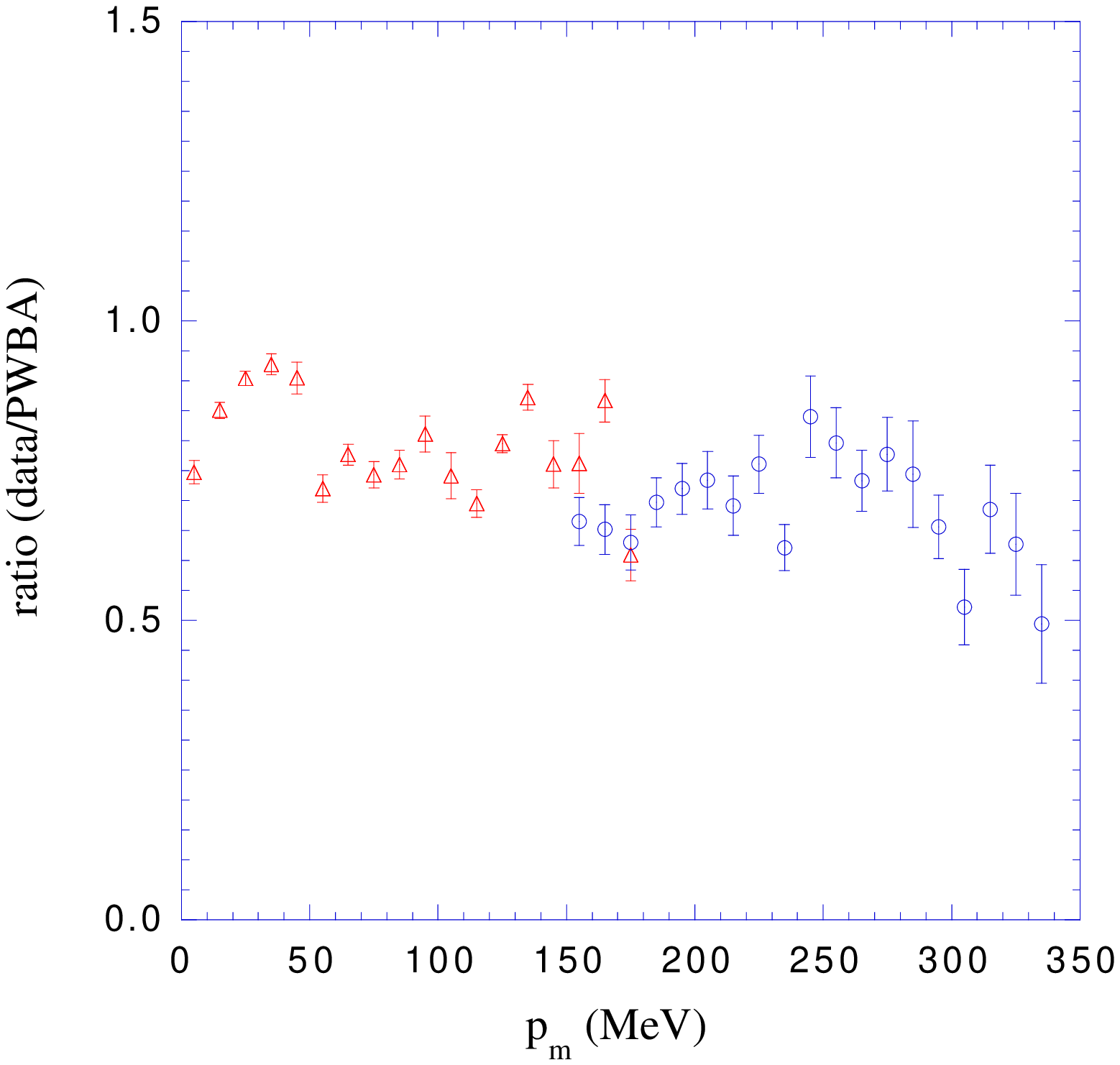}}
%\vspace{0.5in}
\caption{\footnotesize\baselineskip=12pt The ratio of the Berheim data
to the acceptance averaged relativistic RIA calculation shown in
Fig.~[7].  The triangles are the low $p_m$ data set, and the circles
the high
$p_m$ set.}
\label{figB3}  
\end{figure}
%-----------------------------------------------------
 
The low energy Bernheim (1981) data \cite{bernheim} are shown 
in  Fig.~\ref{figB1}.  This figure shows (i) that the
relativistic effects  are large and (ii) at these low
missing momenta  there is no evidence for relativistic
effects of higher order in
$v/c$, nor for model dependencies coming from the difference between
the Argonne V18 and Model IIB wave functions.  The importance of
relativistic effects in the $d(e,e'p)n$ reaction has already been
emphasized in Ref.~\cite{Beck}, and is confirmed in our
calculations.  For this reason we will dispense with further
nonrelativistic calculations, and present only
calculations with relativistic effects included.  

Figure \ref{figB1}, and the accompanying Figs.~\ref{figB2} and
\ref{figB3}, also show that the low energy cross section is reasonably
well approximated by the (relativistic) RIA.  The agreement
is at the level of $\pm50\%$ for a drop in the cross section of four
orders of magnitude.  We take this as evidence that the RIA is
sufficient for the kind of crude survey carried out in this paper.

\subsection{Covariant RIA predictions for high $Q^2$}

\begin{table}
\caption{Overview over the kinematics employed in the calculation of the
differential cross section and asymmetry $A_{\phi}$. }
\label{kintab}
\begin{ruledtabular}
\begin{tabular}{cccccccc}
$Q^2$ (GeV $^2$) & $x$ & $q_{_L}$ (GeV) & $\nu$ (GeV) &
${\rm p}_2^{min}$ (GeV) &
 ${\rm p}_2^{max}$ (GeV) & $E_{np}$ (GeV) & $T_{pn}^{Lab}$ (GeV)  \cr
\tableline
0.5 & 0.5 & 0.885 &  0.533 & 0.214 &  1.099 & 0.362 & 0.793 \cr
0.5 & 1.0 & 0.756 &  0.266 & 0.004 &  0.752 & 0.126 & 0.261 \cr
0.5 & 1.25 & 0.738 &  0.213 & 0.081 &  0.657 & 0.076 & 0.155 \cr
0.5 & 1.5 & 0.729 &  0.178 & 0.152 & 0.577  & 0.041 & 0.084 \cr
0.5 & 1.8 & 0.722 &  0.148 & 0.246 &  0.477 & 0.012 & 0.025 \cr
0.5 & 1.89 & 0.721 & 0.141 & 0.285 & 0.436 & 0.005 &  0.011 \cr
0.5 & 1.97 & 0.720 & 0.136 & 0.360 & 0.360 & 0 &  0 \cr
\hline
1 &  0.5 &  1.461 &  1.065 & 0.265 &  1.726 & 0.674 & 1.591 \cr
1 &  1.0 & 1.133 &  0.533 & 0.003 & 1.130 &  0.247 & 0.527 \cr
1 &  1.25 & 1.087 &  0.426 & 0.108 &  0.979 & 0.151 & 0.314 \cr
1 &  1.5 & 1.061 &  0.355 & 0.207 &  0.854 & 0.084 & 0.172 \cr
1 &  1.8 & 1.043 &  0.296 & 0.340 & 0.703 & 0.027 & 0.054 \cr
1 &  1.95 & 1.037 &  0.273 & 0.446 &  0.590 & 0.004 & 0.009 \cr
1 &  1.98 & 1.035 &  0.269 & 0.518 &  0.518 & 0 & 0 \cr
\hline
2 & 0.5 &  2.557 &  2.130 &  0.310 &  2.867 & 1.206 & 3.186 \cr
2 & 1.0 &  1.770 &  1.065 & 0.003 &  1.768 &  0.470 & 1.058\cr
2 & 1.25 &  1.651 & 0.852 & 0.141 &  1.510 & 0.293 & 0.633 \cr
2 & 1.5 & 1.582 &  0.710 & 0.280 & 1.303 & 0.167 & 0.349 \cr
2 & 1.8 & 1.533 &  0.592 & 0.473 & 1.060 & 0.055 & 0.112 \cr
2 & 1.97 & 1.514 &  0.541 & 0.667 & 0.847 & 0.005 & 0.010 \cr
2 & 1.99 & 1.512 &  0.535 & 0.756 & 0.756 & 0 & 0 \cr
\hline
3 &  0.5 & 3.634 & 3.195 & 0.331 & 3.966 & 1.658 & 4.781 \cr
3 &  1.0 & 2.356 & 1.598 & 0.002 & 2.354 & 0.673 & 1.589 \cr
3 &  1.25 & 2.153 & 1.278 & 0.162 & 1.991 & 0.427 & 0.951 \cr
3 &  1.5 & 2.033 &  1.065 & 0.329 & 1.704 & 0.247 & 0.526 \cr
3 &  1.8 & 1.946 &   1.946 & 0.574 & 1.372 & 0.084 & 0.171 \cr
3 &  1.98 &  1.911 &  0.807 & 0.859 & 1.052 & 0.005 & 0.010 \cr
3 &  1.99 &  1.909 &  0.802 & 0.954 & 0.954 & 0 & 0 \cr
\end{tabular}
\end{ruledtabular}
\end{table}

We surveyed the $d(e,e'p)n$ reaction over the wide range of
kinematical conditions summarized in Table \ref{kintab}.  We
did calculations at 4 different four-momentum transfers $Q^2=$
0.5, 1, 2, and 3 GeV$^2$, and 6 different values of the
Bjorken variable $x=$ 0.5, 1.0, 1.25, 1.5, 1.8, and just below
2, near the highest value accessible in the $d(e,e'p)n$
reaction at a given Q$^2$. [The value of x=2 can be reached
only in the elastic reaction, not in electrodisintegration.] 
In addition to $Q^2$ and $x$, Table \ref{kintab} gives the
transferred energy $\nu$, the magnitude of the transferred
3-momentum $q_{_L}= |{\bf q}_{_L} |$, and the range in missing
momentum, where ${\rm p}_2^{min}$ corresponds to a value of
$\theta^* = 0^\circ$, and ${\rm p}_2^{max}$ corresponds to a
value of
$\theta^* = 180^\circ$. In addition, we list the value of
the final state $np$ relative energy (in the c.m.~system),
$E_{np}$, which is
\begin{equation} 
E_{np} = \sqrt{(\nu + M_d)^2 - q_{_L}^2} - m_p - m_n\, ,
\end{equation} 
and the kinetic energy of the $pn$ system in the Lab
frame, 
\begin{equation} 
T_{pn}^{Lab} = \frac{M_d^2}{2 m} - 2 m + \frac{Q^2}{2 m}
\left(\frac{M_d}{m \,x } - 1\right)\, . 
\end{equation}
The highest accessible value of $x$ for a certain
$Q^2$ is given by 
\begin{equation} 
x_{max} = \frac{Q^2 \, M_d}{m \, ( 4 m^2 - M_d^2 +
Q^2)}\, , 
\end{equation}
which comes from the requirement that $T_{pn}^{Lab}>0$. In
Table \ref{kintab}, we list the kinematic variables for $x =
x_{max}$ in the last line for each $Q^2$. The closest $x$
value to that is the highest one for which we present
calculations later on; it is characterized by a kinetic energy
in the Lab frame of roughly 10 MeV. 
 
Table \ref{kintab} shows that the values of the transferred 
energy and transferred three-momentum are closest for low
$x$.  In nonrelativistic reduction schemes one often assumes 
$\nu<<q_{_L}$, which is clearly not the case for low $x$. 
Note also that the $np$ relative energy is highest for low
$x$.  These imply that relativistic effects should be very
strong in this region.  This is interesting as it allows for a
description of final state interaction by Glauber theory
\cite{Glauber}.  Although final state interactions (FSIs) are
not considered in this present study, it is useful to know 
that they can be calculated reliably in this kinematic region. 

At high $x$, the $np$ relative energy is very small, on the
order of a few tens of MeV. Under these kinematic conditions,
the system is reminiscent of a bound system, and one might
realistically expect wave function physics to be important
here, e.g.~the presence of the relativistic $P$-waves. The
calculation of final state interactions proceeds by including
the lowest partial waves. So, we have reliable methods for the
calculation of FSIs both at high $x$ and at low $x$. In the
region in between, the description of FSIs is more involved and
accordingly more difficult.

The case of $x=1$ roughly corresponds to the quasi-free case.
Strictly speaking, the quasi-free case corresponds to
\begin{equation}
\nu = \frac{Q^2}{2 m_N} - E_{b}\, , 
\end{equation}
where the binding energy $E_b$ leads to a small deviation 
from $x = 1$. However, the binding energy for the deuteron is
small and we will refer to $x = 1$ as the quasi-free case in
the following discussion.

%-------------------------------------------------------------
% Figure 9  The differential cross section at Q^2 = 0.5 GeV^2
\begin{figure}%[!h,t]
\begin{center}
\mbox{\epsfysize=20cm \epsfbox{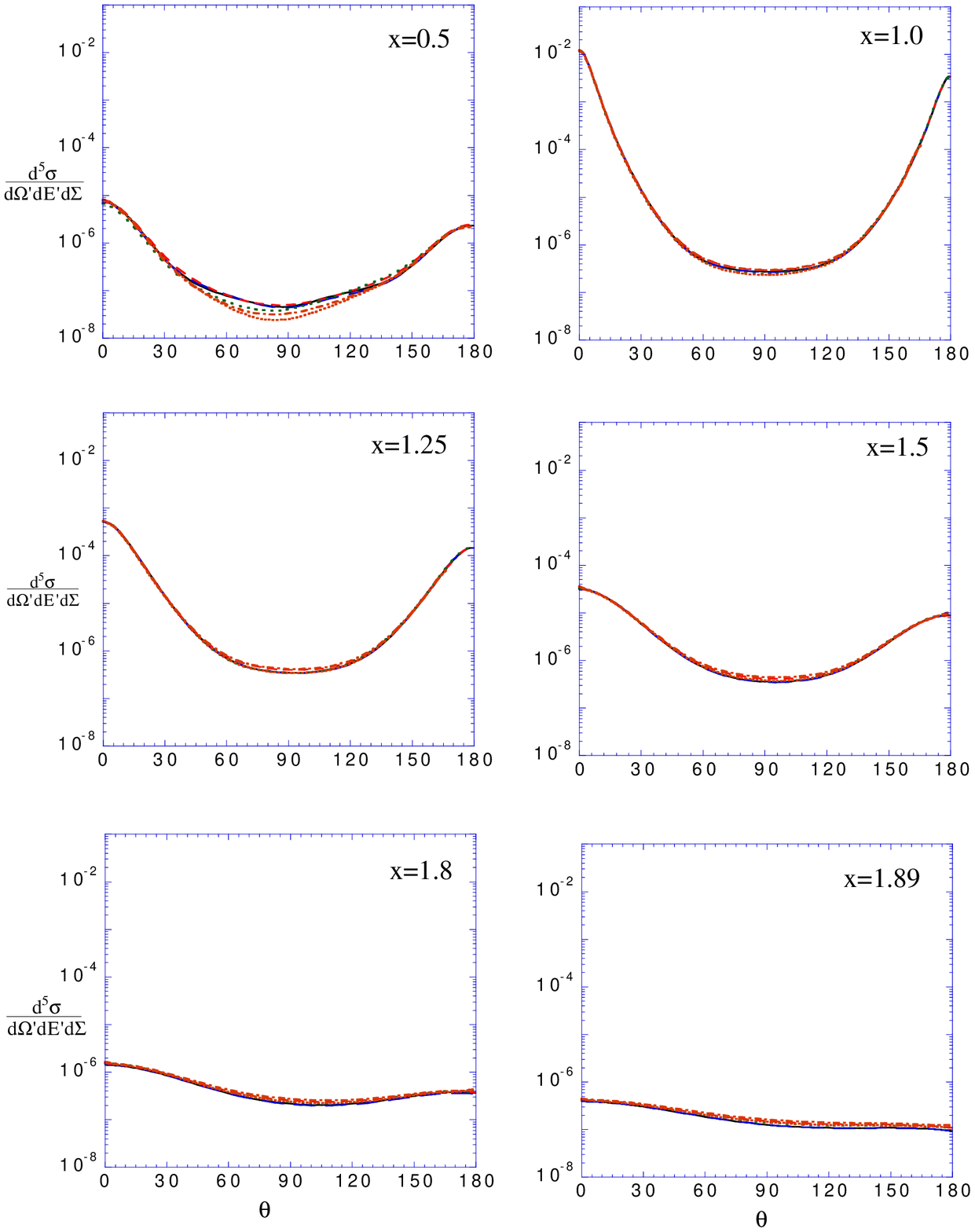}}
%xsec.q05.ps}}
%\leavevmode
%\epsfig{file = /u/home/sabine/cures/res2/xsec_q05.ps, height=12cm}
\end{center}
\vspace{0cm}
\caption{\footnotesize\baselineskip=12pt The differential cross
section {\it in the c.m.~system\/} at $Q^2$ = 0.5 GeV$^2$ and for
$x=0.5, 1, 1.25, 1.5, 1.8$, and 1.89. Each panel shows the six
calculations described in the text: C-IIB (long-dashed line),
C-IIB-noP (solid line), C-AV18 (dashed line), AA-$v/c$ (widely
dotted line), JD-full (dash-dotted line), and JD-1st (closely
dotted line).  Note that the different approximations are hard
to distinguish, as discussed in the text.} 
\label{figdsigmaq05}  
\end{figure} 
%-------------------------------------------------------------

%-------------------------------------------------------------
% Figure 10  The differential cross section at Q^2 = 1 GeV^2
\begin{figure}%[!h,t]
\begin{center}
\mbox{\epsfysize=20cm \epsfbox{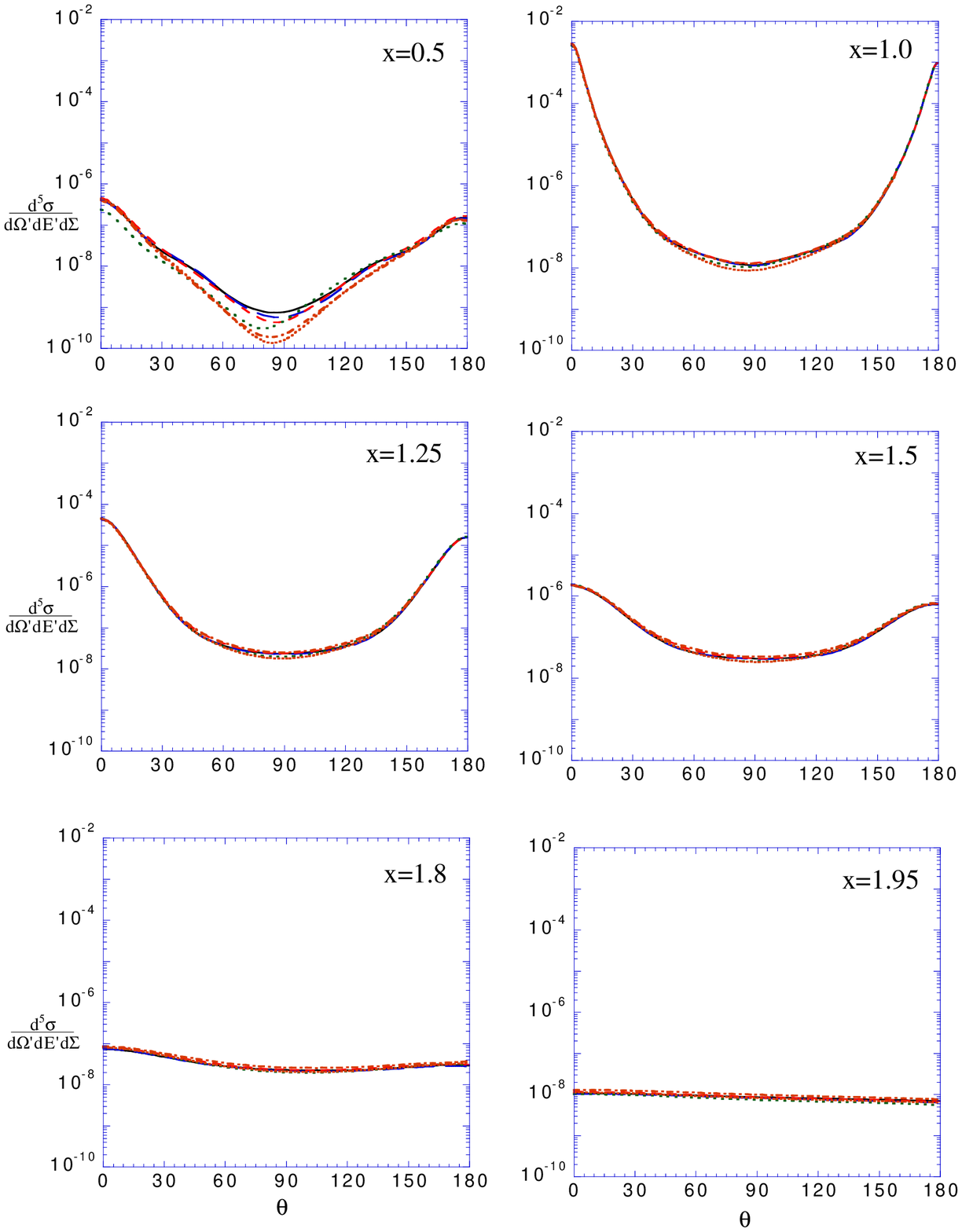}}
%xsec.q1.ps}}
%\leavevmode
%\epsfig{file = /u/home/sabine/cures/res2/xsec_q1.ps, height=12cm}
\end{center}
%\vspace{-2cm}
\caption{\footnotesize\baselineskip=12pt The differential cross
section at $Q^2$ = 1 GeV$^2$  and for various $x$. The meaning of the
curves is the same as in Fig. \protect{\ref{figdsigmaq05}}.}
\label{figdsigmaq1}  
\end{figure}
%-------------------------------------------------------------

%-------------------------------------------------------------
% Figure 11  The differential cross section at Q^2 = 2 GeV^2 
\begin{figure}%[!h,t]
\begin{center}
\mbox{\epsfysize=20cm \epsfbox{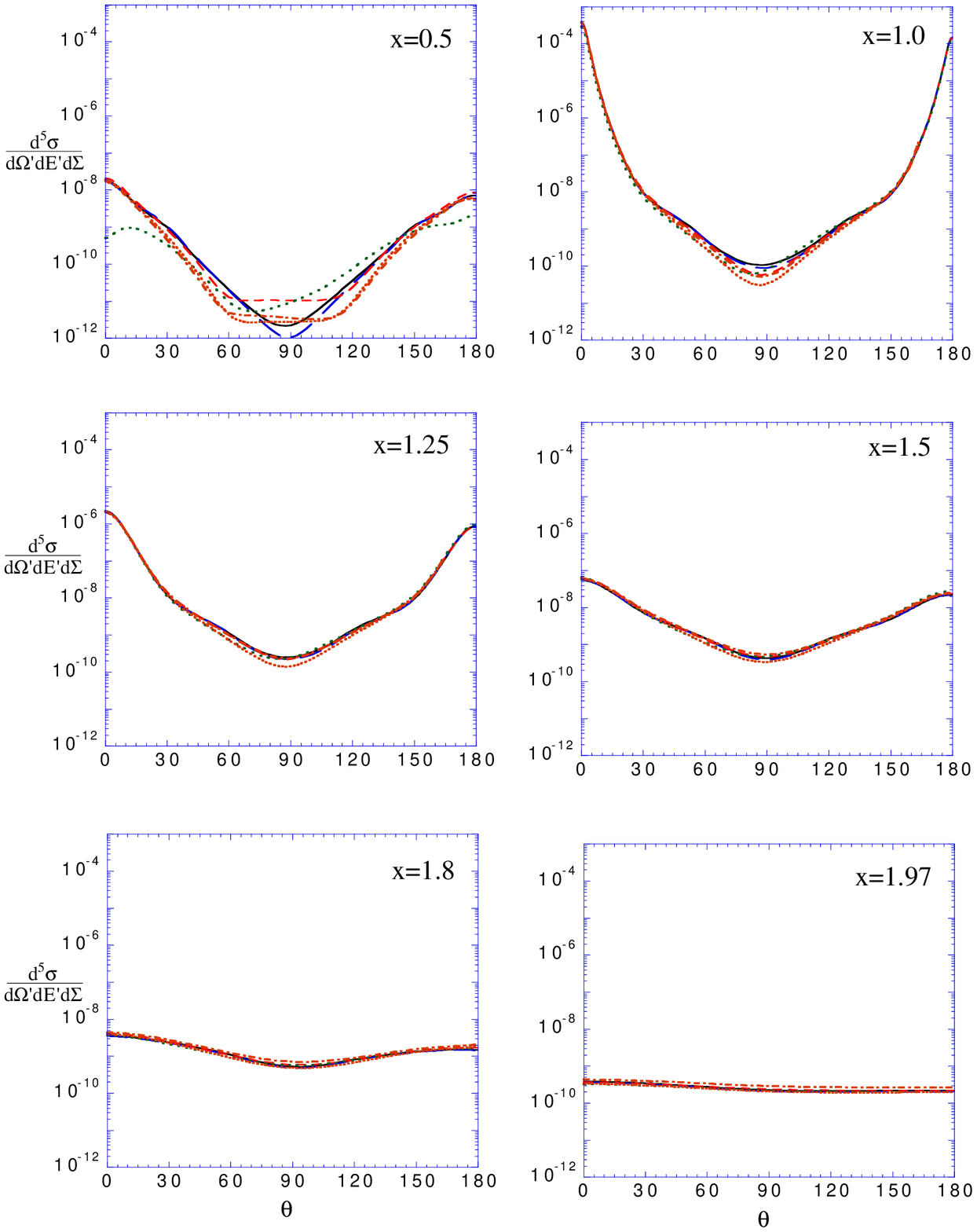}}
%xsec.q1.ps}}
%\leavevmode
%\epsfig{file = /u/home/sabine/cures/res2/xsec_q1.ps, height=12cm}
\end{center}
%\vspace{-2cm}
\caption{\footnotesize\baselineskip=12pt The differential cross
section at $Q^2$ = 2 GeV$^2$  and for various $x$. The meaning of the
curves is the same as in Fig. \protect{\ref{figdsigmaq05}}.}
\label{figdsigmaq2}  
\end{figure}
%-------------------------------------------------------------

%-------------------------------------------------------------
% Figure 12  The differential cross section at Q^2 = 3 GeV^2
\begin{figure}%[!h,t]
\begin{center}
\mbox{\epsfysize=20cm \epsfbox{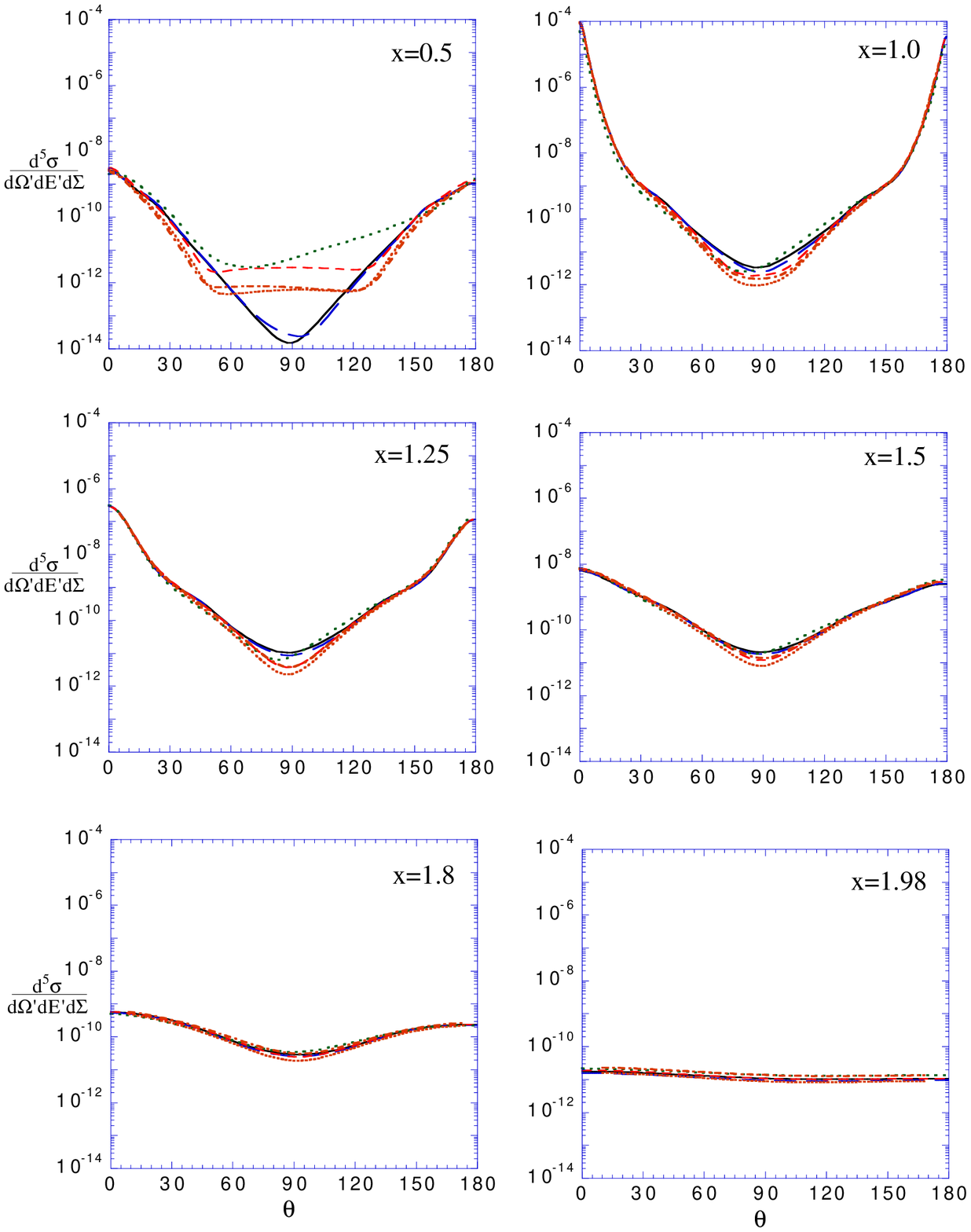}}
%xsec.q3.ps}}
%\leavevmode
%\epsfig{file = /u/home/sabine/cures/res2/xsec_q3.ps, height=12cm}
\end{center}
%\vspace{-2cm}
\caption{\footnotesize\baselineskip=12pt The differential cross
section at $Q^2$ = 3 GeV$^2$  and for various $x$. The meaning of the
curves is the same as in Fig. \protect{\ref{figdsigmaq05}}. }
\label{figdsigmaq3}  
\end{figure}
%-------------------------------------------------------------

\subsection{Six approximations}
  
Six different theoretical approximations will be discussed in the
following.  The first three are based
on the covariant spectator RIA presented in detail in this paper, and
will be denoted ``C-IIB'', ``C-IIB-noP'', and ``C-AV18''.      

\begin{itemize} 
 
\item{\bf C-IIB} is calculated using the covariant IIB deuteron wave
function obtained from the successful IIB $NN$
interaction \cite{GVOH92}.   This wave function and the relativistic
spectator model have been previously used to successfully explain the
elastic deuteron form factors \cite{dff}.  The full wave function
has 4 components: the familiar S- and D-states, and two small
P-states of relativistic origin.   

\item{\bf C-IIB-noP}  is calculated using covariant IIB S- and
D-state wave functions, but setting the small relativistic
P-state components to zero. 

\item{\bf C-AV18}   is calculated from the  covariant spectator RIA
formulae using the S- and D-state Argonne V18 deuteron wave
functions \cite{AV18} (instead of the wave functions derived
from the IIB one boson exchange model) and setting the
P-state components to zero. This is included among the
covariant models even though, strictly speaking, the wave
function is not consistent with the covariant formalism.  This
model is very similar to the ``covariant model'' previously
discussed by Arenh\"ovel \cite{Beck}, but they used ordinary
spin instead of helicity, and the Paris deuteron wave
functions instead of the (very similar) AV18 wave functions.

\end{itemize}

\noindent The next three calculations are not consistently 
covariant, but they do use relativistic current operators.  
They all use the nonrelativistic Argonne V18 wave function. 
The first of these is based on the work of Adam and Arenh\"ovel
\cite{AdAr95}.  
 
\begin{itemize}

\item {\bf AA-$v/c$} uses a current operator that results from a
$v/c$-expansion of the intrinsic current \cite{FrGP,AdAr95}.
Matrix elements of this current are made frame independent by
replacing the approximate  noninvariant effective 3-momentum
transfer derived in
\cite{FrGP,AdAr95} by its invariant extension, defined to be:
\begin{equation}
  \vec{q}_{eff}^{\, \, 2} = \frac{Q^2}{1+ \frac{Q^2}{16m_N^2}} +
                               (\epsilon_d- W)^2  \, .
\end{equation} 
where $\epsilon_d, W$ are energies of $pn$ pair in its 
respective c.m.\ frame in the initial and final states. This
prescription is  similar to one proposed long ago for the
elastic scattering by  J.~Friar \cite{Friar}, but differs from
calculations by  Arenh\"ovel et al \cite{Beck}.  

\end{itemize}

\noindent The next two versions are from Jeschonnek and
Donnelly \cite{DonnSab}. 

\begin{itemize}

\item {\bf JD-full} uses a fully relativistic, positive energy 
current  operator.  This  covariant current differs from the
spectator one by certain off-mass-shell extensions studied in a
recent paper by two of the authors \cite{jvo}. 

\item{\bf JD-1st} uses a current operator expanded to first 
order in the initial nucleon momentum, with all other terms
retained fully.  This  approximate ``first order'' form should
be closer to the covariant one than the traditional $v/c$ 
current mentioned above, since an expansion is made only in
terms of the moderate momenta of nucleons in the initial
nucleus.  

\end{itemize}

The relativistic one-nucleon current used here in the JD-full
calculation has been recently employed by Donnelly
\cite{DonnSab} in studies of $(e,e'N)$ reactions. In these 
studies relativistic  models appeared to be far more
successful than nonrelativistic ones \cite{Beck,DonnSab}. It
is, however, a non-trivial task to extend them beyond RIA. 

Final state interaction and meson-exchange currents (MECs) 
have been so far included into realistic calculations mostly
within approximate frameworks based on various expansions of
the nuclear operators in terms of supposedly small
momenta \cite{DonnSab,Arenhovel,KrF}. We do not intend to give an 
exhaustive survey of those techniques, neither do we dare to
compete in completeness and consistency with recent elaborate
calculations \cite{Arenhovel,Amaro}. We only show in our
figures the results obtained with the various one-nucleon
currents introduced above.  
 
While a much more comprehensive study of relativistic effects,
including relativistic expansions of $\pi$ exchange currents 
and heavy meson exchange currents including boost terms,
$\gamma \pi \rho$ and $\gamma \pi \omega$ currents, and isobar
contributions, was performed by Ritz et al \cite{Arenhovel}
for lower energies, we focus on high energies. Here, high
energies mean the GeV region, accessible by CEBAF at Jefferson
Lab and even the new kinematic regime opening up with the
planned 12 GeV upgrade of CEBAF.  Our C-IIB calculation given
here is fully covariant, and part of a consistent treatment of
the nuclear dynamics and the one-body current that does not
rely on any kind of nonrelativistic expansion.

None of our calculations is complete, the purpose is rather to 
explore various experimentally feasible kinematical regions to find 
those for which  the complete microscopic calculations and precise
measurements would be  worthwhile. Nevertheless, the
variations between results obtained with versions of covariant currents and
their approximations, as well as those between two covariant or two approximate
formulations themselves, should provide some insight on the region of validity
of the expansions and approximations used.

\subsection{Differential cross section}

The differential cross section (\ref{crosssec}) is given in
Figs.~\ref{figdsigmaq05}--\ref{figdsigmaq3}.  The six panels in each
figure all have the same scale, so the relative size of the 
cross section may be seen at a glance.  Examination of these
figures shows that the magnitude of the cross section depends 
strongly on $Q^2$, $x$, and $\theta^*$ . The bulk
feature of the differential cross section consists of the two
peaks at $\theta^* = 0^\circ$ and $\theta^* =
180^\circ$. The first peak corresponds to the impulse approximation
contribution, where the photon couples to the proton which is
detected later on, and the second peak at $180^\circ$ corresponds
to the Born contribution, where the photon interacts with the
neutron. For $x\leq$ 1.5 the two peaks are well separated
because (cf.~Fig.~\ref{circles}) the nucleons have very
different momenta at $\theta^*\simeq0$ and 180$^\circ$.  In
this case one of the two RIA contributions [recall
Eq.~(\ref{PWBA0}) and Fig.~\ref{fig_diag}] is much larger than
the other.  

%%%%%%%%%%%%%%%%%%%%%%%%%%%%%%%%%%%%%%%%%%%
\begin{table}
\begin{minipage}{6.5in}
\caption{\label{tab:errors} Sensitivity to relativistic 
effects (R) compared to estimated
measurement errors (M) for selected
$Q^2$ and $x$.} 
\begin{ruledtabular}
\begin{tabular}{c|cccc|cccc|cccc}
$Q^2$ (GeV)$^2$ & &&1&&&&2&&&&3& \cr
$x$& $\;$0.5 $\;$& $\;$1.0 $\;$ &  $\;$1.5 $\;$ &
 $\;$1.95 $\;$ & $\;$0.5 $\;$ &  $\;$1.0 $\;$ &  $\;$1.5 $\;$ &
 $\;$1.97 $\;$ & $\;$0.5 $\;$ &  $\;$1.0 $\;$ &  $\;$1.5 $\;$ &
 $\;$1.98  
\cr  & && & &&&&&&&&  \cr
$d\sigma$ R (in \%) & 50 & 20& 10 & 10 & 100 & 50 & 20 &
10 & 100+& 50 & 50 & 20 
\cr
$d\sigma$ M (in \%)  & 1 &1&1-3 &5 & 5 & 1-10&
1-10&&10+&1-20&1-20&50 
\cr & && & &&&&&&&&  \cr
$A_\phi$ R (absolute)   & 0.1 & $<$0.1 &$<$0.1 &$<$0.05  &0.2 &
0.1 & 0.1 & 0.01&0.5 & 0.2 & 0.1 & $<$0.05
\cr
$A_\phi$ M (absolute)   &0.1-0.2 & & &0.1  &0.2-0.5&&&0.1-0.2
&$<$0.1&&& $>$0.5
\cr
\end{tabular}
\end{ruledtabular}
\end{minipage}
\end{table}
%%%%%%%%%%%%%%%%%%%%%%%%%%%%%%%%%%%%%%%%%%%%

However, if we wish to probe the deuteron wave function at high
momentum, we will seek the region near
$\theta^*\simeq90^\circ$, where both nucleons have nearly the same
momenta and only high momentum components of the wave function can
contribute. In this region the cross section is very small  (and FSI
will be large), reflecting the small size of the
deuteron wave function at large momenta.  At large $x$, and in
particular near
$x\simeq2$, the two diagrams will always have large momenta (because
the relative momentum of the final state is low), and the cross
section shows no sharp forward or backward peak.  Here high momentum
components of the deuteron contribute over the whole angular range.

%-------------------------------------------------------------
% Figure 13  Ratios at Q^2 = 0.5 GeV^2
\begin{figure}%[!h,t]
\begin{center}
\mbox{\epsfysize=20cm \epsfbox{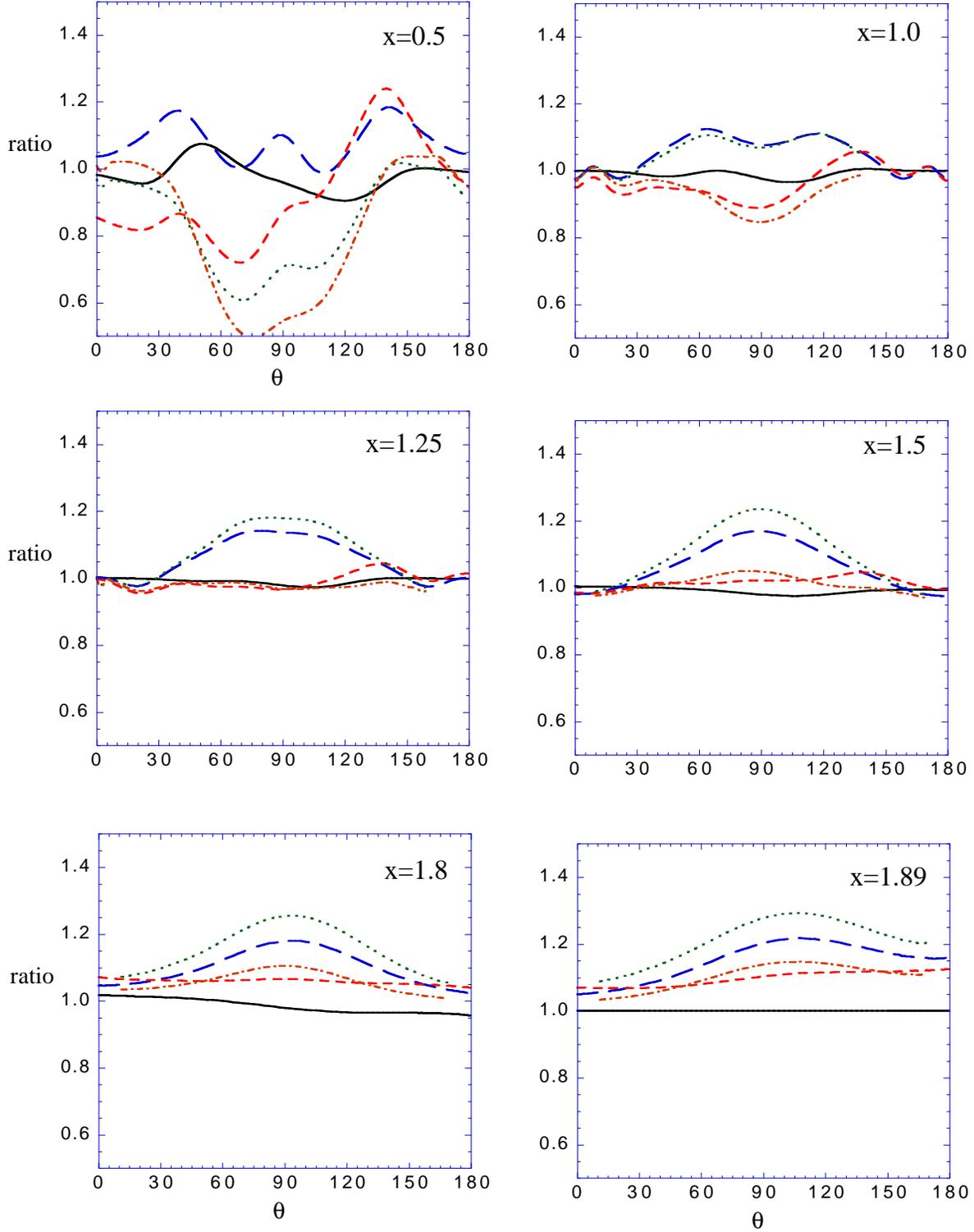}} 
%xsec.q05.ps}}
%\leavevmode
%\epsfig{file = /u/home/sabine/cures/res2/xsec_q05.ps, height=12cm}
\end{center}
\vspace{0cm}
\caption{\footnotesize\baselineskip=12pt Ratios of the differential
cross section {\it in the c.m. system\/} for the cases shown in
Fig.~\protect{\ref{figdsigmaq05}} ($Q^2$ =0.5 GeV$^2$). Here five of
the calculations are divided by the C-IIB-noP calculation: C-IIB
(solid line), C-AV18 (long dashed line), AA-$v/c$ (short
dashed line), JD-full (dotted line), and JD-1st (dash-dotted
line).}  
\label{figdsigmaq05r}  
\end{figure}
%-------------------------------------------------------------

%-------------------------------------------------------------
% Figure 14  Ratios at Q^2 = 1 GeV^2
\begin{figure}%[!h,t]
\begin{center} 
\mbox{\epsfysize=20cm \epsfbox{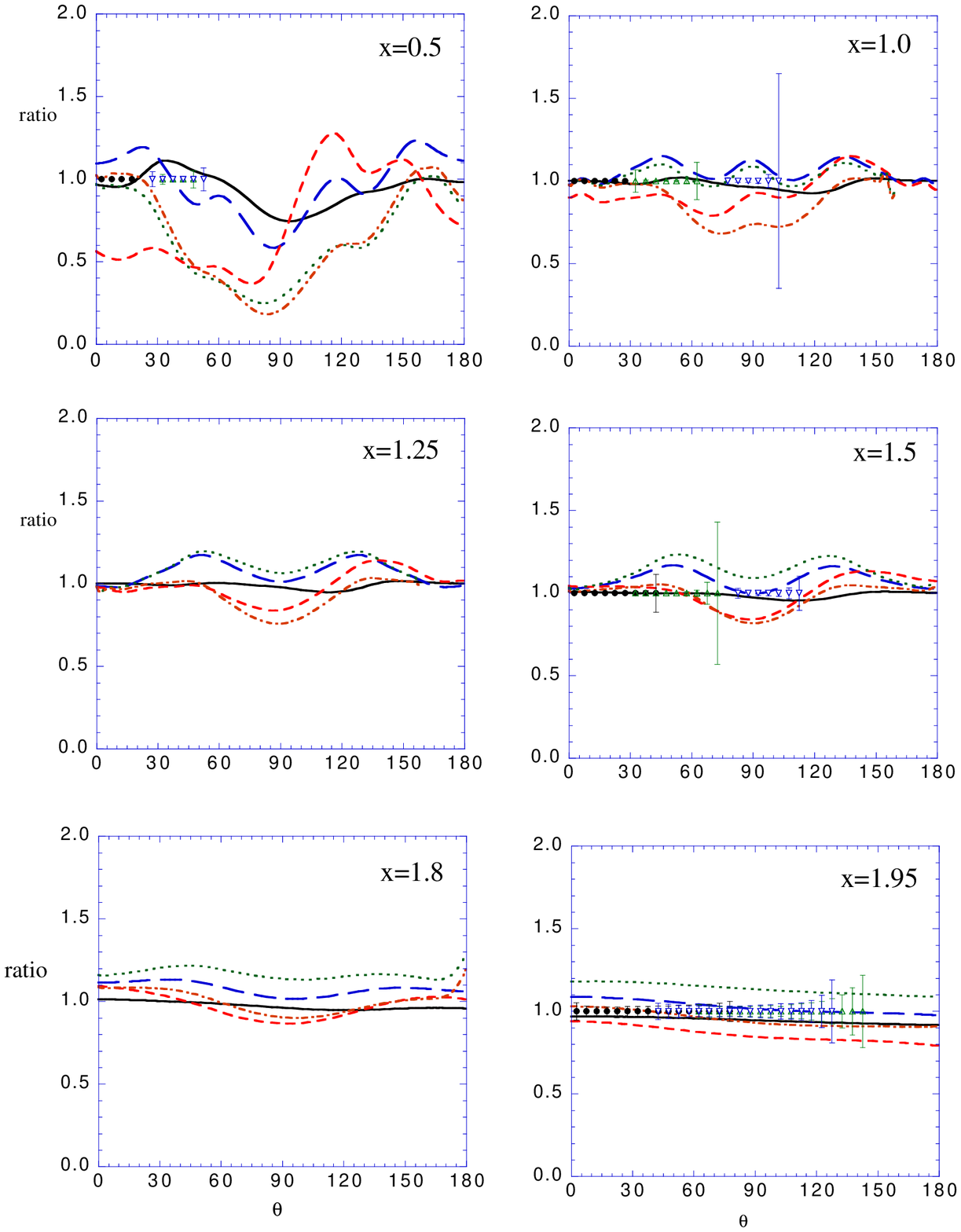}} 
%xsec.q1.ps}}
%\leavevmode
%\epsfig{file = /u/home/sabine/cures/res2/xsec_q1.ps, height=12cm}
\end{center}
%\vspace{-2cm}
\caption{\footnotesize\baselineskip=12pt Ratios of the differential
cross section at $Q^2$ = 1 GeV$^2$  and for various $x$. The meaning
of the curves is the same as in Fig. \protect{\ref{figdsigmaq05r}}. 
The estimated errors for a JLab measurement using existing equipment,
shown on panels for $x=0.5, 1.0, 1.5$, and
$1.95$, are discussed in the text.}
\label{figdsigmaq1r}  
\end{figure}
%-------------------------------------------------------------

%-------------------------------------------------------------
% Figure 15  Ratios at Q^2 = 2 GeV^2
\begin{figure}%[!h,t]
\begin{center} 
\mbox{\epsfysize=20cm \epsfbox{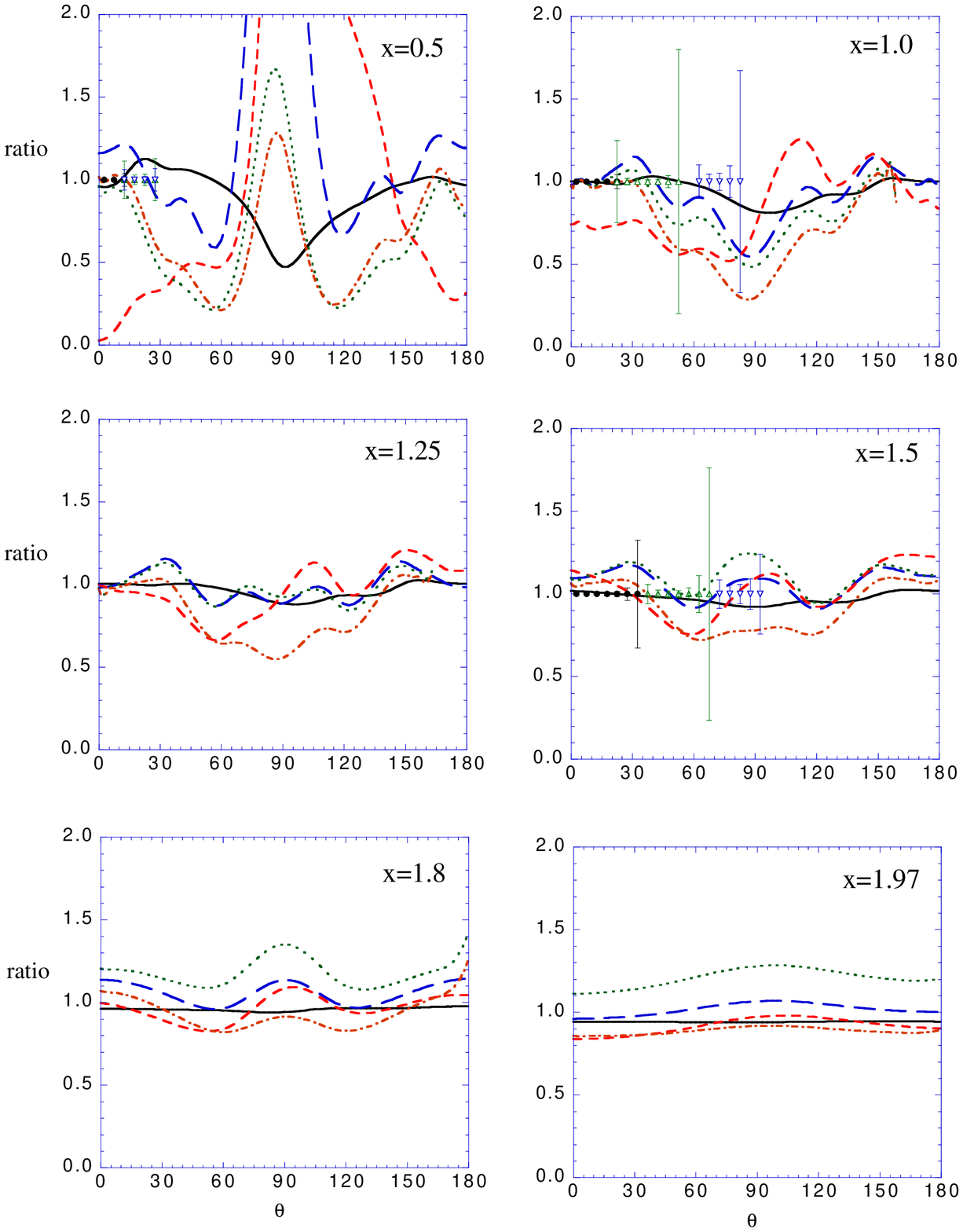}} 
%xsec.q1.ps}}
%\leavevmode
%\epsfig{file = /u/home/sabine/cures/res2/xsec_q1.ps, height=12cm}
\end{center}
%\vspace{-2cm}
\caption{\footnotesize\baselineskip=12pt Ratios of the differential
cross section at $Q^2$ = 2 GeV$^2$  and for various $x$. The meaning
of the curves is the same as in Fig. \protect{\ref{figdsigmaq05r}}. 
The estimated errors for a JLab measurement using existing equipment,
shown on panels for $x=0.5, 1.0$, and
$1.5$ are discussed in the text.}
\label{figdsigmaq2r}  
\end{figure}
%-------------------------------------------------------------

%-------------------------------------------------------------
% Figure 16  Ratios of the differential cross section at Q^2 =
%3 GeV^2
\begin{figure}%[!h,t]
\begin{center}
\mbox{\epsfysize=20cm \epsfbox{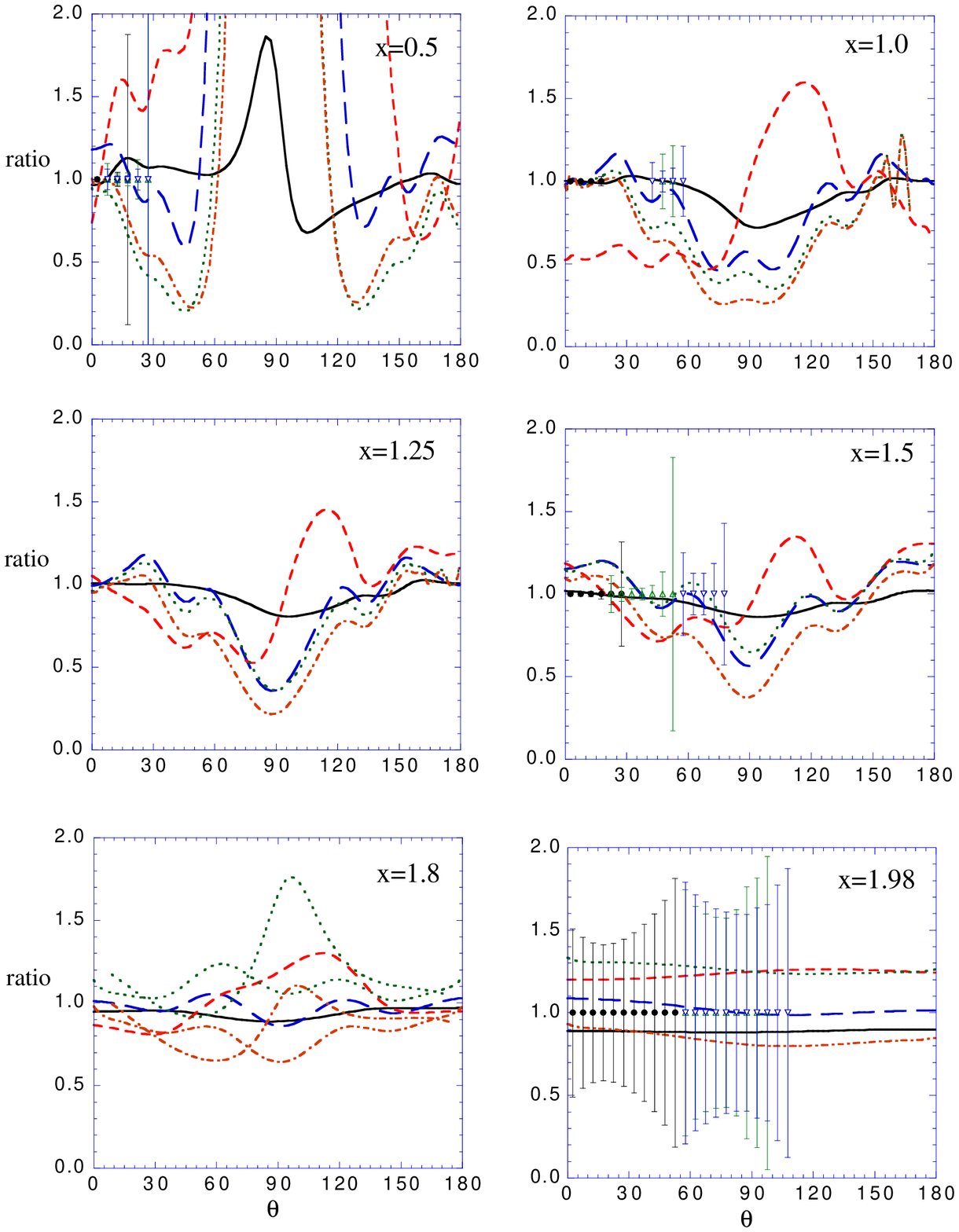}}
%xsec.q3.ps}}
%\leavevmode
%\epsfig{file = /u/home/sabine/cures/res2/xsec_q3.ps, height=12cm}
\end{center}
%\vspace{-2cm}
\caption{\footnotesize\baselineskip=12pt Ratios of the differential
cross section at $Q^2$ = 3 GeV$^2$  and for various $x$. The meaning
of the curves is the  same as in Fig. \protect{\ref{figdsigmaq05r}}. 
The estimated errors for a JLab measurement using existing equipment,
shown on panels for $x=0.5, 1.0, 1.5$, and $1.98$, are discussed in
the text.} 
\label{figdsigmaq3r}  
\end{figure}
%-------------------------------------------------------------
%-------------------------------------------------------------
% Figure 17  The asymmetry A_phi at Q^2 = 0.5 GeV^2
\begin{figure}%[!h,t]
\begin{center}
\mbox{\epsfysize=20cm \epsfbox{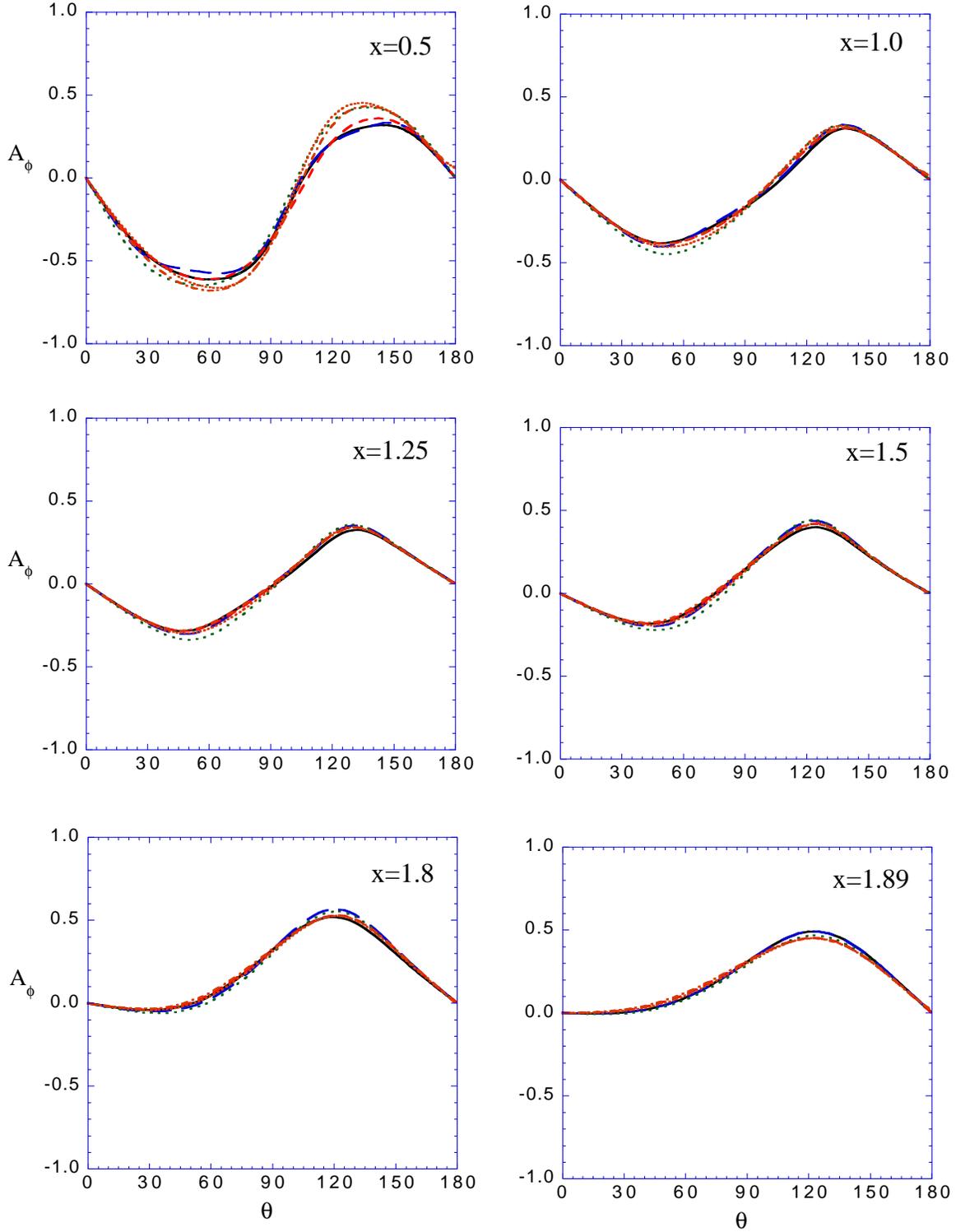}}
%aphi.q05.ps}}
%\leavevmode
%\epsfig{file = /u/home/sabine/cures/res2/aphi_q05.ps, height=12cm}
\end{center}
%\vspace{-2cm}
\caption{\footnotesize\baselineskip=12pt The asymmetry $A_{\phi}$ at
$Q^2$ = 0.5 GeV$^2$  and for various $x$. The meaning of the curves is
the same as in Fig. \protect{\ref{figdsigmaq05}}. } 
\label{figaphiq05}
\end{figure}
%-------------------------------------------------------------

%-------------------------------------------------------------
% Figure 18  The asymmetry A_phi at Q^2 = 1 GeV^2
\begin{figure}%[!h,t]
\begin{center}
\mbox{\epsfysize=20cm \epsfbox{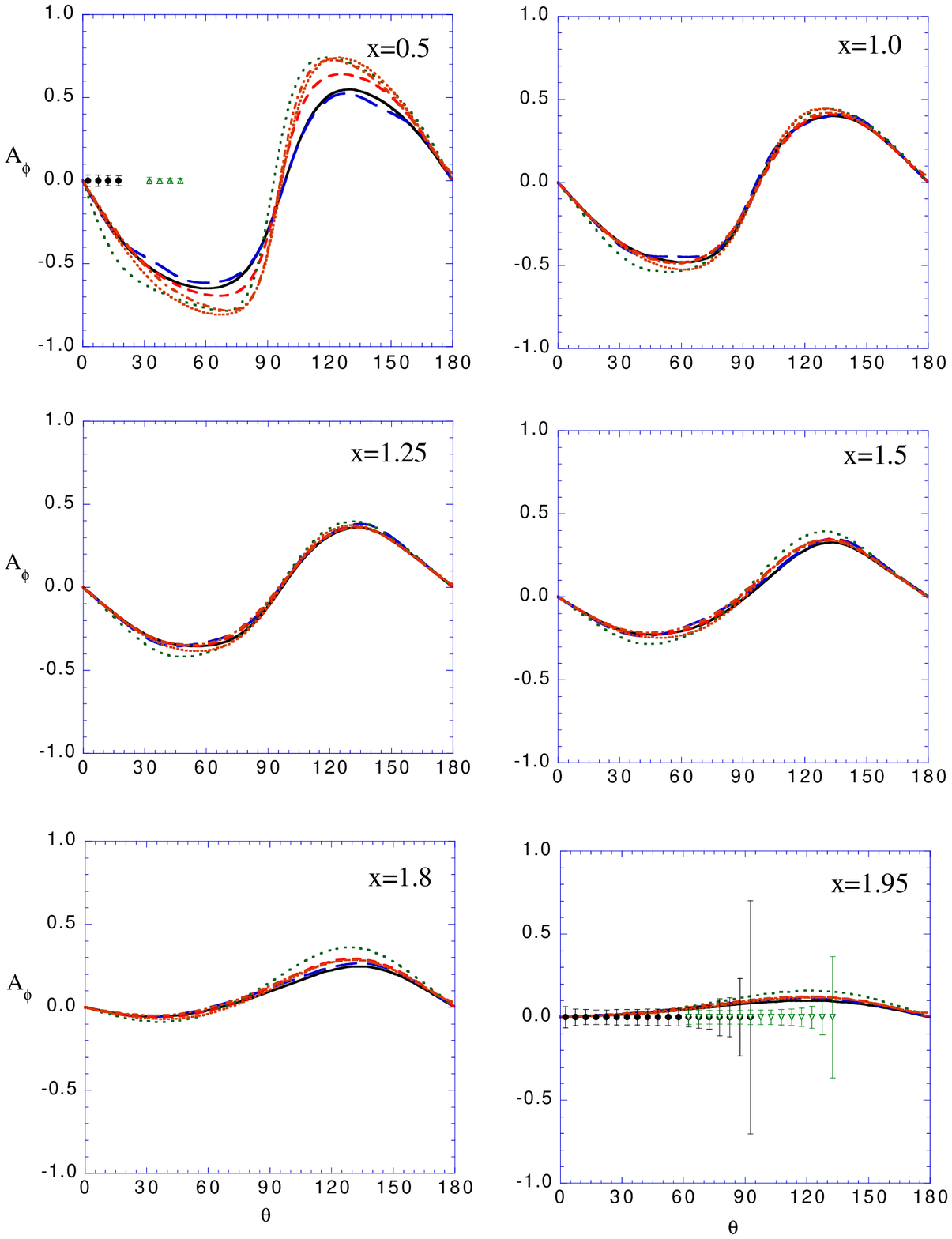}}
%aphi.q1.ps}}
%\leavevmode
%\epsfig{file = /u/home/sabine/cures/res2/aphi_q1.ps, height=12cm}
\end{center}
%\vspace{-2cm}
\caption{\footnotesize\baselineskip=12pt The $A_{\phi}$ at $Q^2$ = 1
GeV$^2$  and for various $x$. The meaning of the curves is the same as
in Fig. \protect{\ref{figdsigmaq05}}.  The estimated errors
for  a JLab measurement using existing equipment, shown on
panels for $x=0.5$ and $1.95$, are discussed in the text.}
\label{figaphiq1}
\end{figure}
%-------------------------------------------------------------

%-------------------------------------------------------------
% Figure 19  The asymmetry A_phi at Q^2 = 2 GeV^2
\begin{figure}%[!h,t]
\begin{center}
\mbox{\epsfysize=20cm \epsfbox{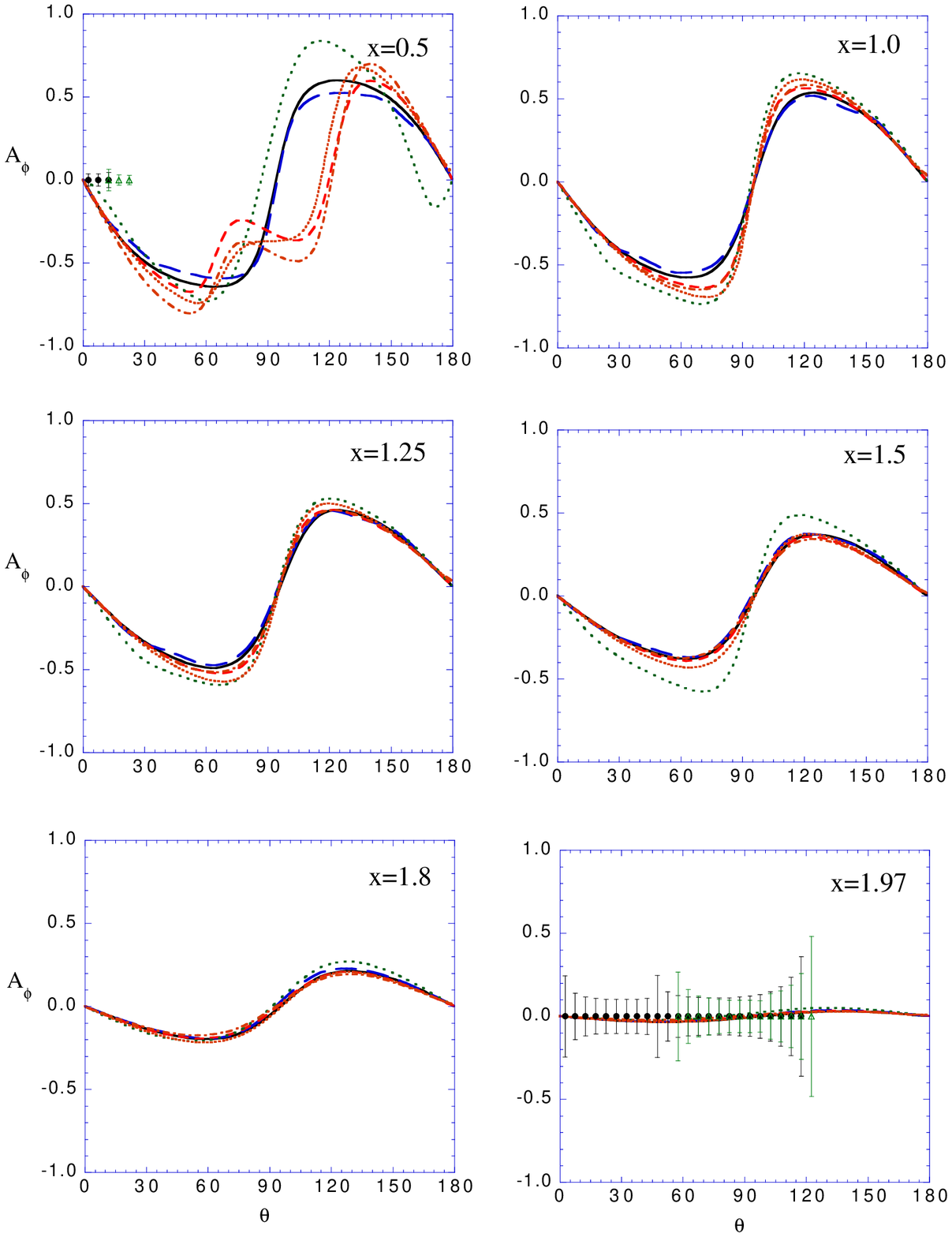}}
%aphi.q2.ps}}
%\leavevmode
%\epsfig{file = /u/home/sabine/cures/res2/aphi_q2.ps, height=12cm}
\end{center}
%\vspace{-2cm}
\caption{\footnotesize\baselineskip=12pt The $A_{\phi}$ at $Q^2$ = 2
GeV$^2$  and for various $x$. The meaning of the curves is the same as
in Fig. \protect{\ref{figdsigmaq05}}. The estimated errors for
a JLab measurement using existing equipment, shown on panels
for $x=0.5$ and $1.97$, are discussed in the text.}  
\label{figaphiq2} 
\end{figure}
%-------------------------------------------------------------
%-------------------------------------------------------------
% Figure 20  The asymmetry A_phi at Q^2 = 3 GeV^2
\begin{figure}%[!h,t]
\begin{center}
\mbox{\epsfysize=20cm \epsfbox{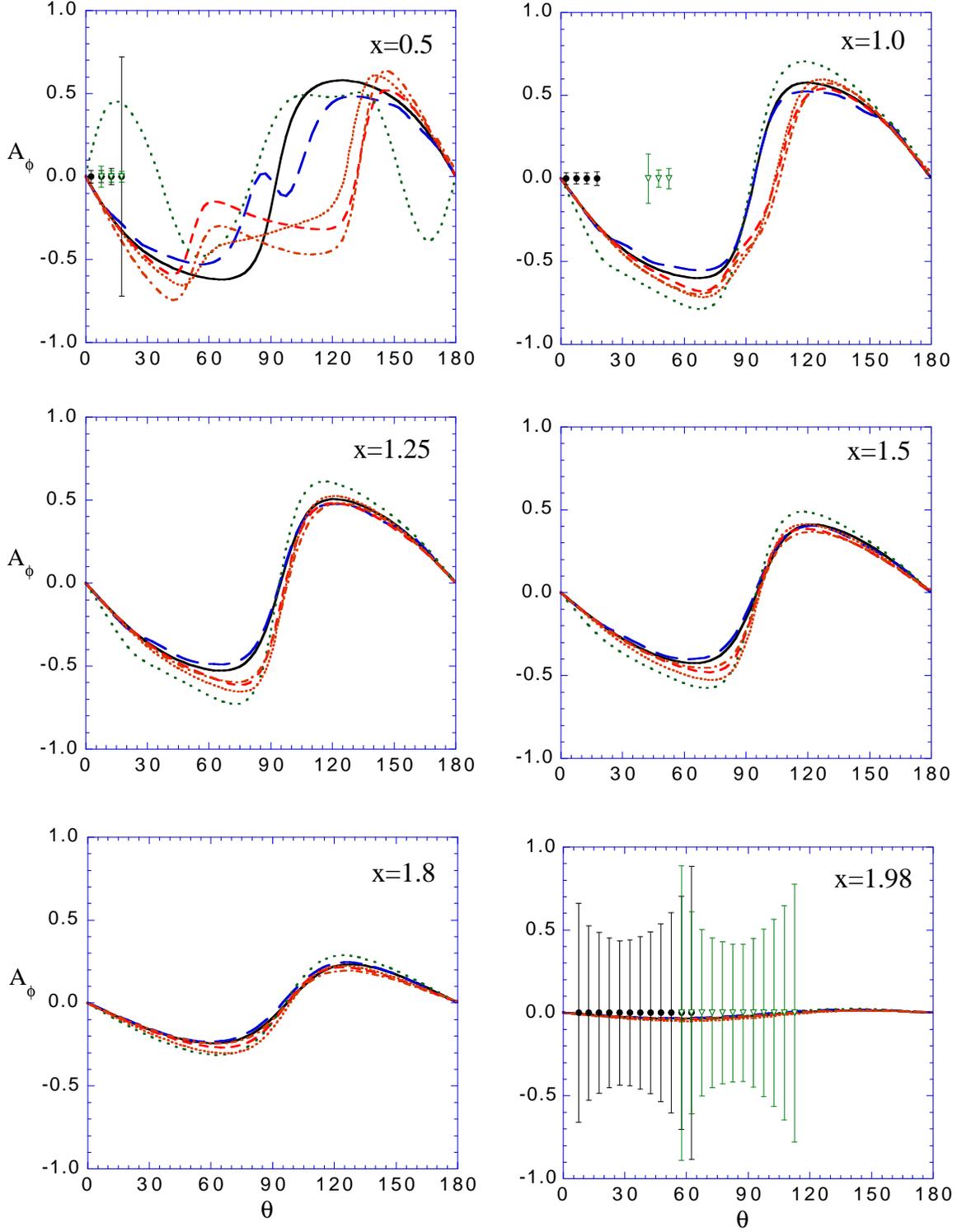}}
%aphi.q3.ps}}
%\leavevmode
%\epsfig{file = /u/home/sabine/cures/res2/aphi.q3.ps, height=12cm}
\end{center}
%\vspace{-2cm}
\caption{\footnotesize\baselineskip=12pt The $A_{\phi}$ at $Q^2$ = 3
GeV$^2$  and for various $x$. The meaning of the curves is the same as
in Fig. \protect{\ref{figdsigmaq05}}.  The estimated errors
for a JLab measurement using existing equipment, shown on
panels for $x=0.5, 1$, and $1.98$ are discussed in the
text.}
\label{figaphiq3}  
\end{figure} 
%-------------------------------------------------------------

Figures~\ref{figdsigmaq05}--\ref{figdsigmaq3} seem to suggest
that, for $x>0.5$, the  differential cross section is
relatively insensitive to the model used.  However, this is
largely an artifact of the log scales used in the figures, and
to show more clearly the {\it relative size\/} of the different
calculations, Figs.~\ref{figdsigmaq05r}--\ref{figdsigmaq3r}
show the {\it ratio\/} of each approximation to the C-IIB-noP
calculation.  The {\it relative\/} variation is largest for low $x$
($x$=0.5), but is significant at all $x$, varying from about
$\pm 10\%$ to $\pm 50\%$, particularly near
$\theta^*\simeq90^\circ$, depending on the values of $Q^2$
and $x$.  These variations are summarized in Table
\ref{tab:errors} for the larger values of $Q^2$.  Since the
cross sections  vary by many orders of magnitude, this model
dependence is not large enough to prevent these
calculations from providing a useful estimate of the size of
the cross section over a wide range of kinematics.  

For a few choices of kinematics, we have estimated the size of
the experimental errors that can be expected from a
measurement of this reaction at JLab using existing
equipment.  We find, for many kinematics, that the
experimental errors would be small enough to distinguish
between the different theoretical models shown in the
figures.  Our estimates of the experimental errors are shown in
Figs.~\ref{figdsigmaq05r}--\ref{figdsigmaq3r} and in Table
\ref{tab:errors}.  In all cases we have examined, except
possibly at the largest values of $x$ at the largest $Q^2$,
we could distinguish these models from one another.  Of
course, the final state interactions and exchange current
contributions must be calculated before one has a complete
picture of this process, but our results suggest that such a
calculation is likely to be worthwhile.

To estimate these errors we assumed the measurement would
be carried out in Hall A with the hadron arm of the HRS
spectrometer pair placed either in the direction of the
momentum transfer vector ${\bf q}_{_L}$ (data points with
solid circles) or to the left (triangles) or right (inverted
triangles) of ${\bf q}_{_L}$.  Each spectrometer setting is
able to measure a range of angles $\theta^*$, with the
settings to the left and the right of ${\bf q}_{_L}$ able to
measure larger $\theta^*$'s than the setting along ${\bf
q}_{_L}$ (which samples angles near $\theta^*\simeq0$). 
The errors grow as $\theta^*$ gets close to the
limit of the acceptance of the spectrometer, and this explains
the large errors at certain angles shown in the figures (see
for example, the case when $x=1.5$ and $Q^2=3$, where there are
large errors at
$\theta^*\simeq25^\circ$,
$\theta^*\simeq50^\circ$, and
$\theta^*\simeq80^\circ$).  This is clearly an
artifact of our crude estimates, and could be removed by
repositioning the spectrometers.  The statistics are based
on running for one day at each setting under normal JLab
operating conditions. 

\begin{table}
\begin{minipage}{3.5in}
\caption{\label{tab:errors2} Cuts used to arrive at the statistical
uncertainties.} 
\begin{ruledtabular}
\begin{tabular}{llll}
$Q^2$ (GeV)$^2$ & $x$ & $Q^2$-cut & $x$-cut \cr
\tableline
   1&   0.5 &  0.8-1.2 &  0.4-0.6\cr
    &   1.0 &  0.8-1.2 &  0.8-1.2\cr
    &   1.5 &  0.8-1.2 &  1.3-1.7\cr
    &   1.95 & 0.8-1.2 &  1.9-2.0\cr
   2 &  0.5  & 1.8-2.2 &  0.4-0.6\cr
     &  1.0 &  1.8-2.2 &  0.8-1.2\cr
      & 1.5 &  1.8-2.2 &  1.3-1.7\cr
      & 1.97 & 1.8-2.2 &  1.9-2.0\cr
   3 &  0.5 &  2.6-3.4 &  0.4-0.6\cr
     &  1.0 &  2.6-3.4 &  0.8-1.2\cr
     &  1.5 &  2.6-3.4 &  1.3-1.7\cr
     &  1.98&  2.6-3.4 &  1.95-2.0\cr
\end{tabular}
\end{ruledtabular}
\end{minipage}
\end{table}

These graphs show statistical errors only.   The estimates of the 
statistical uncertainties were made by acceptance averaging 
\cite{mceep} and radiatively  folding \cite{bd71}  the PWBA model of
Jeschonnek and Donnelly  \cite{DonnSab}.  An alternate 3-pole
parameterization  of the MMD nucleon  form factors \cite{mmd96} and
the Argonne  V18 {\sl NN} interaction \cite{AV18}  were used.  The
simulations  were done  using  a realistic  acceptance  model for  the
JLab-Hall A high resolution  spectrometer pair.  A maximum beam energy
of 4 GeV (except 6 GeV was used for the $Q^2$ = 3 GeV$^2$, $x=0.5$
case), beam  current of  100
$\mu$A  on a  15 cm  liquid deuterium target, and  measurement time
per  kinematic setting of 24  hours were assumed.   The cuts  shown  in
Table
\ref{tab:errors2}  were used  to restrict  the simulations to
reasonable intervals around the desired kinematics.

\subsection{The Asymmetry $A_{\phi}$} 

Next, we present our results for the asymmetry $A_{\phi}$, 
which is closely related to the transverse-longitudinal
response $\tilde R_{LT}$; see Eq. (\ref{Aphi}).
The numerical results are shown in Figs. \ref{figaphiq05} -
\ref{figaphiq3}. The asymmetry is zero for $\theta^* =
0^\circ$,  then becomes negative, with an extremum around
$\theta^* = 60^\circ$ for $Q^2$ = 0.5 GeV$^2$ and $x = 0.5$.
The extremum shifts to smaller angles for increasing $x$ and to
larger angles for increasing $Q^2$. Then, the asymmetry
changes sign and exhibits another peak around $\theta^*
= 140^\circ$ for $Q^2$ = 0.5 GeV$^2$ and $x = 0.5$.  The positive
peak shifts to lower angles both for increasing $x$ and $Q^2$. 
The appearance of the positive valued part of $A_{\phi}$
depends on the presence of the Born graph contribution, the
impulse approximation alone would only lead to one negative
peak. Accordingly, when both processes start to interfere,
i.e.~for the highest $x$ values, the minimum tends to wash out,
especially for the situations where the peak around
$\theta^* = 180^\circ$ has vanished in the cross section.

One can see at first glance that the asymmetry is less 
sensitive to the differences in the calculations than is the
differential cross section, except near $x=0.5$. 
Perhaps the most interesting  feature of these calculations is
the irregular shape of the asymmetry at $x=0.5$ for $Q^2=2$
and 3 GeV$^2$. At $Q^2=2$ GeV$^2$ both versions of the JD
and the C-AV18 calculation develop an extra dip. At $Q^2=3$
GeV$^2$, the results for $x$ = 0.5 develop even more
structure, with the AA-$v/c$
calculation having the opposite sign near $\theta^*$
near 0$^\circ$, and the C-IIB-noP showing an extra peak around 
$\theta = 90^\circ$. At $Q^2=3$ and $x=0.5$, $A_{\phi}$ could give
unique insight both into the effects of relativity and
different wave functions.  By contrast, at large $x$ and large
$Q^2$ the asymmetry is very small and not measurable with sufficient
accuracy.

The uncertainties in $A_\phi$ were generated by propagating the errors
in the  cross sections, where  the latter included  statistical errors
folded  in  quadrature with  an  overall  5\% systematic  uncertainty.
Further, to simplify the  procedure, the integrated yields for protons
within  the right  and left  hemispheres about  the  momentum transfer
direction  were  assumed to  correspond  to  $\phi=0$ and  $\phi=\pi$.
Finally, the  values of  the cross sections,  needed to  propagate the
errors  for  $A_\phi$, were  taken  from  the  point ({\sl  i.e.}  not
acceptance  averaged)  values; the  statistical  uncertainties in  the
cross sections were, of course, determined using the full acceptance.

\section{conclusions}
 
In this paper we have estimated the $d(e,e'p)n$ coincidence
cross section using the relativistic impulse approximation
(RIA).  Our calculations span the range $0.5\le Q^2\le 3$
GeV$^2$ and $x$ from 0.5 to just less than 2. In this
kinematic region, we find that the results are sensitive to
different approximate treatments of the single nucleon
current, and conclude the following:  

\begin{itemize}

\item{} Using equipment already in existence at JLab, it is
feasible to measure the unpolarized coincidence cross
section over this entire kinematic range.  The asymmetry
$A_\phi$ can be measured at small $x$ where it is large.   

\item The coincidence cross section is sensitive to the
theory over the entire kinematic range, and it appears that
measurements can be done to an accuracy sufficient to
distinguish a large variety of relativistic models from each
other, except possibly when {\it both\/} $x$ and $Q^2$ are very
large.

\item The asymmetry is less sensitive to the theory, except
at the smallest value of $x=0.5$ where measurements can
easily distinguish between different theoretical models. 

\end{itemize}

To complete this preliminary study, we must add relativistic
final state interactions and interaction currents that are
consistent with the RIA.  This is certainly feasible at large
values of $x$, where the low relative momentum in the final
state makes it possible to use existing relativistic $NN$
interaction models.  It may also be feasible at low $x$, where
the large excitation energy deposited into the final state
may justify the use of a relativistic generalization of the
Glauber approximation.

\acknowledgements

This work was supported in part by the US  Department of Energy. The
Southeastern  Universities Research Association (SURA) operates the
Thomas Jefferson National Accelerator Facility under DOE contract
DE-AC05-84ER40150. J.A. was also supported by the grant GA CR 202/00/1669.

%%%%%%%%%%%%%%%%%%%%%%%%%%%%%%%%%%%%%%%%%%%%%%%%%%%%%%%%%%%%%%%%%%%%%%%%%%%
%\newpage

\appendix

\section{The deuteron wave function}
\label{Appen:A}

%%%%%%%%%%%%%%%%%%%%%%%%%%%%%%%%%%%%%%%%%%%%%%%%%%%%%%%%%%%%%%%%%%%%%%%%%%%
\subsection{Helicity spinors}   

Following the conventions of Jacob and Wick \cite{jw}, the helicity
spinors for a particle with four-momentum $p=\{E, {\bf p}\}$ are
obtained from spinors with four-momentum $\{m, {\bf 0}\}$ and spins up
or down in the $\hat z$ direction.  The state is first boosted along
the $\hat z$-axis until its momentum is $\{E,0,0,p\}$  and then
is rotated it in the direction of ${\bf p}$.  The Lorentz
transformation that does this is therefore
\begin{equation}
S(\hat p,\zeta_p)={\cal R}(\hat p){\cal B}(\zeta_p)
=S(\Lambda(\hat p,\zeta_p))\label{lortrans}
\end{equation}
where ${\cal B}$ is the Dirac operator for a pure boost along
the $z$-axis
\begin{equation}
{\cal B}(\zeta_p)=e^{\alpha_3\,\zeta_p/2}
\end{equation}
with $\tanh\zeta_p=p/E$, and ${\cal R}$ is the Dirac rotation operator
\begin{equation}
{\cal R}(\hat p)={\cal R}(\phi,\theta,-\phi)=
e^{-i\,\Sigma_3\,\phi/2}
e^{-i\,\Sigma_2\,\theta/2}e^{i\,\Sigma_3\,\phi/2}
\end{equation}
that takes the momentum from along the $\hat z$-axis into its final
direction. Using this transformation in the $\hat x\hat z$ plane
($\phi=0$), the helicity spinors for particle 1 are  defined as in
Refs.~\cite{GVOH92,SGF}  
\begin{eqnarray}
u({\bf p},\lambda)&\equiv& u_1({\bf p},\lambda) =S(\hat
p,\zeta_p)\,u({\bf 0},\lambda)\equiv u_1({\rm p},\lambda, \theta_1)
\nonumber\\ &=& {\cal R}_y(\theta_1)\,
u_1({\rm p},\lambda, 0) = 
\left(\begin{array}{c} \cosh{1\over2}\zeta_p \\
2\lambda\sinh{1\over2}\zeta_p
\end{array}\right)\chi_{_\lambda}(\theta_1)
\nonumber\\ 
v(-{\bf p},\lambda)&\equiv& v_1({\bf p},\lambda)
=-(-1)^{\frac{1}{2}-\lambda}
\,{\cal C}\, u^*({\bf p},-\lambda) \equiv v_1({\rm p},\lambda,
\theta_1) \nonumber\\ 
&=& {\cal R}_y(\theta_1)\,
v_1({\rm p},\lambda, 0)={\cal R}_y(\theta_1)\, \gamma^5\gamma^0
u_1({\rm p},\lambda, 0)=
\left(\begin{array}{c} -2\lambda\sinh{1\over2}\zeta_p \\
\cosh{1\over2}\zeta_p 
\end{array}\right)\chi_{_\lambda}(\theta_1) \qquad \label{spinor1}
\end{eqnarray}
where ${\cal R}_y(\theta)={\cal R}(0,\theta,0)$, $\theta_1$ is the
%
%-------------------------------------------------------------
% Figure 21  The momentum variables
\begin{figure}
\begin{leftline}
\mbox{\epsfysize=5cm \epsfbox{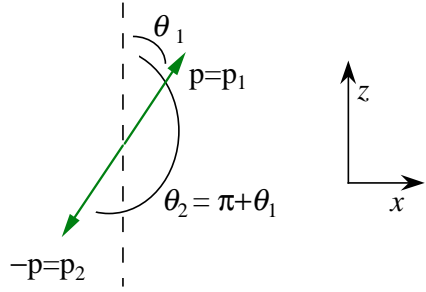}}
\end{leftline}
%\vspace{-2cm}
\caption{\footnotesize\baselineskip=12pt The definitions of the
momenta and angles $\theta_1$ and
$\theta_2$.}
\label{fig:mom}  
\end{figure} 
%-------------------------------------------------------------
% 
polar angle of the vector ${\bf p}={\bf p}_1$ (see
Fig.~\ref{fig:mom}), and
\begin{equation}
\chi_{_\frac{1}{2}}(\theta)=\left(\begin{array}{c} 
\cos{1\over2}\theta \\
\sin{1\over2}\theta \end{array}\right) \qquad
\chi_{_{-\frac{1}{2}}}(\theta)=\left(\begin{array}{c}
-\sin{1\over2}\theta \\ \phantom{-} \cos{1\over2}\theta 
\end{array}
\right) \, .
\end{equation}
Note that the $u({\bf p},\lambda)$ and $v(-{\bf p},\lambda)$ of
Ref.~\cite{SGF} are identical to the $u_1({\bf p},\lambda)$ and
$v_1({\bf p},\lambda)$ of Ref.~\cite{GVOH92}.  Following Jacob and
Wick
\cite{jw}, the helicity spinors for particle 2 are defined as 
\begin{eqnarray}
u(-{\bf p},\lambda)&\equiv& u_2({\bf p},\lambda)
=(-1)^{\frac{1}{2}-\lambda}\, {\cal
R}_y(\theta_1)\,{\cal R}_y(\pi)\, {\cal B}(\zeta_p)\, u({\bf
0},\lambda) \equiv u_2({\rm p}, \lambda, \theta_2) 
\nonumber\\ &=& {\cal R}_y(\theta_2)\,
u_2({\rm p},\lambda, 0) = {\cal R}_y(\theta_2)\,
(-1)^{\frac{1}{2}-\lambda}\,u_1({\rm p},\lambda, 0) =  
\left(\begin{array}{c} \cosh{1\over2}\zeta_p \\
2\lambda\sinh{1\over2}\zeta_p
\end{array}\right)\chi_{_{-\lambda}}(\theta_1)
\nonumber\\
v({\bf p},\lambda)&\equiv& v_2({\bf p},\lambda)
=(-1)^{\frac{1}{2}-\lambda}\,{\cal C}\,
u^*(-{\bf p},-\lambda) \equiv v_2({\rm p}, \lambda, \theta_2)
\nonumber\\ &=& {\cal R}_y(\theta_2)\,
v_2({\rm p},\lambda, 0) = {\cal R}_y(\theta_2)\,
\gamma^5\gamma^0 u_2({\rm p},\lambda, 0) = \left(\begin{array}{c} -
2\lambda\sinh{1\over2}\zeta_p \\ \cosh{1\over2}\zeta_p
\end{array}\right)\chi_{_{-\lambda}}(\theta_1)\qquad \label{spinor2}
\end{eqnarray}
where $\theta_2=\pi+\theta_1$ is the polar angle of the vector  $-{\bf
p}={\bf p}_2$ (see Fig.~\ref{fig:mom}) and
$u(-{\bf p},\lambda)$ and $v({\bf p},\lambda)$ are identical to
the $u_2({\bf p},\lambda)$ and $v_2({\bf p},\lambda)$ of
Ref.~\cite{GVOH92}.  Note that the angular conventions have $0\le
\theta_1\le \pi$ and have $\pi\le
\theta_2\le 2\pi$.   Useful relations, valid on the two-component
subspace, are   
\begin{eqnarray}
{\cal R}_y(\theta_2)\,\chi_{_{\lambda}}(0)&=&
2\lambda\,\chi_{_{-\lambda}}(\theta_1)\nonumber\\
{\cal R}_y(\pi)\,\chi_{_{\lambda}}(\theta)&=&
2\lambda\,\chi_{_{-\lambda}}(\theta) \, ,
\label{tworot}
\end{eqnarray}
and, on the full four-component space, $u_1$ and $u_2$ are related
by 
\begin{eqnarray}
u_2({\rm p},\lambda,\theta_2)=2\lambda\, {\cal
R}_y(\pi)\,u_1({\rm p},\lambda,\theta_1)\, .
\label{spinorrel}
\end{eqnarray}
[Note that $2\lambda\equiv (-1)^{\frac{1}{2}-\lambda}$.] This last
formula is useful for the applications in this paper.   

%%%%%%%%%%%%%%%%%%%%%%%%%%%%%%%%%%%%%%%%%%%%%%%%%%%%%%%%%%%%%%%%%%%%%%%%%%%
\subsection{Deuteron wave functions in the rest frame}

The deuteron wave function (\ref{wf}) is
manifestly covariant, and we use this feature to simplify
the treatment. Applying Eqs.~(\ref{spinor1}) and (\ref{spinor2}) in
the deuteron rest frame, the  spinor for either particle 1 or 2 can
be written in terms of the spinor with the momentum in the $\hat z$
direction   
\begin{eqnarray}
u_i({\bf p},\lambda)=u_i({\rm p},\lambda,\theta_i)={\cal
R}_y(\theta_i)\, u_i({\rm p},\lambda, 0)\, .
\label{B1}
\end{eqnarray}
Note that this equation holds for particle 2 even though, in
applications, we restrict $\theta_2\ge \pi$. The active rotation
of the deuteron helicity vector from an initial direction along the
$\hat z$-axis to an angle $\theta$ with respect
to the $\hat z$-axis is given in terms of the spin 1 rotation
matrices 
\begin{equation}
\xi_{_{\lambda_d}}(\theta)=\sum_{\lambda'_d}
d^{(1)}_{\lambda_d\,\lambda'_d}(\theta)\;
\xi_{_{\lambda'_d}}(0)
\, . \label{B2}
\end{equation}
Substituting (\ref{B1}) into (\ref{wf}), working the operator
${\cal R}$ through the rest of the expression, and then using
(\ref{B2}) to realign the deuteron helicity vector in the
$\hat z$ direction by rotating it through angle $-\theta$,
gives
\begin{equation}
\psi^{(i)}_{\lambda,\lambda_d}(p,{P})\equiv
\psi^{(i)}_{\lambda\mu}({\rm p},\theta_i)\; 
\xi^{\mu}_{_{\lambda_d}}(0)= {\cal R}_y(\theta_i)\, \sum_{\lambda'_d}
\psi^{(i)}_{\lambda \mu}({\rm p},0)\;\xi^{\mu}_{_{\lambda'_d}}(0)\,
d^{(1)}_{\lambda'_d\,\lambda_d}(\theta_i) \label{relrot} 
\end{equation}
where we used $d^{(1)}_{\lambda_d\,\lambda'_d}(-\theta) =
d^{(1)}_{\lambda'_d\,\lambda_d}(\theta)$.  This argument
works only in the deuteron rest frame where there is no
total three-momentum to be rotated by ${\cal R}_y$.     

In the {\it original\/} reference \cite{bg} the on-shell particle 
was taken to be particle 1 with four-momentum $p_1$, and the
wave function in the deuteron rest frame was expanded in terms of
(on-shell) particle 2 spinors \cite{GVOH92}.  [Also, be aware that the
spinors used in Ref.~\cite{bg} were quantized along the fixed $\hat z$
axis, in for $v$ spinors the notation $-s$ corresponded to spin
projection
$+s$ in the $\hat z$ direction.]  In the notation of Eq.~(\ref{spinor2})
this becomes 
\begin{equation}
\psi^{(1)}_{\lambda_1,\lambda_d}(p_1,{P})=
\sum_{\lambda_2}
\left[u_2({\rm p}_1,\lambda_2, \theta_2)\,
\psi^+_{\lambda_1\lambda_2,\lambda_d}({\bf p}_1) +  
v_2({\rm p}_1,\lambda_2,\theta_2)\,
\psi^-_{\lambda_1\lambda_2,\lambda_d}({\bf p}_1)\right]\, .  
\end{equation}
Using the fact that $\theta_2=\pi+\theta_1$, so that $\theta_2=\pi$
when $\theta_1=0$, and using Eq.~(\ref{relrot}) gives 
\begin{eqnarray}
\psi^{(1)}_{\lambda_1,\lambda_d}(p_1,{P}) &=&
\sum_{\lambda_2,\lambda'_d}
\Bigl[u_2({\rm p}_1,\lambda_2,\theta_2)\,
\psi^+_{\lambda_1\lambda_2,\lambda'_d}({\rm p}_{1z})\nonumber\\
&&\qquad\quad+   v_2({\rm p}_1,\lambda_2,\theta_2)\,
\psi^-_{\lambda_1\lambda_2,\lambda'_d}({\rm p}_{1z})\Bigr]\,
d^{(1)}_{\lambda'_d\,\lambda_d}(\theta_1) \, ,\qquad \label{B5}  
\end{eqnarray}
where the components
$\psi^\pm_{\lambda_1\lambda_2,\lambda'_d}({\rm p}_{1z})$ have the
relative momentum vector (same as the momentum of particle 1 in the
rest system) aligned along the
$+\hat z$ direction.  The $\pm$ components of the wave
functions follow from the helicity spinors defined above and the
expansions given in Ref.~\cite{bg}:
\begin{eqnarray}
\psi^+_{\lambda_1\lambda_2,\lambda_d}({\bf p}) &=&
{1\over\sqrt{4\pi}} \chi^\dagger_{_{-\!\lambda_2}}\left[u({\rm p})\,
{\bf{\sigma}\cdot \xi}_{_{\lambda_d}} +
{w({\rm p})\over\sqrt{2}}\left(3\,{\bf \hat p\cdot 
\xi}_{_{\lambda_d}} {\bf{\sigma}\cdot \hat
p}- {\bf{\sigma}\cdot\xi}_{_{\lambda_d}}
\right)\right] {i\sigma_2\over\sqrt{2}}\,\chi_{_{\lambda_1}}
\nonumber\\
\psi^-_{\lambda_1\lambda_2,\lambda_d}({\bf p}) &=&
\sqrt{{3\over 4\pi}} \chi^\dagger_{_{-\lambda_2}}\left[v_s({\rm p})\;
{\bf\hat p\cdot \xi}_{_{\lambda_d}}
-{v_t({\rm p})\over\sqrt{2}} \left({\bf \sigma\cdot\hat p\,
\sigma\cdot \xi}_{_{\lambda_d}} -  {\bf\hat p}\cdot
{\bf\xi}_{_{\lambda_d}} \right)
\right]{i\sigma_2\over\sqrt{2}}\,\chi_{_{\lambda_1}}\, .
\label{B6}
\end{eqnarray}
Here $u({\rm p})$ and $w({\rm p})$ are the momentum space
radial wave functions for the S and D states and $v_t({\rm p})$ and
$v_s({\rm p})$ are triplet and  singlet P state wave functions, which
appear in a manner similar to the lower component wave functions
for the Dirac equation. These wave functions are functions of the
variable ${\rm p}=|{\bf p}|$ and satisfy  the normalization condition
\cite{bg}
\begin{equation}
\int_0^\infty {\rm p}^2d{\rm p}\left\{ u^2({\rm p})+w^2({\rm p})+ 
v^2_t({\rm p}) +v^2_s({\rm p})\right\}=1\, . \label{norm}
\end{equation}

The wave functions (\ref{B6}) can be simplified by specifying
the helicity states of the deuteron (in the rest frame). 
Since the deuteron is a particle 2 in the sense of Jacob and
Wick, its helicity states are 
\begin{equation}
{\bf \xi}_{_{\lambda_d}}\equiv {\bf \xi}^2_{_{\lambda_d}}  =
(-1)^{s-\lambda_d}\,e^{-i\,\pi J_y} \,
{\bf \xi}^1_{_{\lambda_d}} = \cases{ {1\over\sqrt{2}}
(\pm1, -i, 0) & if $\lambda_d=\pm$ \cr
\qquad(0,0,1) & if $\lambda_d=0$ \, .\cr }
\end{equation}
Note, for future reference, that 
\begin{equation}
{\cal R}_\pi\,{\bf \xi}_{_{\lambda_d}} =(-1)^{1-\lambda_d}\,
\xi_{_{-\lambda_d}} \label{A18}
\end{equation}
where ${\cal R}_\pi\equiv{\cal R}_y(\pi)$.
If ${\bf p}$ is in the $+\hat z$ direction, it is not
difficult to evaluate (\ref{B6}), giving 
\begin{eqnarray}
\psi^+_{\lambda_1\lambda_2,\lambda_d}({\rm p}_{z}) &=&
{1\over\sqrt{8\pi}}\,\delta_{\lambda_d,\,\lambda_2-\lambda_1}
\, f^+_{|\lambda_d|}({\rm p}) \nonumber\\
\psi^-_{\lambda_1\lambda_2,\lambda_d}({\rm p}_{z}) &=&
-{2\lambda_1\over\sqrt{8\pi}}\,
\delta_{\lambda_d,\,\lambda_2-\lambda_1}
\, f^-_{|\lambda_d|}({\rm p}) \, , \label{B8}
\end{eqnarray}   
where 
\begin{eqnarray}
\begin{array}{ll}
f^+_0({\rm p}) = u({\rm p}) +\sqrt{2}\,w({\rm p})\qquad  & 
f_0^-({\rm p})=\sqrt{3}\,v_s({\rm p}) \cr \\
f^+_1({\rm p}) = \sqrt{2}\,u({\rm p}) - w({\rm p})\qquad & 
f_1^-({\rm p})=\sqrt{3}\,v_t({\rm p}) \, , \cr
\end{array}
\end{eqnarray}   
Combining the expressions (\ref{B5}) and (\ref{B8}) gives
\begin{eqnarray}
\psi^{(1)}_{\lambda_1,\lambda_d}(p_1,{P})&=&{1\over\sqrt{8\pi}}
\sum_{\lambda_2}
\Bigl[u_2({\rm
p}_1,\lambda_2,\theta_2)\,f^+_{|\lambda_2-\lambda_1|}({\rm p}_1)
\nonumber\\
&&\qquad \qquad -   2\lambda_1\,v_2({\rm
p}_1,\lambda_2,\theta_2)\, f^-_{|\lambda_2-\lambda_1|}({\rm
p}_1)\Bigr]\, d^{(1)}_{\lambda_2-\lambda_1,\,\lambda_d}(\theta_1) \, .
\label{B10}
\end{eqnarray}

One disadvantage of the expansion (\ref{B10}) is that the
four-momentum of particle 2, which is
$p_2=\{M_d-E_{p_1}, -{\bf p}_1\}$ in the rest system, is {\it not\/}
the same as the four-momentum of the spinor $u_2$, which is
$\tilde p_2=\{E_{p_1}, -{\bf p}_1\}$. This difference can lead to
confusion, especially because the four-momentum $\tilde p_2$ is not
one of the four-momenta that naturally occurs in the problem. 
In {\it this\/}  paper we avoid this confusion, exploit
the freedom to expand the off-shell particle in terms spinors
with {\it any\/} four-momentum, and choose the four-momentum of the
on-shell particle (i.e. the spectator) for the expansion.  The
advantage of this choice is that four-momentum used to describe the
off-shell particle is now one of the naturally occurring momenta in
the problem.  Also, we find that the formalism is simplified if we
use $\gamma^5\,u$ instead of $v$ spinors to describe the negative
energy states (as in Ref.~\cite{SGF}).  With this choice, we find the
following expansion for the wave function     
\begin{eqnarray}
\psi^{(1)}_{\lambda_1,\lambda_d}(p_1,{P})&=&\sqrt{{3\over8\pi}}
\sum_{\lambda_2}
\Bigl[u_1({\rm p}_1,\lambda_2, \theta_1)\,
\phi^+_{|\lambda_2+\lambda_1|}({\rm p}_1) 
\nonumber\\
&&\qquad \qquad -  
2\lambda_2\,\gamma^5\,u_1({\rm p}_1,\lambda_2, \theta_1)\,
\phi^-_{|\lambda_2+\lambda_1|}({\rm p}_1)\Bigr]\,
d^{(1)}_{-\lambda_2-\lambda_1,\,\lambda_d}(\theta_1) \, ,
\label{psi1a}
\end{eqnarray}
where and extra factor of $\sqrt{3}$ has been introduced for
convenience.  Projecting out the independent components, gives
\begin{eqnarray}
\begin{array}{ll}
\phi^+_0({\rm p}) ={\displaystyle{E_p\over\sqrt{3}\,
m}}f^+_0({\rm p})\qquad  & 
\phi^-_0({\rm p}) ={\displaystyle{1\over\sqrt{3}}\left({{\rm p}\over
m}f^+_0({\rm p})-f_0^-({\rm p})\right)} \cr \\
\phi^+_1({\rm p}) ={\displaystyle{E_p\over\sqrt{3}\,
m}}f^+_1({\rm p})\qquad & 
\phi^-_1({\rm p}) ={\displaystyle{1\over\sqrt{3}}\left({{\rm p}\over
m}f^+_1({\rm p})+f_1^-({\rm p})\right)}\, , \cr \label{exp}
\end{array}
\end{eqnarray}   
where the new wave functions were given in Table
\ref{tab:wave}. 

Comparison of the expansions (\ref{B10}) and (\ref{psi1a}) 
underline the fact that {\it the separation of the wave
function into positive and negative energy parts is a
matter of convention\/}; only the total result is
independent of this separation.       

By a similar argument, we expect the expansion for $\psi^{(2)}$ to
be 
\begin{eqnarray}
\psi^{(2)}_{\lambda_2\,\lambda_d}(p_2,{P})&=&\sqrt{{3\over8\pi}}
\sum_{\lambda_1}
\Bigl[u_2({\rm
p}_2,\lambda_1,\theta_2)\,\phi^{'+}_{|\lambda_2+\lambda_1|}({\rm
p}_2) 
\nonumber\\ &&\qquad\qquad-  
2\lambda_1\,\gamma^5\,u_2({\rm p}_2,\lambda_1,\theta_2)\,
\phi^{'-}_{|\lambda_2+\lambda_1|}({\rm p}_2)\Bigr]\,
d^{(1)}_{\lambda_2+\lambda_1,\,\lambda_d}(\theta_2-\pi) \, . \qquad
\label{B11}
\end{eqnarray}
One can prove that $\phi^\pm=\phi^{'\pm}$ by finding the
expansions for $\psi^{(2)}$ directly from those for
$\psi^{(1)}$.  Using (\ref{B11}) and orthogonality relations for the
spinors gives  
\begin{eqnarray}
(3/8\pi)^{1/2}\,\phi^{'+}_{|\Lambda|}({\rm p}_2)\,
d^{(1)}_{\Lambda,\,\lambda_d}(\theta_1) &=&
\bar u_2({\rm p}_2,\lambda_1,\theta_2)\,
\psi^{(2)}_{\lambda_2,\lambda_d}(p_2,{P}) \nonumber\\
&=&\bar u_2({\rm p}_2,\lambda_1,\theta_2)\,
\Gamma_{\lambda_d}(p_2,P)\,{\cal C}\,\bar{u}_2^T({\rm
p}_2,\lambda_2,\theta_2)
\nonumber\\
&=&4\lambda_1\lambda_2\,\bar u_1({\rm p}_2,\lambda_1,\theta_1)\,
{\cal R}_\pi\,\Gamma_{\lambda_d}(p_2,P)\,{\cal C}\,{\cal R}^{-1}_\pi\,
\bar{u}_1^T({\rm p}_2,\lambda_2,\theta_1)
\nonumber\\
&=&4\lambda_1\lambda_2(-1)^{1-\lambda_d}\,\bar u_1({\rm
p}_2,\lambda_1,\theta_1)\,\Gamma_{-\lambda_d}(p_1,P)\,
{\cal C}\,\bar{u}_1^T({\rm p}_2,\lambda_2,\theta_1)
\nonumber\\
&=&(-1)^{\Lambda-\lambda_d}\,\bar u_1({\rm
p}_2,\lambda_1,\theta_1)\,\Gamma_{-\lambda_d}(p_1,P)\,
{\cal C}\,\bar{u}_1^T({\rm p}_2,\lambda_2,\theta_1)
\, , \qquad
\label{B10a}
\end{eqnarray}
where $\Lambda=\lambda_1+\lambda_2$, $\Gamma_{\lambda_d} (p,P) \equiv
\Gamma_\mu(p,P)\,\xi^\mu_{\lambda_d}(P)$,
$R_y(\pi)\,p_2=p_1$, and we used the relations (\ref{spinorrel}) and
(\ref{A18}). However, from (\ref{psi1a}) 
\begin{eqnarray}
(3/8\pi)^{1/2}\,\phi^{+}_{|\Lambda|}({\rm p}_1)\,
d^{(1)}_{\Lambda,\,\lambda_d}(\theta_2-\pi)&=&
(3/8\pi)^{1/2}\,(-1)^{\Lambda-\lambda_d}\,
\phi^{+}_{|\Lambda|}({\rm p}_1)\,
d^{(1)}_{-\Lambda,\,-\lambda_d}(\theta_1) \nonumber\\
 &=& (-1)^{\Lambda-\lambda_d}\,
\bar u_1({\rm p}_1,\lambda_2,\theta_1)\,
\psi^{(1)}_{\lambda_1,-\lambda_d}(p_1,{P}) \nonumber\\
&=&(-1)^{\Lambda-\lambda_d}\,\bar u_1({\rm p}_1,\lambda_2,\theta_1)\,
\Gamma_{-\lambda_d}(p_1,P)\,
{\cal C}\,\bar{u}_1^T({\rm p}_1,\lambda_1,\theta_1)
\, . \qquad
\label{B10b}
\end{eqnarray}
Since this equation is symmetric under the interchange of $\lambda_1$
and $\lambda_2$, and ${\rm p}_1={\rm p}_2$ in the c.m. system, the
two equations (\ref{B10a}) and (\ref{B10b}) are equal, and
$\phi^+=\phi^{'+}$. A similar argument holds of $\phi^-$.

The wave functions used in this paper were obtained by solving 
the Spectator equation using a kernel adjusted to fit the $NN$
data below 350 MeV lab energy \cite{GVOH92}.

%%%%%%%%%%%%%%%%%%%%%%%%%%%%%%%%%%%%%%%%%%%%%%%%%%%%%%%%%%%%%%%%%%%

\subsection{Boosting helicity spinors}

In order to boost the spectator equation wave functions it is 
necessary to have expressions for the pure boosts of the
helicity spinors. Indeed, for the applications to elastic
deuteron electromagnetic form factors and the  response
functions for deuteron electrodisintegration, it is only
necessary to study the case where the boosts are made along
the $z$-axis. Since the Dirac spinors for arbitrary momentum
are defined in terms of a Lorentz transformation  of the rest
frame spinors, it is useful to define the four-momentum in the
particle rest frame as
\begin{equation}
\tilde p=(m,{\bf 0}).
\end{equation}
Three basic Lorentz transformations are required in the 
following discussion.  In the notation of Eq.~(\ref{lortrans}), these
are: 
\begin{eqnarray}
p&=&S(\hat p,\zeta_p)\,\tilde p\equiv {\cal R}(\hat
p){\cal B}(\zeta_p)\,\tilde p 
\nonumber\\
p'&=&S(\hat p',\zeta_{p'})\,\tilde p\equiv {\cal R}(\hat
p'){\cal B}(\zeta_{p'})\,\tilde p 
\nonumber\\
p'&=&{\cal B}(\zeta_z)\, p
\end{eqnarray}
The first Lorentz transformation connects the particle rest 
frame with the initial frame of the particle and is composed
of a pure boost along the $z$-axis
${\cal B}(\zeta_p)$ followed by a rotation into the direction of the
initial particle direction ${\cal R}(\hat p)$.
The second Lorentz transformation connects the particle rest 
frame with the final frame of the particle and is also
composed of a pure boost  along the $z$-axis followed by a
rotation. The third Lorentz transformation connects the
initial and final particle frames with a pure boost along the 
$z$-axis.

Now consider the boost of a helicity spinor [either $u_1$ of
Eq.~(\ref{spinor1}) or $u_2({\bf p},\lambda)$ of
Eq.~(\ref{spinor2})] along the
$z$-axis
\begin{eqnarray}
{\cal B}(\zeta_z)\,u_i({\bf p},\lambda)&=&{\cal B}(\zeta_z)\,   
S(\hat p,\zeta_p)\, u_i({\bf 0},\lambda)  \nonumber\\ 
&=&S(\hat p',\zeta_{p'})\,S^{-1}(\hat p',\zeta_{p'})\,
{\cal B}(\zeta_z)\,S(\hat p,\zeta_p)\,u_i({\bf 0},\lambda) 
\nonumber\\ 
&=&S(\hat p',\zeta_{p'})\,S\left(\Lambda^{-1}(\hat p',\zeta_{p'})
B(\zeta_z)\Lambda(\hat p,\zeta_p)\right)u_i({\bf 0},\lambda)
\end{eqnarray}
where $S(\hat p,\zeta_{p})=S(\Lambda(\hat p,\zeta_{p}))$ is the
representation of the Lorentz transformation $\Lambda(\hat
p,\zeta_{p})$ and the group composition property of the Lorentz
transformations has been used in writing the final step. Since
\begin{equation}
\Lambda^{-1}(\hat p',\zeta_{p'})B(\zeta_z)
\Lambda(\hat p,\zeta_p)\,\tilde p=\tilde p,
\end{equation}
this combination of Lorentz transformations must be equivalent 
to a rotation,and is referred to as the Wigner rotation. In this
case, where the boost is along the $z$-axis and the momenta are in
the $\hat x\hat z$ plane, the Wigner rotation is a rotation about
the $\hat y$ axis, denoted ${\cal R}_y(\omega_i)$. The
boosted spinor can therefore be written 
\begin{eqnarray}
{\cal B}(\zeta_z)\,u_i({\bf p},\lambda)&=&S(\hat
p',\zeta_{p'})\,{\cal R}_y(\omega_1)\,u_i({\bf
0},\lambda)
= S(\hat p',\zeta_{p'}) \sum_{\lambda'}u_i({\bf 0},\lambda')
\,d_{\lambda'\lambda}^{\left({1/2}\right)}(\omega_i)
\nonumber\\
&=& \sum_{\lambda'}u_i({\bf p}',\lambda')
\,d_{\lambda'\lambda}^{\left({1/2}\right)}(\omega_i)\, .
\label{bstspin1}
\end{eqnarray}
Since $\gamma^5$ and ${\cal C}\,\gamma^0$ both commute with the
boost, and since the boost is real, (\ref{bstspin1})
also holds for $\gamma^5\,u_i({\bf p},\lambda)$ and ${\cal C}\,
\bar u_i^T({{\bf p}},\lambda)$.  

In this paper ${\cal B}(\zeta_z)$ will be chosen to be the boost from
the c.m.~frame where $P^*=\{D_0, 0,0,-q_0\}$ to the lab frame where
$P=\{M_d,0,0,0\}$.  This is accomplished by a boost in the positive
$\hat z$ direction with $\tanh\zeta_z= q_0/D_0$.  In the notation
we have introduced, the basic equations are   
\begin{eqnarray}
{\cal B}(\zeta_z)\,{\cal C}\,\bar u^T_i({\bf p}_i^*,\lambda) =
\sum_{\lambda'} {\cal C}\,\bar u^T_i({\bf p}_i,\lambda')
\,d_{\lambda'\lambda}^{\left({1/2}\right)}(\omega_i)\, ,
\label{bstspina}
\end{eqnarray}
and the inverse relations
\begin{eqnarray}
{\cal B}^{-1}(\zeta_z)\,{\cal C}\,\bar u^T_i({\bf p}_i,\lambda) =
\sum_{\lambda'} {\cal C}\,\bar u^T_i({\bf p}^*_i,\lambda')
\,d_{\lambda'\lambda}^{\left({1/2}\right)}(-\omega_i)\, ,
\label{bstspinb}
\end{eqnarray}
where the unstarred variables are in the deuteron rest frame
(the lab frame) and the starred variables are in the c.m.~frame of
the electrodisintegration process.

The Wigner rotation angles can be computed from the standard relations
\begin{equation}
{\cal B}^{-1}(\zeta_z)\,{\cal R}_y(\theta_i)\,{\cal B}(\zeta_{p_i})
= {\cal R}_y(\theta^*_i)\,{\cal B}(\zeta_p) {\cal R}_y(\mp \omega_i)
\, , \label{wig}
\end{equation}
where the upper sign holds for $i=1$ and the lower for $i=2$ [the
change in sign is a consequence of the phase relation in
Eq.~(\ref{spinorrel})], and
$\theta^*_1=\theta^*$ and $\theta^*_2=\theta^*+\pi$ [recall Table
\ref{tab:variables}; the form of $\theta_2^*$ is a
consequence of the extra rotation by $\pi$ in Eq.~(\ref{spinorrel})]. 
Writing the rotation and boost operators in closed form,
Eq.~(\ref{wig}) can be written   
\begin{eqnarray}
\left[\cosh{\textstyle{1\over2}} \zeta_z - 
\alpha_3\sinh {\textstyle{1\over2}}\zeta_z\right]
\left(\cos{\textstyle{1\over2}}\theta_i-
i\Sigma_2\sin{\textstyle{1\over2}}\theta_i\right)
\left[\cosh{\textstyle{1\over2}}\zeta_{p_i}-\alpha_3\sinh
{\textstyle{1\over2}}\zeta_{p_i}\right]\qquad
\nonumber\\ =\left(\cos{\textstyle{1\over2}}\theta_i^*-
i\Sigma_2\sin{\textstyle{1\over2}}\theta_i^*\right)
\left[\cosh{\textstyle{1\over2}}\zeta_{p}-\alpha_3\sinh
{\textstyle{1\over2}}\zeta_{p}\right]
\left(\cos{\textstyle{1\over2}}\tilde\omega+
i\Sigma_2\sin{\textstyle{1\over2}}\tilde\omega\right)\, ,
\end{eqnarray}
where $\tilde\omega$ is a shorthand for either $\omega_1$ or
$-\omega_2$, depending on the case under consideration.  This operator
relation can be separated into four independent equations relating
$\tilde\omega$ to the lab variables
$\{p_i,\,\theta_i\}$ to the c.m.\
variables $\{p,\theta^*_i\}$. A
convenient form of these equations is
\begin{eqnarray}
C_{-+}\,\cos{\textstyle{1\over2}}\theta_i &=&
\sqrt{2M_d(E+m)}
\,\cos{\textstyle{1\over2}}(\theta^*_i-\tilde\omega)
\nonumber\\
C_{--}\,\cos{\textstyle{1\over2}}\theta_i  &=&
\sqrt{2M_d(E-m)}
\,\cos{\textstyle{1\over2}}(\theta^*_i+\tilde\omega)
\nonumber\\
C_{++}\,\sin{\textstyle{1\over2}}\theta_i &=&
\sqrt{2M_d(E+m)}
\,\sin{\textstyle{1\over2}}(\theta^*_i-\tilde\omega)
\nonumber\\
C_{+-}\,\sin{\textstyle{1\over2}}\theta_i &=&
\sqrt{2M_d(E-m)}
\,\sin{\textstyle{1\over2}}(\theta^*_i+\tilde\omega)
\label{wigangles}
\end{eqnarray}
where
\begin{equation} 
C_{ab}=\sqrt{(D_0+ M_d)(E_i+am)} +b
\sqrt{(D_0-M_d)(E_i-am)}\, ,
\end{equation}
with $E=\sqrt{m^2+p^2}$ and
$E_i=\sqrt{m^2+p_i^2}$. The following identities, derived from these
equations, are very useful:
\begin{eqnarray}
{E\over m} \sin\theta^*_i\cos\tilde\omega - 
\cos\theta^*_i\sin\tilde\omega  &=& {E_i\over m}\sin\theta_i
\nonumber\\ 
{E\over m} \sin\theta^*_i\sin\tilde\omega +
\cos\theta^*_i\cos\tilde\omega  &=&
\cos\theta_i
\nonumber\\
-{E\over m} \cos\theta^*_i\sin\tilde\omega +
\sin\theta^*_i\cos\tilde\omega  &=& {D_0\over M_d}\sin\theta_i
\nonumber\\ 
{E\over m} \cos\theta^*_i\cos\tilde\omega +
\sin\theta^*_i\sin\tilde\omega  &=& {D_0\,E_i\over M_d\,m}
\cos\theta_i - {q_0\,p_i\over M_d\,m}
\nonumber\\ 
-p\cos\tilde\omega &=& {q_0\over M_d}E_i\cos\theta_i -  
{D_0\over M_d}\, p_i
\nonumber\\ 
p\sin\tilde\omega &=& {m\over M_d}\,q_0\sin\theta_i
\label{wigidentities}
\end{eqnarray}

%%%%%%%%%%%%%%%%%%%%%%%%%%%%%%%%%%%%%%%%%%%%%%%%%%%%%%%%%%%%%%%%%%%%%%%

\subsection{Boosting the deuteron wave function}

In order to calculate the response functions for deuteron
electrodisintegration it is necessary boost the deuteron wave
functions from the center of momentum frame of the final state
proton-neutron pair to the rest (lab) frame of the deuteron (where
the decomposition of the wave functions onto S, D, and P states has
been defined).  If the system is quantized such that the
three-momentum transfer ${\bf q}$ lies along the $z$-axis, then the
deuteron wave functions in the c.m.~must be boosted to the rest frame
by a pure active boost ${\cal B}(\zeta_z)$ in the $\hat z$ direction
with $\tanh\zeta_x=q_0/D_0$, as defined in the previous section.

The rest frame wave functions are obtained by applying the operator
${\cal B}={\cal B}(\zeta_z)$ to the wave functions (\ref{wf}), which
gives
\begin{eqnarray}
{\cal B}\,\psi^{(i)}_{\lambda\lambda_d}(p^*_i,\,{P^*})&=&
{\cal B}\,\frac{m+\not{\!P}^*-\not{\!p}^*_i}{m^2-(P^*-p^*_i)^2}\,
N_d\,\Gamma_{\lambda_d}(p^*_i,{P}^*) \,{\cal C}\,
\bar{u}_i^T({\bf p}^*_i,\lambda)
\nonumber\\ 
&=& \left\{{\cal B}\,
\frac{m+\not{\!P}^*-\not{\!p}^*_i}{m^2-(P^*-p^*_i)^2}\,{\cal
B}^{-1}\right\}\left\{{\cal
B}\,N_d\,\Gamma_{\lambda_d}(p^*_i,{P^*})\,{\cal
B}^{-1}\right\}\left[{\cal B}\, {\cal C}\,\bar{u}_i^T({\bf
p}^*_i,\lambda)\right]
\nonumber\\ 
&=&\sum_{\lambda'}\frac{m+\not{\! P}
-\not{\! p}_i}{m^2-(P- p_i)^2}
\,N_d\,\Gamma_{\lambda_d}(p_i,P)\, {\cal C}\,\bar{u}_i^T({\bf
p}_i,\lambda')\;
d^{(1/2)}_{\lambda'\,\lambda}(\omega_i) 
\nonumber\\
&=&\sum_{\lambda'} \psi^{(i)}_{\lambda'\lambda_d}(p_i, P)\; 
d^{(1/2)}_{\lambda'\,\lambda}(\omega_i)\, ,
\end{eqnarray}
where, in the next to last step, we used the boost  properties
of the  helicity spinors (\ref{bstspina}) and 
fact that the propagator and $\Gamma_{\lambda_d}(p_i,P)$
are Lorentz scalars.  Note that there is no Wigner rotation of the
deuteron helicity vector because the boost is in the same direction
as its momentum (but the components of the vector $\xi_0$ do
change).  The wave function in the c.m.~frame can therefore be
written in terms of the rest frame wave function
\begin{equation}
\psi^{(i)}_{\lambda\lambda_d}(
p^*_i, P^*)=\sum_{\lambda'}{\cal B}^{-1}\,
\psi_{\lambda'\lambda_d}(p,{P})
\;d^{(1/2)}_{\lambda'\,\lambda}(\omega_i)\, . \label{wfboost}
\end{equation}
Using the representations 
(\ref{psi1a}) or (\ref{B11}), and the boost formula
(\ref{wfboost}) and (\ref{bstspin1}), the wave function in the
c.m.~frame becomes 
\begin{eqnarray}
\psi^{(i)}_{\lambda_i,\lambda_d}(p^*_i,P^*)
&=&\sqrt{{3\over8\pi}}
\sum_{\lambda'_j\lambda'_i}
{\cal B}^{-1}\,\left[u_i({\bf
p}_i,\lambda'_j)\,\phi^+_{|\Lambda|}({\rm p}_i) -  
2\lambda'_j\,\gamma^5\,u_i({\bf p}_i,\lambda'_j)\,
\phi^-_{|\Lambda|}({\rm p}_i)\right]
\nonumber\\ &&\qquad\qquad\qquad\times
d^{(1)}_{\mp\Lambda,\,\lambda_d}(\tilde\theta_i)
\;d^{(1/2)}_{\lambda'_i\,\lambda_i}(\omega_i)
\nonumber\\
&=& \sqrt{{3\over8\pi}}\sum_{\lambda_j'\lambda_i'\lambda} 
\left[u_i({\bf p}^*_i,\lambda)\,
\phi^+_{|\Lambda|}({\rm p}_i) -  
2\lambda'_j\,\gamma^5\,u_i({\bf p}^*_i,\lambda)\,
\phi^-_{|\Lambda|}({\rm p}_i)\right]\nonumber\\
&&\qquad\qquad\qquad\times
d^{(1)}_{\mp\Lambda,\,\lambda_d}(\tilde\theta_i)
\;d^{(1/2)}_{\lambda'_i\,\lambda_i}(\omega_i)
\;d^{(1/2)}_{\lambda'_j\,\lambda}(\omega_i)\, ,
\label{psib}
\end{eqnarray}
where $\Lambda=\lambda'_i+\lambda'_j$, $\tilde\theta_1=\theta_1$ and
$\tilde\theta_2=\theta_2-\pi$, and the upper(lower) sign in the
$\Lambda$ index of $d^{(1)}$ is for $i=1$(2).  Note that the
notation is mixed in the last equation : the  momentum of the spinors
is expressed in the c.m. frame and the variables of the
$\phi$'s and $d^{(1)}$ are in the deuteron rest frame.

%%%%%%%%%%%%%%%%%%%%%%%%%%%%%%%%%%%%%%%%%%%%%%%%%%%%%%%%%%%%%%%%%%%%%%

\section{The hadronic matrix element} 
\label{Appen:B}

\subsection{The plane wave matrix elements}

Using Eq.~(\ref{psib}), the current matrix
element (\ref{feynman}), in the c.m.~frame, becomes  
\begin{eqnarray}
&&\left< \lambda_1\lambda_2\left|
J_{\lambda_\gamma}(q)\right|\lambda_d\right>
=\sqrt{{3\over16\pi}}\,{1\over N_d}\sum_{\lambda_1'\lambda_2'\lambda}
\nonumber\\
&&\quad\Biggl\{
\bar u_1({\bf
p}_1^*,\lambda_1)\,j^{(1)}_{\lambda_\gamma}(p_1^*,p_1^*-q^*)  
\left[u_2({\bf p}^*_2,\lambda)\,
\phi^+_{|\Lambda|}({\rm p}_2) -  
2\lambda'_1\,\gamma^5\,u_2({\bf p}_2^*,\lambda)\,
\phi^-_{|\Lambda|}({\rm p}_2)\right]\,\Xi_2
\nonumber\\
&&\quad    -\bar u_2({\bf p}_2^*,\lambda_2)\, 
j_{\lambda_\gamma}^{(2)}(p^*_2,p^*_2-q^*)\,
\left[u_1({\bf p}_1^*,\lambda)\,
\phi^+_{|\Lambda|}({\rm p}_1) -  
2\lambda'_2\,\gamma^5\,u_1({\bf p}_1^*,\lambda)\,
\phi^-_{|\Lambda|}({\rm p}_1)\right]\,\Xi_1 \Biggr\} \, ,
\qquad
\label{curr1}
\end{eqnarray}
where
\begin{eqnarray}
\Xi_2&=&
d^{(1)}_{\Lambda,\,\lambda_d}(\theta_2-\pi)
\;d^{(1/2)}_{\lambda'_2\,\lambda_2}(\omega_2)
\;d^{(1/2)}_{\lambda'_1\,\lambda}(\omega_2)
\nonumber\\
\Xi_1&=& 
d^{(1)}_{-\Lambda,\,\lambda_d}(\theta_1)
\;d^{(1/2)}_{\lambda'_1\,\lambda_1}(\omega_1)
\;d^{(1/2)}_{\lambda'_2\,\lambda}(\omega_1)\, .
\label{curr2}
\end{eqnarray}
and $\Lambda$ and the Wigner rotation angles $\omega_i$ were defined
above.

The matrix elements of the single nucleon current operator, in
the c.m.~system, are defined to be 
\begin{eqnarray}
j^{(i)\,\rho}_{\lambda_i\,\lambda; \lambda_\gamma}
({\rm p},\theta^*,q_0)  =\cases{\bar u_i({\bf p}^*_i,\lambda_i)\,
j_{\lambda_\gamma}^{(i)} (p^*_i,p^*_i-q^*)\, 
u_j({\bf p}^*_j,\lambda) &  if $\rho=+$ \cr
\bar u_i({\bf p}^*_i,\lambda_i) \,j_{\lambda_\gamma}^{(i)}
(p^*_i,p^*_i-q^*) \,\gamma^5\, u_j({\bf p}^*_j,\lambda)
&if $\rho=-$ \, .}
\end{eqnarray}
These are calculated in the next section. Using this notation, the
current matrix elements (\ref{curr1}) can be written in the following
compact notation
\begin{eqnarray}
&&\left< \lambda_1\lambda_2\left|
J_{\lambda_\gamma}(q)\right|\lambda_d\right>=
(J_{\lambda_\gamma})^{\lambda_d}_{\lambda_1\,\lambda_2}
(p^*_1,p^*_2,q^*)\nonumber\\ &&\qquad=
\sqrt{{3\over16\pi}}\,{1\over N_d}\sum_{\lambda\,\rho}
\sum_{\lambda_1'\lambda_2'}
\Biggl\{\eta_\rho(2\lambda'_1)\,
j^{(1)\,\rho}_{\lambda_1\,\lambda; \lambda_\gamma}({\rm
p},\theta^*,q_0)\,
\phi^\rho_{|\Lambda|}({\rm p}_2)\,
d^{(1)}_{\Lambda,\,\lambda_d}(\theta_2-\pi)
\;d^{(1/2)}_{\lambda'_2\,\lambda_2}(\omega_2)
\;d^{(1/2)}_{\lambda'_1\,\lambda}(\omega_2)
\nonumber\\
&&\qquad\qquad-\eta_\rho(2\lambda_2')\,
j^{(2)\,\rho}_{\lambda_2\,\lambda;
\lambda_\gamma}({\rm p},\theta^*,q_0)\,
\phi^\rho_{|\Lambda|}({\rm p}_1)\,
d^{(1)}_{-\Lambda,\,\lambda_d}(\theta_1)
\;d^{(1/2)}_{\lambda'_1\,\lambda_1}(\omega_1)
\;d^{(1/2)}_{\lambda'_2\,\lambda}(\omega_1) \Biggr\} \, ,
\qquad
\label{curr3}
\end{eqnarray}
where $\eta_\rho(x)$ is the phase defined in
Eq.~(\ref{phase1}). The unpolarized cross section will be calculated
from this matrix element after the matrix elements of the nucleon
current have been  discussed.

%%%%%%%%%%%%%%%%%%%%%%%%%%%%%%%%%%%%%%%%%%%%%%%%%%%%%%%%%%%%%%%%%%%%%%%%%%%%%%
\subsection{The single nucleon current}

In the spectator formalism used in this paper, the $NN$ interaction
kernel has a form factor $h(p^2)$ ($h(m^2)=1$) attached to each
off-shell nucleon which enters of leaves the interaction. 
Alternatively, this form factor can be removed from the kernel and
attached to the nucleon propagators, which then have the form
\begin{equation}
\tilde{S}_F(p)=\frac{h^2(p^2)}{m-\not{\!p}}
\end{equation}
Gauge invariance \cite{RG} will be insured if we introduce a 
{\it reduced\/} nucleon current  $j_R^\mu(p',p)$
\begin{eqnarray}
j^\mu(p',p) = && h'j_R^\mu(p',p)h \,  , 
\label{jr}                    
\end{eqnarray}
where $h=h(p^2)$ and $h'=h(p'^2)$, which
satisfies the Ward-Takahashi identity using the dressed propagator
\begin{equation}
q_\mu j_R^\mu(p',p)= \tilde{S}^{-1}_F(p) - \tilde{S}^{-1}_F(p')  
\, . 
\end{equation}
A simple choice for the reduced current which satisfies this
identity is \cite{dff,RG,SG}
\begin{eqnarray}
j_R^\mu(p',p) = && F_0 (F_1(Q^2)-1) \tilde{\gamma}^\mu + 
F_0 F_2(Q^2)\,\frac{i \,\sigma^{\mu \nu}
q_\nu }{2m} + F_0\,\gamma^\mu \nonumber\\
&&+ G_0 (F_3(Q^2)-1) \, \Lambda_- (p') 
\tilde{\gamma}^\mu 
\Lambda_- (p) + G_0 \, \Lambda_- (p') 
\gamma^\mu 
\Lambda_- (p)  \,  ,  \label{d15a}                    
\end{eqnarray}
where $F_{1,2}(Q^2)$ are the on-shell nucleon form factors, $F_3(Q^2)$
is a completely unknown form factor describing the off-shell
structure of the nucleon (subject to the constraint that $F_3(0)=1$),
$\Lambda_- (p) =(m-\not\!\!p)/(2m)$, 
$\tilde{\gamma}^\mu=\gamma^\mu-q^\mu \rlap/q /q^2$ and $F_0$ and
$G_0$ are functions of $p^2$ and $p'^2$ completely determined by the
WT identity:
\begin{eqnarray}
F_0=&& \frac{1}{h'^2} \;\frac{m^2 - p'^2}{p^2 - p'^2} +
\frac{1}{h^2}\;
\frac{m^2 - p^2}{p'^2 - p^2} \nonumber\\
G_0=&&\left( \frac{1}{h'^2} - \frac{1}{h^2} \right) 
\frac{4 m^2}{p'^2 -p^2} \,  .  \label{d17}
\end{eqnarray}
Since the final nucleon is
always on-shell in the RIA approximation, $p'^2=m^2$ and the terms
multiplied by $\Lambda_-(p')$ vanish, giving
\begin{eqnarray}
j_R^\mu(p',p) = && \frac{1}{h^2}\left\{ F_1(Q^2) \tilde{\gamma}^\mu + 
F_2(Q^2)\,\frac{i \,\sigma^{\mu \nu}
q_\nu }{2m} + \frac{q^\mu\,
\rlap/q }{q^2} \right\} \,  . 
\label{ff2}                    
\end{eqnarray}
Furthermore, the terms proportional to $q^\mu$ vanish when the
current is contracted with the photon helicity vectors.  Hence, the
current for use in the RIA reduces to the traditional current divided
by $h$
\begin{eqnarray}
j^\mu(p',p) = \frac{1}{h(p^2)}\left\{ F_1(Q^2) \gamma^\mu + 
F_2(Q^2)\,\frac{i \,\sigma^{\mu \nu}
q_\nu }{2m} \right\}\qquad\left({\rm when}\; p'^2=m^2\right)\,  .
\label{ff3}                    
\end{eqnarray}
{\it Even when one of the particles is off-shell\/} the only
modification to the on-shell current which survives is the
appearance of the factor of $1/h$.   

The photon helicity states in the c.m.~frame, as defined in
Ref.~\cite{dg89}, are  
\begin{eqnarray}
\varepsilon_{\pm 1}&=& {1 \over \sqrt 2} \left( 0 , \mp 1 , -i , 0
\right)
\nonumber\\
\varepsilon_0&=& {1 \over Q} \left(q_0 , 0 , 0 , \nu_0\right) \, .
\end{eqnarray}
Hence, the single nucleon current operator, defined in
Eq.~(\ref{current}) and (\ref{ff3}), is
\begin{eqnarray}
j_{\lambda_\gamma}^{(i)} (p^*_i,p^*_i-q^*) = F_1(Q^2) 
\not\!\varepsilon_{_{\lambda_\gamma}} -F_2(Q^2)
\frac{\not\!\varepsilon_{_{\lambda_\gamma}}\!\not\!q}{2m}\, ,
\end{eqnarray}
where the factor of $1/h$ has been omitted (it is absorbed into
the wave function).  As discussed above and in Ref.~\cite{dg89}, the
results are simplified if we consider the symmetric and
anti-symmetric combinations of the transverse helicity amplitudes,
which are found from
\begin{eqnarray}
\varepsilon_{s}&=&
\frac{1}{2}\left(\varepsilon_{1}-\varepsilon_{-1}\right)=  
{1 \over\sqrt 2} \left( 0 , - 1 , 0 , 0\right)
\nonumber\\
\varepsilon_{a}&=& 
\frac{1}{2}\left(\varepsilon_{1}+\varepsilon_{-1}\right)=
{1 \over\sqrt 2} \left( 0 , 0 , -i , 0\right)\, . \label{B14}
\end{eqnarray}

Using the explicit form of the spinors given in Eqs.~(\ref{spinor1})
and (\ref{spinor2}) we can show that the current matrix elements have
the form
\begin{eqnarray}
j^{(i)\,\rho}_{\lambda'\,\lambda;\, g}({\rm
p},\theta^*,q_0)=\cases {\delta_{\lambda',\;-\lambda}\;
j_{1g}^{(i)\rho} +
2\lambda'\,\delta_{\lambda',\lambda}\;j_{2g}^{(i)\rho} & for
$\{\rho,g\}=(+,0),(+,s),(-,a)$  \cr 
2\lambda'\,\delta_{\lambda',\;-\lambda}\;
j_{1g}^{(i)\rho} + \delta_{\lambda',\lambda}\;j_{2g}^{(i)\rho} & for
$\{\rho,g\}=(-,0),(-,s),(+,a)$\, .\cr} \label{curent4}
\end{eqnarray}
where $g=\{0,s,a\}$ replaces the helicity.
If the proton (particle 1) matrix elements are
calculated first, the neutron elements follow from 
\begin{eqnarray}
j^{(2)\,+}_{\lambda'\,\lambda;\, g}({\rm
p},\theta^*,q_0)&=&\bar u_2({\rm p}, \lambda',\theta^*+\pi)\,
j_{g}^{(2)} (p^*_2,p^*_2-q^*)\,u_1({\rm p},\lambda,\theta^*) 
\nonumber\\
&=&-
4\lambda'\lambda\,\bar u_1({\rm p}, \lambda',\theta^*)\, 
{\cal R}_\pi 
j_{g}^{(2)} (p^*_2,p^*_2-q^*)\,
{\cal R}^{-1}_\pi\,u_2({\rm p},\lambda,\theta^*+\pi)
\nonumber\\ &=&-
4\lambda'\lambda\,\eta'(g)\,\bar u_1({\rm p}, \lambda',\theta^*)\, 
j_{g}^{(2)} (p^*_1,p^*_1-\hat q^*)
\,u_2({\rm p},\lambda,\theta^*+\pi) 
\nonumber\\ &=&
-4\lambda'\lambda\,\eta'(g)\,j^{(1)\,+}_{\lambda'\,\lambda;\, g}({\rm
p},\theta^*,-q_0)
\, ,
\end{eqnarray}
where the rotation by $\pi$ about the $\hat y$ axis has changed
$p^*_2\to p^*_1$, $q^*\to \hat q^*=(\nu_0,0,0,-q_0)$, $\varepsilon_0
\to -(-q_0,0,0,\nu_0)$ [so that the effect on the $g=0$ amplitude
is to change the phase and to change $q_0\to -q_0$], and
therefore the phase $\eta'(g)$, arising from the rotation of the
photon helicity vectors, is negative (for
$g=0$ or $s$) or positive (for $g=a$).  The change in the operator
$j_g^{(2)} \to j_g^{(1)}$ corresponds to the replacement of the
neutron form factors with proton form factors.  The effect of the
phase $-4\lambda'\lambda$ is an additional change in the sign of all
$j_2$ type elements.  Combining all of these effects allows us to
obtain $j^{(2)\,+}_{\lambda'\,\lambda;\, g}({\rm p},\theta^*,q_0)$
from $j^{(1)\,+}_{\lambda'\,\lambda;\, g}({\rm p},\theta^*,q_0)$ by
changing proton to neutron form factors, $q_0\to-q_0$, $j_{1(0,s)}\to
-j_{1(0,s)}$, $j_{2(0,s)}\to j_{2(0,s)}$, $j_{1a}\to j_{1a}$, and
$j_{2a}\to -j_{2a}$.  These phase changes are recorded through the
factor $\delta$ shown in Table~\ref{tab:current}. 

%%%%%%%%%%%%%%%%%%%%%%%%%%%%%%%%%%%%%%%%%%%%%%%

\subsection{The hadronic structure functions}

Substituting the form (\ref{curent4}) for the current into the
expressions (\ref{curr3}) allows the sum over $\lambda$ to be carried
out.  If $g=0$ or $s$ the result is
\begin{eqnarray}
&&(J_g)^{\lambda_d}_{\lambda_1\,\lambda_2}
(p^*_1,p^*_2,q^*)\nonumber\\ 
&&=\sqrt{{3\over16\pi}}\,{1\over N_d}\sum_{\rho}
\sum_{\lambda_1'\lambda_2'}
\Biggl\{\eta_\rho(4\lambda_1\lambda_1')\left[
j_{1g}^{(1)\rho}\,\phi^\rho_{|\Lambda|}({\rm p}_2)\,
d^{(1/2)}_{\lambda'_1\,-\lambda_1}(\omega_2) +
2\lambda_1\,
j_{2g}^{(1)\rho}\,\phi^\rho_{|\Lambda|}({\rm p}_2)\,
d^{(1/2)}_{\lambda'_1\,\lambda_1}(\omega_2)\right]\,{\cal Y}_2
\nonumber\\ 
&&\qquad\qquad-\eta_\rho(4\lambda_2\lambda_2')\left[
j_{1g}^{(2)\rho}\,\phi^\rho_{|\Lambda|}({\rm p}_1)\,
d^{(1/2)}_{\lambda'_2\,-\lambda_2}(\omega_1) +
2\lambda_2\,
j_{2g}^{(2)\rho}\,\phi^\rho_{|\Lambda|}({\rm p}_1)\,
d^{(1/2)}_{\lambda'_2\,\lambda_2}(\omega_1)\right]\,{\cal Y}_1 
\Biggr\} \, ,\qquad\qquad
\label{curr4}
\end{eqnarray}
where 
\begin{eqnarray}
{\cal Y}_2&=&d^{(1/2)}_{\lambda'_2\,\lambda_2}(\omega_2)\,
d^{(1)}_{\Lambda,\,\lambda_d}(\theta_2-\pi) 
\nonumber\\
{\cal Y}_1&=&d^{(1/2)}_{\lambda'_1\,\lambda_1}(\omega_1)\,
d^{(1)}_{-\Lambda,\,\lambda_d}(\theta_1) \, .
\end{eqnarray}
If $g=a$ the phase $2\lambda_1$ in the first square bracket
multiplies $j_{1g}^{(1)\rho}$ instead of $j_{2g}^{(1)\rho}$, and the
phase $2\lambda_2$ in the second square bracket
multiplies $j_{1g}^{(2)\rho}$ instead of $j_{2g}^{(2)\rho}$.  This
different phase insures that there is no interference between
$J_a$ and the other components of the current; $J_a$
contributes only quadratically in the term $J_a\,J_a^\dagger$.  

The unpolarized hadronic structure functions are now obtained by
squaring the current (\ref{curr4}), summing over final hadron
helicities and averaging over the initial deuteron helicity, and
multiplying by the kinematic factors given in
Eq.~(\ref{bornanalytic}).  The structure functions $R^{{\rm (II)}}$
are identically zero in the RIA.  The others are proportional to 
\begin{eqnarray}
\left<J_g\,J_{g'}^\dagger\right>
&=&\sum_{\lambda_1\lambda_2\lambda_d} 
\,(J_{g})^{\lambda_d}_{\lambda_1\,\lambda_2}
(p^*_1,p^*_2,q^*)
(J_{g'}^\dagger)^{\lambda_d}_{\lambda_1\,\lambda_2}
(p^*_1,p^*_2,q^*) 
\end{eqnarray}
This generates three terms: the proton contribution
(proportional to $[j^{(1)}]^2$), the neutron
contribution (proportional to $[j^{(2)}]^2$), and an interference
term (proportional to $j^{(1)}\times j^{(2)}$).  

The proton and neutron terms simplify easily.  The sum over
$\lambda_d$ and $\lambda_2$ (for the proton term) or $\lambda_1$ (for
the neutron term) collapses the sum over $\lambda'_1$ and
$\lambda'_2$ (associated with $J_g$ and the sum over
$\lambda''_1$ and $\lambda''_2$ (associated with 
$J^\dagger_{g'}$) reducing the ``diagonal'' terms to
\begin{eqnarray}
\left<J_g\,J_{g'}^\dagger\right>\Biggr|_i
&=&\frac{3}{{16\pi\,N_d^2}}\sum_{\stackrel{\lambda'_1\lambda'_2}
{\lambda}} 
\Bigg\{\left[-4\lambda'_1\lambda\,j_{1g}^{(i)+}\,
\phi^+_{|\Lambda|}({\rm p}_j)+j_{1g}^{(i)-}\,\phi^-_{|\Lambda|}({\rm
p}_j)\right]\,
d^{(1/2)}_{-\lambda'_1\,\lambda}(\omega_j)\nonumber\\
 &&\qquad\quad + \left[
2\lambda\,j_{2g}^{(i)+}\,\phi^+_{|\Lambda|}({\rm p}_j)-2\lambda'_1
\,j_{2g}^{(i)-}\,\phi^-_{|\Lambda|}({\rm p}_j)
\right]\,d^{(1/2)}_{\lambda'_1\,\lambda}(\omega_j) \Bigg\}\times
\Bigg\{g\to g'\Bigg\}\qquad\quad
\end{eqnarray}
where $i=1$ (for the proton) or 2 (for the neutron), $j=1$ or 2
(but $j\ne i$), and we used
$d^{(1/2)}_{\lambda'_1\,-\lambda}(\omega_i)
=-4\lambda'_1\lambda\,d^{(1/2)}_{-\lambda'_1\,\lambda}(\omega_i)$. 
Next, using the identities
\begin{eqnarray}
\sum_{\lambda} d^{(1/2)}_{-\lambda'_1\,\lambda}(\omega_j)
d^{(1/2)}_{-\lambda'_1\,\lambda}(\omega_j) &=& 1 =
\sum_{\lambda} d^{(1/2)}_{\lambda'_1\,\lambda}(\omega_j)
d^{(1/2)}_{\lambda'_1\,\lambda_1}(\omega_j) 
\nonumber\\
\sum_{\lambda} 2\lambda\,
d^{(1/2)}_{a\,\lambda}(\omega_j)
d^{(1/2)}_{b\,\lambda}(\omega_j) &=&
d^{(1/2)}_{-a\,b}(\pi-2\omega_j) =  2b\, d^{(1/2)}_{a\,b}(2\omega_j)
\nonumber\\
\sum_{\lambda}
d^{(1/2)}_{\lambda'_1\,\lambda}(\omega_j)
d^{(1/2)}_{-\lambda'_1\,\lambda}(\omega_j) &=& 
0 \, , 
\end{eqnarray}
and $(2\lambda)^2=1$, gives 
\begin{eqnarray}
\left<J_g\,J_{g'}^\dagger\right>\Biggr|_i
&=&\frac{3}{16\pi\,N_d^2}\sum_{\lambda'_1\lambda'_2} 
\Bigg\{J^{(i)+}_{gg'}
\left[\phi^+_{|\Lambda|}({\rm p}_j)\right]^2
+ J^{(i)-}_{gg'}
\left[\phi^-_{|\Lambda|}({\rm p}_j)\right]^2 
\nonumber\\
&&\qquad\qquad + \left(J^{(i)c}_{gg'}\,\cos \omega_j 
 + J^{(i)s}_{gg'}\,\sin \omega_j\right)
\left[\phi^+_{|\Lambda|}({\rm p}_j)\,
\phi^-_{|\Lambda|}({\rm p}_j)\right]
\Bigg\} \qquad
\end{eqnarray}
where the currents $J_{gg'}$ were given in Eq.~(\ref{Bigjs}).  Since
these currents do not depend on the helicities, we may complete the
sums using
\begin{eqnarray}
\sum_{\lambda'_1\lambda'_2}
\phi^\rho_{|\Lambda|}({\rm p}_j)\,
\phi^{\rho'}_{|\Lambda|}({\rm p}_j)=2\left[
\phi^\rho_{0}({\rm p}_j)\,
\phi^{\rho'}_{0}({\rm p}_j) + \phi^\rho_{1}({\rm p}_j)\,
\phi^{\rho'}_{1}({\rm p}_j)\right]\, .
\end{eqnarray}
This gives the result reported in Eq.~(\ref{bornanalytic}).

The interference term does not simplify as nicely.  The sums over
$\lambda_d, \lambda_1$, and $\lambda_2$ can be carried out, but there
are no delta functions to collapse the remaining four sums.  The
result is
\begin{eqnarray}
\left<J_g\,J_{g'}^\dagger\right>\Biggr|_{12}
&=&\frac{3}{{16\pi \, N_d^2}}\sum_{\stackrel{\lambda'_1\lambda'_2}
{\lambda''_1\lambda''_2}}
d^{(1)}_{\Lambda\,-\Lambda'}(\theta_2-\pi-\theta_1) \, \left[   
J_{g}^{(1)}\,J_{g'}^{(2)} + J_{g'}^{(1)}\,J_{g}^{(2)} \right]
\end{eqnarray}
where $\Lambda=\lambda'_1+\lambda'_2$,
$\Lambda'=\lambda''_1+\lambda''_2$, and
\begin{eqnarray}
J_{g}^{(1)}&=& \left[-2\lambda'_1\,j_{1g}^{(1)+}\,
d^{(1/2)}_{\lambda'_1\,\lambda''_1}(\delta_+)
+j_{2g}^{(1)+}\,
d^{(1/2)}_{-\lambda'_1\,\lambda''_1}(\delta_+)\right]\,
\phi^+_{|\Lambda|}({\rm p}_2) \nonumber\\
&&+\left[ j_{1g}^{(1)-}\,
d^{(1/2)}_{-\lambda'_1\,\lambda''_1}(\delta_-)-
2\lambda_1'\,j_{2g}^{(1)-}\,
d^{(1/2)}_{\lambda'_1\,\lambda''_1}(\delta_-) \right] \,
\phi^-_{|\Lambda|}({\rm p}_2) 
\nonumber\\
J_{g'}^{(2)}&=& \left[-2\lambda''_2\,j_{1g'}^{(2)+}\,
d^{(1/2)}_{\lambda''_2\,\lambda'_2}(\delta_+)
+j_{2g'}^{(2)+}\,
d^{(1/2)}_{-\lambda''_2\,\lambda'_2}(\delta_+)\right]\,
\phi^+_{|\Lambda'|}({\rm p}_1) \nonumber\\
&&+\left[ j_{1g'}^{(2)-}\,
d^{(1/2)}_{-\lambda''_2\,\lambda'_2}(-\delta_-)-
2\lambda''_2\,j_{2g'}^{(2)-}\,
d^{(1/2)}_{\lambda''_2\,\lambda'_2}(-\delta_-) \right] \,
\phi^-_{|\Lambda'|}({\rm p}_1) \, .
\end{eqnarray}
%

%%%%%%%%%%%%%%%%%%%%%%%%%%%%%%%%%%%%%%%%%%%%%%%%%%%%%%%%%%%%%%%%%%%%%%%%%%%%%%

\section{kinematic singularity in the lab cross sections} 
\label{Appen:C} 

In Sec.~\ref{Sec:kin} it was shown that, if $x>1$, the lab angle,
$\theta_1$, reaches a maximum value in the first quadrant, leading to
the condition (\ref{crit1}).  In this Appendix we show that
this condition generates a true singularity in the lab cross section
[suggested by Eq.~(\ref{crit2})], but that, because of the finite
resolution of any detector, all observable cross sections remain
finite.

Mathematically, this singularity arises from a zero in the recoil
factor ${\cal R}$ defined in Eq.~(\ref{defrecoil}). The denominator of
Eq.(\ref{defrecoil}) vanishes if
\begin{equation}
  E_W\,{\rm p}_1 = E_1\, q_{_L} \cos \theta_{1}   \,  ,
\label{singcond}
\end{equation}
where $E_W=M_d+\nu=\sqrt{W^2+q_{_L}^2}$ is the energy of the final
$np$ pair in the lab frame. We will first show that the condition that
this denominator vanish is identical to the condition (\ref{crit1}).

To see this, it is convenient differentiate $\cos\theta_1$ with
respect to $\theta^*$.  Using Eq.~(\ref{p1and2}) gives
\begin{eqnarray}
{d\cos\theta_1\over d\theta^*}= 
{d\over d\theta^*}\left({{\rm p}_1^z\over {\rm p}_1}\right) &=& {{\rm
p}E_W\sin\theta^*\over {\rm p}_1 W} - {{\rm p}_1^z\over 2{\rm p}^3_1}
\left[{E_W\over W} q_{_L}{\rm p}\sin\theta^*+{q_{_L}^2\over
W^2} 2{\rm p}^2\cos\theta^*\sin\theta^*\right]\nonumber\\
&=& {{\rm p}\sin\theta^*\over {\rm p}_1^2W} \left\{E_W{\rm
p}_1 - q_{_L}\cos\theta_1\left[{1\over2}E_W +
{q_{_L}\over W}{\rm p}\cos\theta^*\right]\right\}
\nonumber\\
&=& {{\rm p}\sin\theta^*\over {\rm p}_1^2W} \left\{E_W{\rm
p}_1 - q_{_L}\cos\theta_1E_1\right\}\, , \label{singcond2}
\end{eqnarray}
where, in the last step we used
\begin{equation}
E_1={1\over2}E_W + {q_{_L}\over W}{\rm p}\cos\theta^*\,
,\label{etrans}
\end{equation}
easily obtained form the same boost that gave Eq.~(\ref{p1and2}). 
Equation (\ref{singcond2}) shows that the two conditions
(\ref{crit1}) and (\ref{singcond}) are equivalent (except when
$\sin\theta^*=0$, when there is no singularity).

%%%%%%%%%%%%%%%%%%%%%%%%%%
\begin{figure}
\vspace*{-0.5in}
\leftline{
\epsfysize=2.5in \epsfbox{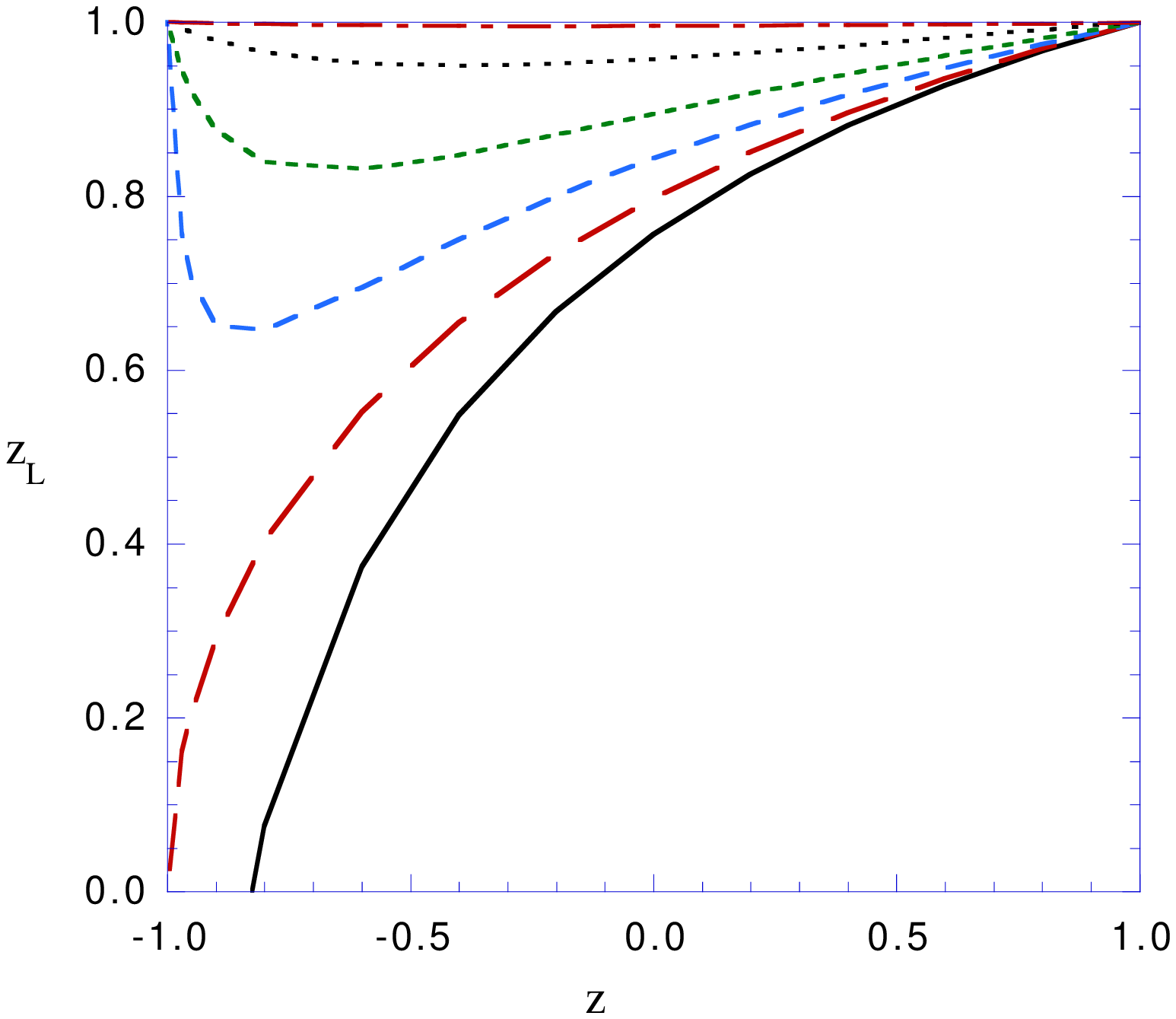}}
\vspace*{-2.5in}
\hspace*{3.0in}
\epsfysize=2.5in \epsfbox{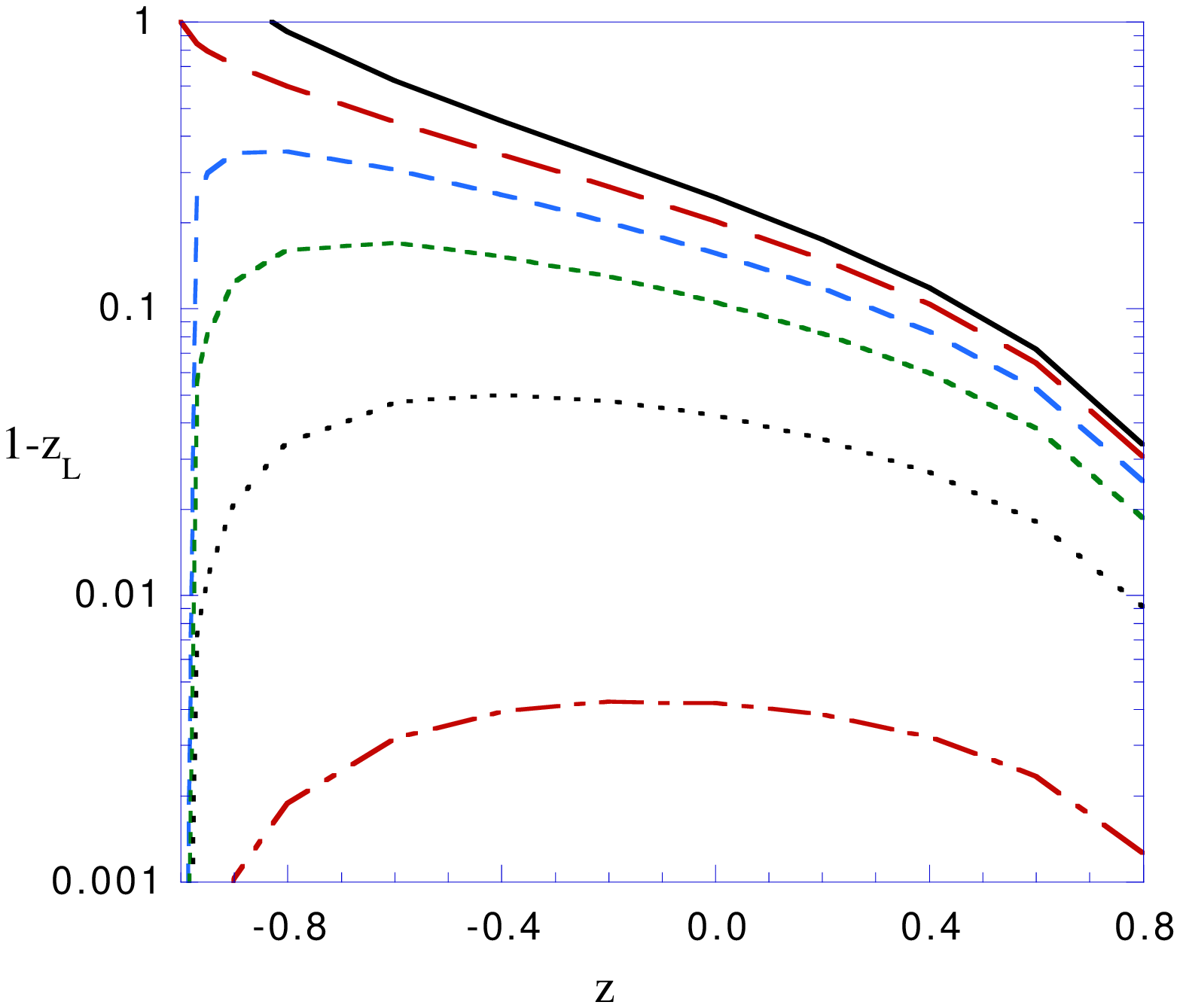}
\vspace*{0.2in}
\caption{\footnotesize\baselineskip=12pt The left panel shows
$z_{_L}=\cos\theta_{1}$ as a function of
$z=\cos\theta^*$  and the right panel shows $1-z_{_L}$.  In all cases
$Q^2=3$ GeV$^2$, $M_d=2$ GeV, $m=1$ GeV, and the lines show $x=0.5$
(solid),
$x=1.0$ (very long dashes), $x=1.25$ (long dashes), $x=1.5$ (dashed), $x=1.8$
(dotted), and $x=1.98$ (dot-dashed).}
\label{angle}
\end{figure}
%%%%%%%%%%%%%%%%%%%%%%%%%%% 
  
It is instructive to see how the cosine,  
\begin{equation}
z_{_L}=
\cos\theta_{1}=\frac{{\rm p}^z_{1}}
{\sqrt{{\rm p}^{z\,2}_{1}+{\rm p}^2\sin^2\theta^*}} ={{q_{_L}W}
+2{\rm p}E_W\,z\over\sqrt{q^2_{_L}W^2 + 4q_{_L}{\rm p}WE_W\,z + 4{\rm
p}^2 q_{_L}^2\,z^2 + 4{\rm p}^2W^2}}
\, ,
\label{cosine}
\end{equation} 
of the proton lab angle, $\theta_{1}$, varies with $z=\cos\theta^*$.
For fixed $x$ and $Q^2$, $z_{_L}$ depends only on
$z$  as given by Eq.~(\ref{cosine}).  As an example,
Fig.~\ref{angle} shows how $z_{_L}$ varies with
$z$ for selected  values of $x$ when $Q^2=3$.  The
value of $z$ at which
$z_{_L}$ is a minimum,  denoted $z_{crit}$ can be computed
from Eq.~(\ref{cosine}).  In the  approximation that $m_n = m_p = m$
we obtain  
\begin{equation}
z_{crit}=-\frac{\sqrt{(W^2+q_{_L}^2)(W^2-4m^2)}}{W q_{_L}}
\, .
\label{cosinemin}
\end{equation}  
Moreover, solving Eq.~(\ref{singcond}) for $z$ gives the {\it same\/}
value, and hence we recover again the observation that the singularity
occurs at the kinematic boundary of $\cos\theta_{1}$.  
  
The values of $E_1$, ${\rm p}_1$ and $z_{_L}$ at the critical
point can be found using the additional constraint that follows
from energy conservation.  Energy conservation gives the following
general formula for $z_{_L}$:
\begin{equation}
z_{_L} = \frac{q_{_L}^2+ m_n^2- m_p^2+ 2 E_{1} (M_d+\nu)- (M_d+\nu)^2}
 {2{\rm p}_{1} q_{_L}} =  \frac{2 E_W E_{1} - {\tilde W}^2 }{2{\rm
p}_{1} q_{_L} } \, ,
\label{protonang}
\end{equation}
where we have defined ${\tilde W}^2= W^2 - m_n^2 + m_p^2$.  For
fixed electron kinematics, Eqs.~(\ref{singcond}) and
(\ref{protonang}), taken together, give the proton energy $E_0$,
proton momentum
${\rm p}_0$ and angle $z_0$ at which the denominator
(\ref{singcond}) is singular:
\begin{equation}
E_0 = \frac{2 m_p^2 \, E_W}{{\tilde W}^2}  \ , \ \ 
{\rm p}_0 = \frac{m_p}{{\tilde W}^2} \sqrt{4 E_W^2\, m_p^2-{\tilde
W}^4}\ ,\ \ 
z_0=  \frac{\sqrt{4 E_W^2\, m_p^2-{\tilde W}^4}}{2 m_p q_{_L}} 
\ . \label{critvals}
\end{equation}
Of course, the value of $E_{0}$ must be physical, i.e. the
denominator  has to be positive and $E_{0} \geq m_p$. The first
condition leads to the  constraint 
\begin{equation}
 W^2 > m_n^2 - m_p^2 \, ,
\end{equation}
which is always satisfied, while the second one requires
\begin{equation}
 Q^2 \geq 2\nu (m_n- \epsilon_d)- \epsilon_d (2 m_n- \epsilon_d) \simeq
      \frac{Q^2}{x} \, , \label{ineq}
\end{equation} 
where $M_d= m_p+ m_n- \epsilon_d$ and in the last step we put
$m_n \simeq m_p \simeq m$, $\epsilon_d\simeq0$, and  $\nu= Q^2/(2 m
x)$. This is the mathematical proof that the differential cross
section is singular if and only if $x>1$.

We conclude this discussion by showing that, even though the cross 
section is singular, the physical observables obtained from it are
not.  Because {\it any\/} measuring apparatus must necessarily have a
{\it finite\/} resolution, all physical measurements must necessarily
{\it average\/} the differential cross section (\ref{crosssec}) over
this finite resolution.  At the kinematic boundary this average is 
\begin{equation}
\left<\frac{d^5\sigma}{d\Omega'dE'd\Omega_1}\right>_\epsilon\equiv
\frac{1}{z_0 \epsilon}
\int_{z_0}^{z_0(1+\epsilon)}
\left(\frac{d^5\sigma}{d\Omega'dE'd\Omega_1}\right)\, d z_{_L}
\, . \label{average}
\end{equation} 
This average will be finite only if the singularity is integrable, and
this will now be shown explicitly.  That it must be so follows from
general physical considerations, and also from the behavior of the
differential cross section in the c.m.~frame,  where there is a
smooth, non-singular behavior for all kinematical conditions.  In the
lab frame, for fixed  electron kinematics, the $z_{_L}$ dependent part
of the integrand is contained in the kinematic factors
\begin{equation}
{\rm p}_{1}\, {\cal R}\,\biggl\{\tilde{R}^{({\rm I})}_L +  \dots
\biggr\} = \frac{W {\rm p}_{1}^2}{E_W\,{\rm p}_{1} - E_{1}\, q_{_L}\,
z_{_L} }\,
\biggl\{\tilde{R}^{({\rm I})}_L +  \dots \biggr\}
\end{equation}
At first glance, if we assume that the factor ${\rm p}_{1}\,{\cal R}$
is an  analytical function of  $z_{_L}$, Eq.~(\ref{singcond})
suggests that the singularity  will be a simple pole, which is not
integrable. However, it is easy to show  that the dependence of
${\rm p}_{1}\, {\cal R}$ on $z_{_L}$ is not analytic. Begin by
using (\ref{etrans}) to rewrite Eq.~(\ref{cosine}) 
\begin{equation}
z_{_L}= \frac{a E_{1}- b}{{\rm p}_{1}} \ , \ a= \frac{E_W}{q_{_L}}
\ ,
\ b=
\frac{{\tilde W}^2}{2 q_{_L}} \, ,
\label{pu} 
\end{equation} 
and solve this equation for ${\rm p}_{1}$ as a function of $z_{_L}$
(recalling that $a$ and $b$ depend only on the electron variables and
are not functions of $z_{_L}$).  The result is
\begin{equation}
 {\rm p}_{1}= \frac{z_{_L}b + a \sqrt{b^2+
(z_{_L}^2-a^2)m_p^2}}{a^2-z_{_L}^2} =
         \frac{z_{_L} b + a m_p
\sqrt{z_{_L}^2-z_{0}^2}}{a^2-z_{_L}^2} \, , \label{p1sol}
\end{equation}
where we used (\ref{critvals}) to express $z_{0}$ in terms of $a$
and $b$
\begin{equation}
z_{0}= \frac{\sqrt{a^2m_p^2-b^2}}{m_p} \ .
\end{equation}
From (\ref{p1sol}) we obtain the following expression for the energy
$E_1$
\begin{equation}
E_1= \frac{m_pz_{_L}\sqrt{z_{_L}^2-z_{0}^2} + ba}{a^2-z_{_L}^2}\ .
\label{e1sol}
\end{equation}
Substituting (\ref{p1sol}) and (\ref{e1sol}) into the recoil factor we
obtain
\begin{equation}
{\rm p}_{1}\, {\cal R}\,= \frac{W {\rm p}_{1}^2}{E_W\,{\rm p}_{1} -
E_{1}\, q_{_L}\, z_{_L} }={W {\rm p}_{1}^2\over m_p\,
q_{_L}\sqrt{z_{_L}^2-z_{0}^2} }\, .
\end{equation}
From this it is clear that in the vicinity of the end point
singularity at $z_{0}$, the cross section Eq.~(\ref{average})
behaves as $1/\sqrt{z_{_L}^2-z_{0}^2}$ and the singularity is
integrable.

%\end{document}

%-----------------------------------------------------------------  

\end{document}